\documentclass[aps,showpacs,preprintnumbers,amsmath,amssymb,twocolumn,superscriptaddress,floatfix,nofootinbib,prx,longbibliography]{revtex4-1}

\usepackage{amsfonts}
\usepackage{amsmath}
\usepackage{amssymb}
\usepackage{algpseudocode}
\usepackage[linesnumbered,lined,boxed,commentsnumbered,ruled,vlined]{algorithm2e}
\usepackage{bm,bbm}
\usepackage{booktabs}
\usepackage{caption}
\DeclareCaptionJustification{justified}{\justifying}
\usepackage{centernot}
\usepackage{colortbl}
\usepackage{comment}
\usepackage{dcolumn}
\usepackage{epstopdf}
\usepackage{float}
\usepackage[bottom]{footmisc}
\usepackage{graphicx}
\usepackage{hyperref}
\hypersetup{colorlinks=true,citecolor=orange,urlcolor=blue,linkcolor=magenta}
\usepackage[capitalise]{cleveref}
\usepackage{mathrsfs}
\usepackage{mathtools}
\usepackage{multirow}
\usepackage{physics}
\usepackage{ragged2e}
\usepackage{subfigure}
\usepackage{tabularx}
\usepackage{tikz}
\usepackage{ifthen}
\usepackage{tikz-network}
\usepackage{verbatim}
\usepackage{xcolor}
\usepackage{xkcdcolors}

\newcommand{\beq}{\begin{equation}}
\newcommand{\eneq}{\end{equation}}

\definecolor{ACol}{HTML}{1F77B4}   
\definecolor{BCol}{HTML}{2CA02C}   
\definecolor{CCol}{HTML}{D62728}   
\definecolor{DCol}{HTML}{9467BD}   
\definecolor{DMNCol}{HTML}{FF7F0E} 

\newcommand{\LabelColor}[1]{%
  \ifthenelse{\equal{#1}{A}}{ACol}{%
  \ifthenelse{\equal{#1}{B}}{BCol}{%
  \ifthenelse{\equal{#1}{C}}{CCol}{%
  \ifthenelse{\equal{#1}{D}}{DCol}{%
  \ifthenelse{\equal{#1}{DMN}}{DMNCol}{black}}}}}%
}

\begin{document}

\title{Assessing imbalance in signed brain networks}

\author{Marzio Di Vece}
\email{marzio.divece@sns.it}
\affiliation{Scuola Normale Superiore, P.zza dei Cavalieri 7, 56126 Pisa (Italy)}
\affiliation{IMT School for Advanced Studies, P.zza San Francesco 19, 55100 Lucca (Italy)}
\author{Emanuele Agrimi}
\affiliation{IMT School for Advanced Studies, P.zza San Francesco 19, 55100 Lucca (Italy)}
\affiliation{Institute for Advanced Study, University of Amsterdam, Oude Turfmarkt 145-147, 1012 GC Amsterdam (The Netherlands)}
\affiliation{Korteweg-de Vries Institute for Mathematics, University of Amsterdam, Science Park 105-107, 1098 XG Amsterdam (The Netherlands)}
\author{Samuele Tatullo}
\affiliation{IMT School for Advanced Studies, P.zza San Francesco 19, 55100 Lucca (Italy)}
\author{Tommaso Gili}
\affiliation{IMT School for Advanced Studies, P.zza San Francesco 19, 55100 Lucca (Italy)}
\author{Miguel Ib\'a\~nez-Berganza}
\affiliation{IMT School for Advanced Studies, P.zza San Francesco 19, 55100 Lucca (Italy)}
\affiliation{Istituto Nazionale di Alta Matematica `Francesco Severi' (INdAM-GNAMPA), P.le Aldo Moro 5, 00185 Rome (Italy)}
\author{Tiziano Squartini}
\affiliation{Scuola Normale Superiore, P.zza dei Cavalieri 7, 56126 Pisa (Italy)}
\affiliation{IMT School for Advanced Studies, P.zza San Francesco 19, 55100 Lucca (Italy)}
\affiliation{Istituto Nazionale di Alta Matematica `Francesco Severi' (INdAM-GNAMPA), P.le Aldo Moro 5, 00185 Rome (Italy)}

\date{\today}

\begin{abstract}
Many complex systems - be they financial, natural, or social - are composed of units - such as stocks, neurons, or agents - whose joint activity can be represented as a multivariate time series. An issue of both practical and theoretical importance concerns the possibility of inferring the presence of a \emph{static} relationship between any two units solely from their \emph{dynamic} behaviour. The present contribution aims at tackling such an issue within the framework of traditional hypothesis testing: Briefly speaking, our suggestion is that of linking any two units if behaving in a sufficiently similar way. To achieve such a goal, we project a multivariate time series onto a signed graph by (1) comparing the empirical properties of the former with those expected under a suitable benchmark and (2) linking any two units with a positive (negative) edge in case the corresponding series shares a significantly large number of concordant (discordant) values. To define our benchmarks, we adopt an information-theoretic approach that is rooted into the constrained maximisation of Shannon entropy, a procedure inducing an ensemble of multivariate time series that preserves some of the empirical properties on average, while randomising everything else. We showcase the possible applications of our method by addressing one of the most timely issues in the domain of neurosciences, i.e. that of determining whether brain networks are frustrated or not, and, if so, to what extent. As our results suggest, this is indeed the case, with the major contribution to the underlying negative subgraph coming from the subcortical regions (and, to a lesser extent, from the limbic ones). At the mesoscopic level, the minimisation of the Bayesian information criterion, instantiated with the signed stochastic block model, reveals that brain regions gather into modules aligning with the statistical variant of the relaxed balance theory.
\end{abstract}


\maketitle

\section{Introduction}\label{sec:I}

Time series are ubiquitous and versatile representations of dynamical complex systems, be they financial, natural, or social. A major challenge is that of inferring the underlying dependencies between units from the (often non-linear) dynamics characterising them. Being capable of transforming a time series into a symbolic sequence, while preserving essential information about the `original' generative process, is of practical relevance, as the \emph{dynamical} properties of the former can be revealed by focusing on the \emph{structural} properties of the latter~\cite{Onnela2004,Sinha2010,Bouchaud2003}. To this aim, sequential data are often represented as traditional graphs, with \emph{nodes} replacing individual units (such as stocks, neurons, or agents) and \emph{edges} reflecting their activity-driven interactions.

This choice plays a major role in the neuroscientific domain, where network-based approaches have revolutionised our understanding of brain functioning, especially through the study of functional connectivity (FC) networks derived from Blood Oxygen Level-Dependent (BOLD) functional Magnetic Resonance Imaging (fMRI) signals. Rooted in graph theory and statistical physics, this perspective, often referred to as \emph{network neuroscience}, aims at revealing the organisational principles of cognition and behaviour~\cite{kulkarni2024principlesbrainnetworkorganization} by modelling the brain as an integrated system of interacting regions~\cite{bassettnetwork2017}. 

In the following subsection, a short overview of the approaches that have been proposed so far to extract relationships between nodes directly from the time series of the original units is provided: in Sec. \hyperlink{sec:IA}{IA}, methods defining an intermediate proximity matrix are discussed; in Secs. \hyperlink{sec:IB}{IB} and \hyperlink{sec:IC}{IC}, methods that map univariate and multivariate time series onto graphs without a proper validation procedure are discussed; in Sec. \hyperlink{sec:ID}{ID}, the need to approach such an issue within the framework of traditional hypothesis testing is motivated.

\subsection{From time series to proximity and from proximity to graphs}\label{sec:IA}

Given a set of time series, the most common approach is that of extracting a \emph{proximity matrix} that is subsequently used for defining a graph. Although several definitions of proximity have been proposed\footnote{We redirect the interested reader to the reviews~\cite{Cliff2023,Liu2024}, where hundreds of proximity measures (such as the dynamic time warping, the mutual information and the transfer entropy) are discussed and compared.}, the commonest choice is identifying it with the notion of \emph{pairwise correlation}, as returned by the calculation of the Pearson correlation coefficient

\begin{equation}\label{eq:1-Pearson Correlation}
r_{ij}=\frac{\sum_{t=1}^T(X_{it}-\underline{X}_i)\times(X_{jt}-\underline{X}_j)}{\sqrt{\sum_{t=1}^T(X_{it}-\underline{X}_i)^2}\times\sqrt{\sum_{t=1}^T(X_{jt}-\underline{X}_j)^2}},
\end{equation}
where

\begin{equation}
\underline{X}_i=\frac{\sum_{t=1}^TX_{it}}{T}\quad\text{and}\quad\underline{X}_j=\frac{\sum_{t=1}^TX_{jt}}{T}
\end{equation}
represent the (sample) average of series $X_i=\{X_{it}\}_{t=1\dots T}$ and $X_j=\{X_{jt}\}_{t=1\dots T}$.

A less popular, but still worth mentioning, definition of proximity is based on the notion of \emph{partial correlation}~\cite{Hampson2002,Fransson2008}. If we keep indicating the matrix of pairwise correlations with $\mathbf{R}$, the generic entry of the matrix of partial correlations, then, reads $\hat{J}_{ik}=J_{ik}/\sqrt{J_{ii}\times J_{kk}}$, with $\mathbf{J}=\mathbf{R}^{-1}$ being the so-called \emph{precision matrix}. Estimations of the FC based on partial correlations are affected by a smaller inter-subject variability~\cite{Brier2015,Rahim2019}, as they have been shown to resemble more closely the so-called structural connectivity patterns~\cite{Huang2010,Salvador2005,Lee2011,Santucci2025}.\\

\noindent\textit{Fixed thresholding algorithms.} The simplest approach of this kind is based on interpreting a correlation matrix (be it partial or not) as a fully-connected, weighted matrix. Non-trivial structures can be, thus, obtained by implementing sparsification procedures, the simplest of which is thresholding correlations~\cite{Bullmore2009}. This procedure ultimately boils down to remove the weaker connections - usually the most affected by experimental noise~\cite{Rubinov2010} - while retaining the stronger ones; still, as reported by the authors of~\cite{MacMahon2015}, thresholding may prevent the detection of a modular organisation, as the correlations within modules may be cut out even if significantly stronger than those between modules.

In the financial literature, the network resulting from the application of a threshold is also known as the \emph{asset graph}~\cite{Sinha2010,Onnela2004,Heimo2009}: in~\cite{Onnela2004}, the authors derived an asset graph from the correlation matrix of a set of stocks by filling an empty graph, edge by edge, starting from the one corresponding to the largest correlation and stopping at the $N$-th largest correlation (with $N$ being the number of nodes); upon doing so, the considered asset graph is comparable in size with the most popular \emph{asset tree}, although being less constrained by requests such as the one of forbidding cycles.

Since letting the network density drive the thresholding procedure appears as arbitrary as setting a threshold directly on correlations, the proposal to determine the resulting network by requiring it to remain connected has been advanced: Such a data-driven procedure removes the weaker correlations until the value at which the resulting network disconnects is reached~\cite{Alexander-Bloch2010,Gallos2012}. Although this kind of analysis has been popularised with the name of \emph{percolation analysis}, the aforementioned value is, technically speaking, the \emph{connectivity threshold}.

If the previous approach prescribes to stop pruning correlations once the connectivity threshold is reached, a second approach focuses on the mesoscopic organisation of the network and prescribes to consider a correlation-induced distance, draw the corresponding dendrogram\footnote{A dendrogram is constructed by clustering objects in a hierarchical fashion: to this aim, a distance between clusters is needed. Some of these `linkages' algorithms induce an ultra-metric distance on the data~\cite{Rammal1986} that has been argued to induce distances progressively less similar to the original ones as higher levels of the taxonomic tree are climbed. According to this intuition, such a method is more reliable to determine the low-level structure of the tree (induced by the strongest correlations) than the high-level one (induced by the weakest correlations)~\cite{MacMahon2015,Milligan1979}.} and cut it in correspondence of a properly-defined threshold~\cite{Mantegna1999,Tola2008}: a possible criterion to individuate it is that of maximising the modular character of the resulting network structure, analysed upon running one of the many algorithms that have been proposed to this aim (Louvain, Infomap, Surprise, etc.)~\cite{Newman2006,Bifone2017,Marchese2022}.\\

\noindent\textit{Filtering algorithms.} A different class of algorithms attempts to filter a correlation matrix by letting its values `dress' peculiar graph topologies, supposedly isolating the most relevant correlations - a threshold is, thus, still individuated, although only indirectly.

The most popular filtering approach returns the so-called \emph{Maximum Spanning Tree} (MST), a technique retaining the $N-1$ largest correlations while ensuring that \emph{i)} loops are discarded; \emph{ii)} each node is reachable from any other node via a path that is the widest between its endpoints; \emph{iii)} the total weight is greater than, or equal to, the total weight of any other spanning tree. This method can be applied upon running the Kruskal's algorithm and is often employed to distinguish the connector regions (i.e. the most central ones) from the provincial regions (i.e. the most peripheral ones)~\cite{Rammal1986,Mantegna1999,Caldarelli2012,Mastrandrea2017,Mastrandrea2021}.

The Kruskal's algorithm can be slightly modified by adding the extra condition that no edge between any two nodes that are already connected is allowed. Upon doing so, a \emph{Maximum Spanning Forest} (MSF) is obtained, i.e. a set of disconnected trees usually interpreted as the backbone of a set of communities. Both the MST and the MSF are rather different from their random counterparts~\cite{Bardella2016}, their structural organisation providing a topological signature of the emergence of diseases such as schizophrenia~\cite{Mastrandrea2021}.

An alternative approach, discarding less information, returns the so-called Planar Maximally Filtered Graph (PMFG)~\cite{Aste2005,Tumminello2005}. The criterion informing such a method prescribes to retain the correlations ensuring that the resulting network is a planar graph, i.e. a graph that can be drawn with no links crossing each other - more precisely, one in which edges may intersect only at common endpoints (nodes); since the PMFG always contains the MST, the former complements the latter by providing additional (and not just different) information. The PMFG has been also described as the simplest instance of a more general procedure embedding high-dimensional data into lower-dimensional manifolds with a controllable number of `holes' called `genus', the PMFG corresponding to the case in which the genus is zero~\cite{Tumminello2005}. Although versatile, it suffers from some degree of arbitrariness too, affecting the properties of the postulated, approximating structure: the value of the genus must, in fact, be fixed a priori; besides, there is no obvious reason why time series should find a natural embedding into a bidimensional plane.\\

\noindent\textit{Testing correlations.} A more principled approach to extract information from a correlation matrix is that of comparing (some of) its properties with the ones expected under a properly defined benchmark, the aim of this class of techniques being that of identifying non-random collective information about the set of time series induced by $\mathbf{C}$. Interestingly, these techniques do not discard negative correlations, which are known to be a consequence of spatial and temporal heterogeneities induced by non-trivial, neural-mediated hemodynamic mechanisms affecting the synchronisation of neural activity, thus contributing to the stability of resting-state brain networks~\cite{Saberi2021}. Although isolated attempts exist - such as the one proposed in~\cite{Salvador2005}, whose authors test the null hypothesis that the average value of the distribution of partial correlations, for each pair of regions and across subjects, is zero - methods of this kind can be split into those based on \emph{Random Matrix Theory}~\cite{Plerou2002,Utsugi2004,Potters2005,Zema2024} and those based on \emph{entropy maximisation}~\cite{Masuda2018,Kojaku2019}.
\
\subsection{From univariate time series to graphs}\label{sec:IB}

Different (classes of) approaches that seek to extract a network from one or more time series while ignoring (the intermediate step represented by) the proximity matrix\footnote{We redirect the interested reader to the reviews~\cite{Silva2021,Zou2019}, discussing the differences between the existing (classes of) algorithms and comparing their performances.}.\\

\noindent\textit{Visibility-induced mapping.} According to such a mapping, the time steps of a univariate time series are interpreted as nodes to be connected if `visible' to each other. Different types of visibility algorithms exist, designed to obtain undirected/directed, binary/weighted graphs.

The first algorithm of this kind was proposed in~\cite{Lacasa2008} and is based on the idea of representing each observation as a vertical bar whose height equals the numerical value of the observation itself: if the tops of any two bars can be \emph{directly} connected via a `visibility line', the corresponding nodes are, then, connected in the resulting network.

Several variants of the so-called \emph{visibility graph} (VG) have been proposed: in~\cite{Luque2009}, a \emph{horizontal visibility graph} (HVG) is constructed by considering (only) horizontal visibility lines - as a consequence, the HVG is always a subgraph of the VG, hence containing less information, although its construction is faster; in~\cite{Lacasa2012}, a \emph{directed horizontal visibility graph} (DHVG) is constructed by letting any pair of time steps $t_i<t_j$ induce an edge $i\rightarrow j$; in~\cite{Zhou2012}, a \emph{limited penetrable visibility graph} (LPVG) is constructed by connecting any two nodes if the visibility line joining the tops of the two associated bars crosses a number of `obstacles' that is smaller than, or equal to, a fixed value $L$ - naturally, the limited penetrable visibility graph (LPVG) reduces to the VG when $L=0$, i.e. when there are no `obstacles'; in~\cite{Supriya2016}, a \emph{weighted visibility graph} is constructed by enriching each edge of a DHVG with a weight matching the `view angle' of the observing node, its sign with respect to the horizontal axis indicating if the values of the time series increase (if positive) or decrease (if negative); in~\cite{Zhu2014}, a \emph{differential visibility graph} is constructed by removing the edge set of the HVG from the edge set of the VG.\\

\noindent\textit{Transition-induced mapping.} According to such a mapping, connections are established between (symbols assigned to) different portions of univariate time series, intended to represent states between which transitions occur: methods of this kind, in fact, represent a time series as a directed, weighted graph.

The mapping introduced in~\cite{Campanharo2011} is based on a partition of the distribution of the values assumed by a time series in quantiles, to each of which a symbol is assigned~\cite{Shirazi2009}; the mapping introduced in~\cite{Gao2009} is named \emph{Coarse-Grained Phase Space Graphs} and requires the definition of a time window $\omega$ to determine the number of values of the series to be represented as an $\omega$-dimensional vector, whose components are numerically determined by adopting a rule similar to the quantiles-based one; a third variant of the same kind of mapping is the one introduced in~\cite{Mutua2015} and named \emph{Visibility Graphlets Networks}: the time steps from $t$ to $t+\omega$ are, then, mapped to a DHVG.\\

\noindent\textit{Proximity-induced mapping.} According to such a mapping, connections are established between (groups of) time steps of a univariate time series, according to measures of distance/similarity between them. 

The first method of this kind was introduced in~\cite{Zhang2006} to represent pseudo-periodic time series. First, it segments a time series into cycles (e.g. dictated by a periodic behaviour) without requiring them to have the same length; then, correlations between pairs of cycles are computed - by shifting the shortest one, if needed - and the largest correlation value is compared against a threshold: if found to be larger, an edge between the two cycles, represented as nodes in the resulting network, is established.

An alternative class of methods is the one gathering the so-called \emph{Recurrence Networks} (RNs), which is based on the idea that the points of a time series are recurrent if their distance, in a properly defined embedding space, is `short enough': first, each time step is mapped to a point of an embedding space and, then, distances are computed by using one of the many norms that can be defined. The way the embedding space is analysed distinguishes RNs in three sub-categories, i.e. \emph{k-Nearest Neighbour Networks}~\cite{Small2009}, \emph{Adaptive Nearest Neighbour Networks}~\cite{Xu2008} and \emph{$\epsilon$-Recurrence Networks}~\cite{Donner2010}.

\subsection{From multivariate time series to graphs}\label{sec:IC}

Methods of this kind aim at extracting a network from a multivariate time series. While some of them output single-layer networks, each time series being represented as a node, others output multi-layer networks, each time series being mapped to a different layer by employing any of the methods reviewed in the previous section.\\

\noindent\textit{Single-layer mapping.} This first kind of mapping collects generalisations of the aforementioned RNs, hereby named \emph{Cross Recurrence Networks} (CRNs), \emph{Inter-System Recurrence Networks} (ISRNs), and \emph{Joint Recurrence Networks} (JRNs)~\cite{Feldhoff2012,Feldhoff2013,Romano2004}. CRNs are defined for pairs of time series that do not need to be of the same length~\cite{Marwan2002,Zbilut1998}; such a recipe can be further extended to deal with more than two time series: in this case, we have the so-called ISRNs, inducing adjacency matrices constituted by (matrices representing the) RNs along the diagonal and by (matrices representing the) CRNs off the diagonal~\cite{Feldhoff2012}. A third alternative is represented by the JRNs~\cite{Romano2004,Feldhoff2013}, inducing adjacency matrices that, similarly to what has been observed for the ISRNs, are block-wise, each diagonal block referring to a single time series and the off-diagonal blocks referring to pairs of time series.\\

\noindent\textit{Multiple-layer mapping.} This second kind of mapping, on the other hand, associates a multivariate time series with a multi-layer network. A first example of such a mapping is provided by~\cite{Lacasa2015}: here, a multiplex visibility graph (MVG) is obtained by mapping each univariate time series into a graph by employing a VG; afterwards, the MVG is aggregated into a single-layer weighted graph, each node representing a layer-specific VG and each weight proxying the similarity between the corresponding layers. The authors of~\cite{Eroglu2018} follow the same pipeline but replace VGs with RNs, thus employing \emph{Multiplex Recurrence Networks} to study paleoclimatic time series.

\subsection{From multivariate time series to\\validated graphs}\label{sec:ID}

The most recent approaches emphasise the importance of probabilistic (null, as well as generative) models to benchmark network features and disentangle fundamental, structural properties from statistical artifacts~\cite{VAROQUAUX2013,vasanull2022}. These models play a crucial role in understanding whether empirical features (e.g. a modular structure, a small-world architecture, a rich-club organisation) are indicators of a genuine degree of self-organisation or simple by-products of lower-order constraints\footnote{More explicitly, the practice of employing graph-theoretical estimations for inter-subject analyses without comparing them with a proper benchmark gives \emph{`[\dots] a fairly unspecific characterisation of the brain, being fragile to noise [\dots] Another caveat is that [\dots] correlation matrices display small-world properties [\dots] by construction [\dots] This observation highlights the need for well-defined null hypotheses [\dots] but also for controlled recovery of brain functional connectivity going beyond empirical correlation matrices [\dots]'}~\cite{VAROQUAUX2013}.}.

In the present contribution, we focus on the issue of representing sequential data as relational data, approaching the question \emph{how can a set of time series be represented as a traditional graph?} from the perspective of traditional hypothesis testing. After comparing the statistical properties of the former against those expected under a suitable benchmark, our suggestion is to project the set of time series onto a signed graph by linking any two series with a positive (negative) edge in case they share a statistically significant number of concordant (discordant) values. More formally, we adopt an information-theoretic approach that is rooted in the constrained maximisation of Shannon entropy, a procedure inducing an ensemble of multivariate time series that preserves some of their empirical properties on average while randomising everything else~\cite{Park2004,Garlaschelli2008,Squartini2011}.

As our method deals with signed data~\cite{gallo2024testing,gallo2024assessing,gallo2025statistically}, one may wonder how relevant the information provided by signs is within the neuroscientific framework. Several studies have recently investigated the role played by the negative correlations characterising resting-state fMRI data: the authors of~\cite{Saberi2021} show that negative correlations are essential for the stability of resting-state brain networks against perturbations or changes; the authors of~\cite{Saberi2022} notice that the interplay between positive and negative correlations induces frustrated patterns preventing the brain from reaching a state of minimum energy and enhancing its adaptability to dynamical changes; finally, the authors of~\cite{Demertzi2022} show that the default mode network (DMN) works `in opposition' with the various task-based networks, a supposedly crucial feature to maintain cognitive functions and consciousness.

\section{Setting up the formalism}\label{sec:II}

Let us start by setting up the formalism. Since a generic times series can be represented as $X_i=\{X_{it}\}_{t=1\dots T}$, a set of $N$ time series whose duration is $T$ will be represented as an $N\times T$ rectangular matrix $\mathbf{X}=\{X_{it}\}_{\substack{i=1\dots N\\t=1\dots T}}$, whose generic entry $X_{it}$ represents the value of the $i$-th time series at time $t$. Since time series are typically pre-processed, let us explicitly distinguish the raw values above from the transformed ones (e.g. after standardisation\footnote{In what follows, we will precisely consider time series of standardised \emph{residuals} (see section \hyperlink{sec:IX}{VIII}).}) and indicate the (value of the) $i$-th random variable at time $t$ with the symbol $w_{it}$. Since $w_{it}$ can be either \emph{positive} or \emph{negative}, to ease mathematical manipulations, let us employ Iverson's brackets - a notation ensuring all quantities of interest are non-negative~\cite{gallo2024testing} - and define

\begin{align}
w_{it}^-&=|w_{it}|\times[w_{it}<0],\quad\forall\:i,t\\
w_{it}^+&=|w_{it}|\times[w_{it}>0],\quad\forall\:i,t
\end{align}
i.e. mutually exclusive variables that induce the non-negative matrices $\mathbf{W}^+$ and $\mathbf{W}^-$ obeying

\begin{align}
w_{it}&=w_{it}^+-w_{it}^-,\quad\forall\:i,t\\
|w_{it}|&=w_{it}^++w_{it}^-,\quad\forall\:i,t
\end{align}
in an entry-wise fashion - or, more compactly, $\mathbf{W}=\mathbf{W}^+-\mathbf{W}^-$ and $|\mathbf{W}|=\mathbf{W}^++\mathbf{W}^-$. Their purely binary counterparts read

\begin{align}
b_{it}^-&=\left[w_{it}<0\right],\quad\forall\:i,t\\
b_{it}^+&=\left[w_{it}>0\right],\quad\forall\:i,t
\end{align}
in turn, inducing the non-negative matrices $\mathbf{B}^+$ and $\mathbf{B}^-$ that obey the analogous relationships

\begin{align}
b_{it}&=b_{it}^+-b_{it}^-,\quad\forall\:i,t\\
|b_{it}|&=b_{it}^++b_{it}^-,\quad\forall\:i,t
\end{align}
in an entry-wise fashion - or, more compactly, $\mathbf{B}=\mathbf{B}^+-\mathbf{B}^-$ and $|\mathbf{B}|=\mathbf{B}^++\mathbf{B}^-$. Adopting the terminology introduced for bipartite networks, we will also refer to the rectangular matrix $\mathbf{B}$ as the \emph{biadjacency matrix} associated with the considered set of time series.

The present analysis solely focuses on binarised time series: besides easing the mathematical tractability of the latter, this simplified representation has been shown to retain the salient features of resting-state brain dynamics - for instance, the \emph{(a)synchrony} of time series - while being robust against inter-subject variability~\cite{Watanabe2013,Ezaki2020}.

\begin{figure*}[t!]
\centering
\includegraphics[width=\textwidth]{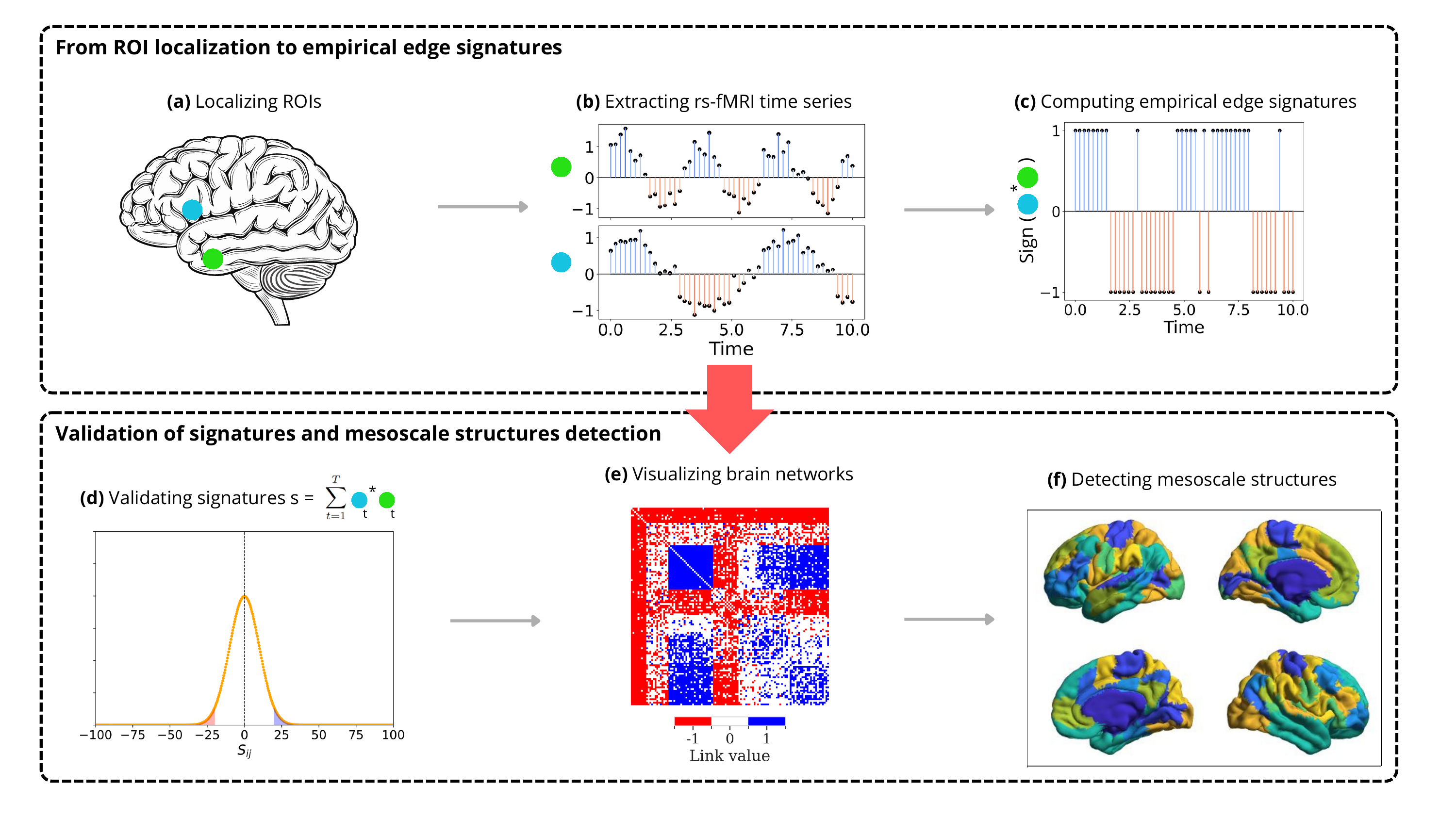}
\caption{\textbf{Infographics illustrating the pipeline of our analysis.} Pictorial representation of the pipeline we follow in the present contribution, from the registration of brain activity to the identification of communities in statistically validated signed projections: (a) rs-fMRI signals are recorded; (b) pairwise interactions between brain regions are considered; (c) the scalar product of each pair of (standardised and binarised) time series is calculated to quantify their \emph{signature}; (d) its empirical value is validated against a null model to remove statistical noise; (e) a significantly large positive (negative) signature induces a positive (negative) link between the two involved brain regions; (f) a BIC-based community detection reveals the modules partitioning our statistically validated signed projections. The first brain image was adapted from Canva (additional elements were added by the authors), while the last one was obtained through the Enigma Toolbox~\cite{Lariviere2021}.}
\label{fig:1}
\end{figure*}

\subsection{Local properties and dependencies of time series}

Let us start by defining a number of quantities of interest for our study. In analogy with the properties traditionally considered to analyse bipartite signed networks, the \emph{total number of positive values} reads

\begin{align}
B^+=\sum_{i=1}^N\sum_{t=1}^Tb_{it}^+=\sum_{i=1}^Nk_i^+=\sum_{t=1}^T\kappa_t^+,
\end{align}
an expression further suggesting the definition of two (sets of) local positive quantities, i.e.

\begin{align}
k_i^+=\sum_{t=1}^Tb_{it}^+,
\end{align}
or \emph{positive degree of series $i$}, capturing the total number of positive entries characterising the $i$-th series, and

\begin{align}
\kappa_t^+=\sum_{i=1}^Nb_{it}^+,
\end{align}
or \emph{positive degree at time $t$}, capturing the total number of positive entries characterising the $t$-th time step and providing information about the correlation between the members of the whole set of time series. Analogously, the \emph{total number of negative values} reads

\begin{align}
B^-=\sum_{i=1}^N\sum_{t=1}^Tb_{it}^-=\sum_{i=1}^Nk_i^-=\sum_{t=1}^T\kappa_t^-
\end{align}
with obvious meaning of the symbols. The density of positive and negative values, reading $c^+=B^+/(N\times T)$ and $c^-=B^-/(N\times T)$ and referred to as \emph{positive} and \emph{negative bipartite connectance}, remain naturally defined.

\subsection{Concordant and discordant temporal motifs}

Let us, now, define the second-order properties known as \emph{dyadic motifs}. In the binary case,

\begin{align}
B_{ij}^{++}&=\sum_{t=1}^Tb_{it}^+b_{jt}^+
\end{align}
counts the number of time steps in correspondence of which both series $i$ and series $j$ are characterised by positive values and

\begin{align}
B_{ij}^{--}&=\sum_{t=1}^Tb_{it}^-b_{jt}^-
\end{align}
counts the number of time steps in correspondence of which both series $i$ and series $j$ are characterised by negative values. It is quite intuitive to ascribe them to the class of \emph{concordant motifs}, i.e. patterns capturing the `agreement' between any two series. Analogously, it is quite intuitive to ascribe

\begin{align}
B_{ij}^{+-}&=\sum_{t=1}^Tb_{it}^+b_{jt}^-
\end{align}
and

\begin{align}
B_{ij}^{-+}&=\sum_{t=1}^Tb_{it}^-b_{jt}^+,
\end{align}
counting the number of time steps in correspondence of which series $i$ is characterised by a positive value and series $j$ is characterised by a negative value or vice versa, to the class of \emph{discordant motifs}, i.e. patterns capturing the `disagreement' between any two series.

Any two nodes whose associated time series exhibit a significantly large number of concordant motifs will be connected by a $+1$, while any two nodes whose associated time series exhibit a significantly large number of discordant motifs will be connected by a $-1$. We would also like to stress that our projection scheme is applicable only in the absence of null values - a condition that is naturally satisfied by BOLD time series, representing the focus of this study.

\subsection{A scheme for the validated projection of multivariate time series}

As depicted in fig.~\ref{fig:1}, our validation algorithm works as follows:

\begin{itemize}
\item[(1)] focus on a specific pair of series, say $i$ and $j$, and measure their (binary) similarity (see section \hyperlink{sec:III}{III});
\item[(2)] quantify the statistical significance of the measured similarity, with respect to a properly-defined benchmark, by computing the corresponding p-value, say $p_{ij}$ (see section \hyperlink{sec:IV}{IV});
\item[(3)] repeat the step above for each pair of series;
\item[(4)] apply a multiple hypothesis testing procedure and insert a (binary) link between the nodes $i$ and $j$ if and only if $p_{ij}$ is found to be less than, or equal to, the threshold value $p_{th}$ (see section \hyperlink{sec:V}{V}).
\end{itemize}

Let us now describe each step in detail.

\section{Computing the similarity of any\\two series}\label{sec:III}

The first step of our method prescribes measuring the degree of similarity of series $i$ and $j$. To this aim, let us consider the quantity named \textit{signature}~\cite{gallo2024testing}, defined as the scalar product of the two series: upon defining $\mathbf{b}_i$ as the $i$-th row of the matrix $\mathbf{B}$, the signature reads

\begin{align}
S_{ij}=\mathbf{b}_i\times\mathbf{b}_j&=\sum_{t=1}^Tb_{it}b_{jt}\nonumber\\
&=\sum_{t=1}^T(b_{it}^+-b_{it}^-)(b_{jt}^+-b_{jt}^-)\nonumber\\
&=\sum_{t=1}^T[(b_{it}^+b_{jt}^++b_{it}^-b_{jt}^-)-(b_{it}^+b_{jt}^-+b_{it}^-b_{jt}^+)];
\end{align}
in words, the signature is the difference between two quantities, i.e. the \textit{concordance} of series $i$ and $j$, reading

\begin{align}
C_{ij}&=\sum_{t=1}^TC_{ijt}=\sum_{t=1}^T(b_{it}^+b_{jt}^++b_{it}^-b_{jt}^-)=B_{ij}^{++}+B_{ij}^{--}
\end{align}
and counting the number of concordant motifs, and the \textit{discordance} of nodes $i$ and $j$, reading

\begin{align}
D_{ij}&=\sum_{t=1}^TD_{ijt}=\sum_{t=1}^T(b_{it}^+b_{jt}^-+b_{it}^-b_{jt}^+)=B_{ij}^{+-}+B_{ij}^{-+}
\end{align}
and counting the number of discordant motifs: in symbols, $S_{ij}=C_{ij}-D_{ij}$, $\forall\:i<j$. A na\"ive way of projecting a multivariate binary time series would prescribe to stop here and pose

\begin{equation}\label{eq:nai}
a^\text{na\"ive}_{ij}=\text{sgn}[S_{ij}],
\end{equation}
i.e. apply the sign function to the signature, hence connecting series $i$ and $j$ with a positive link if $S_{ij}>0$, i.e. $C_{ij}>D_{ij}$, and with a negative link if $S_{ij}<0$, i.e. $C_{ij}<D_{ij}$.

\section{Testing the similarity of any\\two series}\label{sec:IV}

The second step of our method prescribes evaluating the statistical significance of the similarity of any two series. To this aim, let us first notice that

\begin{equation}
-T\leq S_{ij}\leq T
\end{equation}
where $S_{ij}=-T$ if $C_{ij}=0$ (i.e. if each dyadic motif is composed by a $-1$ and a $+1$) and $S_{ij}=T$ if $D_{ij}=0$ (i.e. if each dyadic motif is composed by either two $-1$s or two $+1$s) and, then, provide the derivation of two kinds of benchmarks, i.e. a homogeneous and a heterogeneous one - in what follows, we will employ \emph{Exponential Random Graphs} (ERGs)~\cite{Park2004, Fronczak2018, Squartini2017book} defined by, either global or local, linear constraints, hence treating links as independent random variables.

\subsection{Homogeneous benchmark}

Let us start by considering the homogeneous benchmark defined by the finite scheme\footnote{A mathematical construction consisting of the list of elementary events and the associated probabilities.}

\begin{equation}\label{eq:nmhomofixed}
b_{it}\sim
\begin{pmatrix}
-1  & +1\\
p^- & p^+
\end{pmatrix},\quad\forall\:i,t
\end{equation}
which, in turn, induces the finite scheme

\begin{align}
b_{it}b_{jt}&\sim
\begin{pmatrix}
-1 & +1\\
2p^+p^- & (p^+)^2+(p^-)^2
\end{pmatrix}\nonumber\\
&=
\begin{pmatrix}
-1 & +1\\
q^- & q^+
\end{pmatrix},\quad\forall\:i,j,t
\end{align}
satisfying the relationships

\begin{align}
1&=q^++q^-,\\
\langle b_{it}b_{jt}\rangle&=q^+-q^-,\\
\text{Var}[b_{it}b_{jt}]&=(q^++q^-)-(q^+-q^-)^2\nonumber\\
&=1-(q^+-q^-)^2\nonumber\\
&=4q^+q^-.
\end{align}

Let us, now, notice that $S_{ij}$ is a sum of independent and identically distributed (i.i.d.) Bernoulli random variables\footnote{The finite scheme obeyed by $C_{ijt}-D_{ijt}$ is the same finite scheme obeyed by $b_{it}b_{jt}$.}, the outcomes of each elementary event being, in fact, $-1$ and $+1$ instead of $0$ and $+1$. Such a consideration allows us to find the probability distribution obeyed by $S_{ij}$ quite straightforwardly, by starting from the one describing $C_{ij}$ alone, i.e.

\begin{align}
P(C_{ij}=k)=\binom{T}{k}(q^+)^k(q^-)^{T-k};
\end{align}
let us, in fact, consider the change of variable $k=(T+s)/2$, leading to the binomial

\begin{align}\label{eq:bino}
P(S_{ij}=s)=\binom{T}{\frac{T+s}{2}}(q^+)^\frac{T+s}{2}(q^-)^\frac{T-s}{2}
\end{align}
that ranges from $-T$ to $T$. Let us, now, convince ourselves that this transformation is indeed the correct one, by providing some explicit examples: if nodes $i$ and $j$ establish $k=0$ concordant motifs, they establish $T$ discordant motifs and the signature reads $s=-T$; if nodes $i$ and $j$ establish $k=T/2$ concordant motifs, they establish the same number of discordant motifs and the signature reads $s=0$; if nodes $i$ and $j$ establish $k=T$ concordant motifs, they establish zero discordant motifs and the signature reads $s=T$. In other words, the aforementioned change of variable transforms a discrete distribution whose support extends from $0$ to $T$ into a discrete distribution whose support extends from $-T$ to $T$: more intuitively, we have moved from considering sequences of $0$s and $+1$s to considering sequences of $-1$s and $+1$s, where each $+1$ can be identified with a concordant motif and each $-1$ can be identified with a discordant motif.

A comment about the parity of $T$ and $s$ is needed: let us notice that, in case $T$ is odd, $s$ itself must be odd, assuming the values $s=-T,-T+2,-T+4\dots T$ (corresponding to $k=0,1,2\dots T$); in case $T$ is even, $s$ itself must be even, assuming the same values as above in correspondence of the same values of $k$.

\subsection{Heterogeneous benchmark}

Let us, now, consider the heterogeneous benchmark

\begin{equation}\label{eq:nmheterofixed}
b_{it}\sim
\begin{pmatrix}
-1  & +1\\
p_{it}^- & p_{it}^+
\end{pmatrix},\quad\forall\:i,t
\end{equation}
which, in turn, induces the finite scheme

\begin{align}
b_{it}b_{jt}&\sim
\begin{pmatrix}
-1 & +1\\
p_{it}^+p_{jt}^-+p_{it}^-p_{jt}^+ & p_{it}^+p_{jt}^++p_{it}^-p_{jt}^-
\end{pmatrix}\nonumber\\
&=
\begin{pmatrix}
-1 & +1\\
q_{ijt}^- & q_{ijt}^+
\end{pmatrix},\quad\forall\:i,j,t
\end{align}
satisfying the relationships

\begin{align}
1&=q_{ijt}^++q_{ijt}^-,\\
\langle b_{it}b_{jt}\rangle&=q_{ijt}^+-q_{ijt}^-,\\
\text{Var}[b_{it}b_{jt}]&=(q_{ijt}^++q_{ijt}^-)-(q_{ijt}^+-q_{ijt}^-)^2\nonumber\\
&=1-(q_{ijt}^+-q_{ijt}^-)^2\nonumber\\
&=4q_{ijt}^+q_{ijt}^-.
\end{align}

\begin{figure*}[t!]
\centering
\includegraphics[trim={0cm 2.5cm 0 2.cm},clip,width=\textwidth]{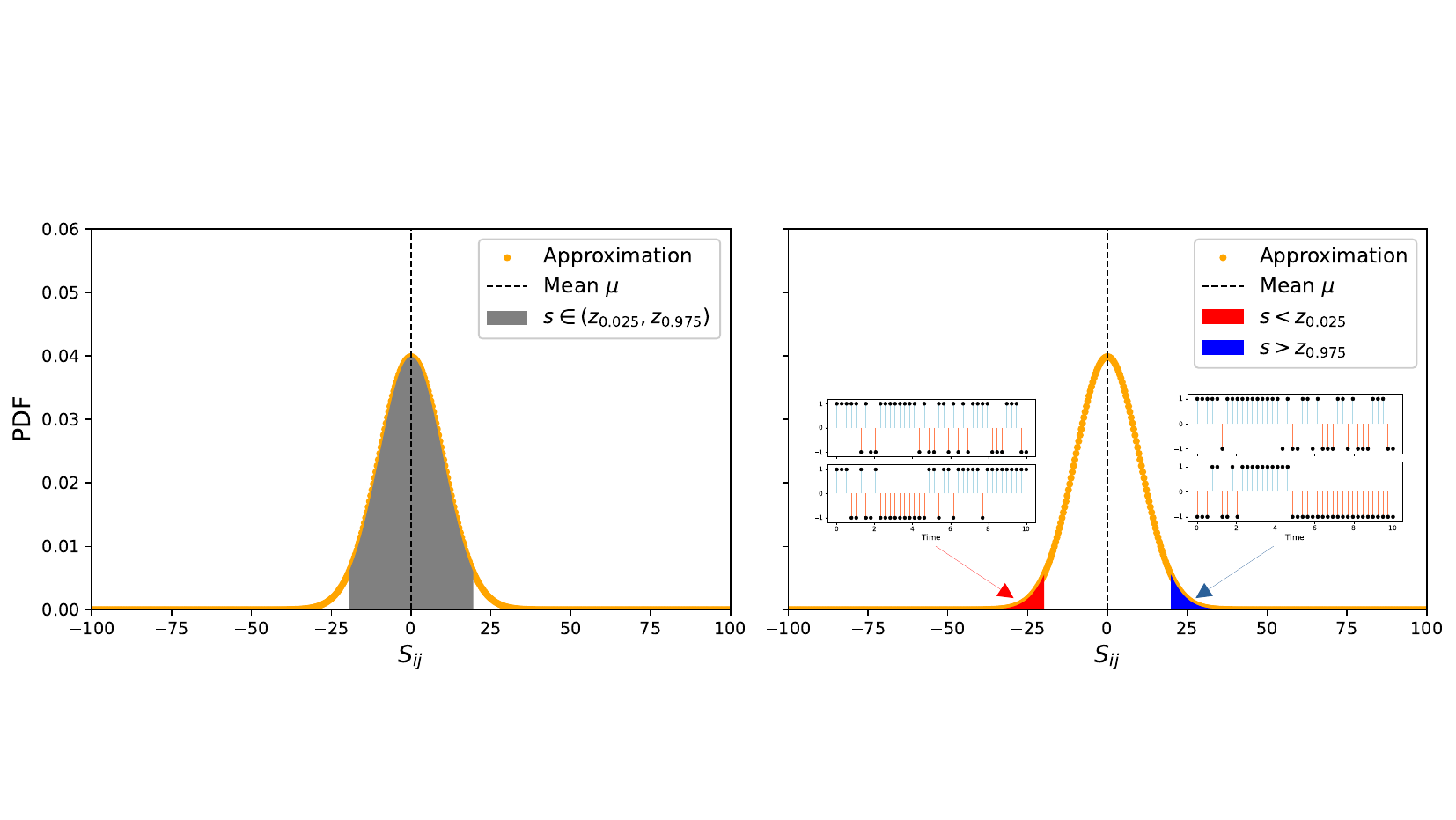}
\caption{\textbf{Infographics illustrating our validation procedure.} Probability distribution of the signature and its Gaussian approximation: while the left panel provides a graphical answer to the question \textit{is the empirical value of the signature significantly different from the one expected under the chosen benchmark?}, the right panel provides a graphical answer to the question \textit{is the deviation negative (hence, the signature is significantly smaller than expected) or positive (hence, the signature is significantly larger than expected)?} The red area corresponds to the region of validation of the negative links, induced by pairs of series `in counterphase' (i.e. \emph{asynchronously} active) most of the time; the blue area corresponds to the region of validation of the positive links, induced by pairs of series `in phase' (i.e. \emph{synchronously} active) most of the time.}
\label{fig:2}
\end{figure*}

Since $S_{ij}$ is, now, a sum of independent and not identically distributed (i.n.i.d.) Bernoulli random variables, the probability distribution obeyed by it is the Poisson-binomial

\begin{align}\label{eq:poibino}
P(S_{ij}=s)&=\sum_{C_k}\left[\prod_{\mu\in C_k}q_{ij\mu}^+\prod_{\nu\notin C_k}q_{ij\nu}^-\right]
\end{align}
where $C_k$ is the set of $k$-tuples of which $\mu$ and $\nu$ are instances. Let us provide some explicit examples:

\begin{align}
&P(S_{ij}=-T)=\prod_{\alpha=1}^Tq_{ij\alpha}^-,\\
&P(S_{ij}=2-T)=\sum_{\beta=1}^T\left[q_{ij\beta}^+\prod_{\substack{\alpha=1\\\alpha\neq\beta}}^Tq_{ij\alpha}^-\right],\\
&P(S_{ij}=4-T)=\sum_{\beta=1}^T\sum_{\substack{\gamma=1\\\gamma>\beta}}^T\left[q_{ij\beta}^+q_{ij\gamma}^+\prod_{\substack{\alpha=1\\\alpha\neq\beta,\gamma}}^Tq_{ij\alpha}^-\right]
\end{align}
and so on. In words, if nodes $i$ and $j$ establish $k=0$ concordant motifs, they establish $T$ discordant motifs and the signature reads $s=-T$; if nodes $i$ and $j$ establish $k=1$ concordant motif, they establish $T-1$ discordant motifs and the signature reads $s=2-T$; if nodes $i$ and $j$ establish $k=2$ concordant motif, they establish $T-2$ discordant motifs and the signature reads $s=4-T$. The same line of reasoning can be repeated for all the admissible values of $k$.

\subsection{The statistical significance of similarity:\\a two-sided test of hypothesis}

Once we have calculated the distribution of the similarity for each pair of series, we must evaluate the statistical significance of the empirical signature: since we have a signed quantity, we need both tails of such a distribution to carry out a two-sided test of hypothesis. In fact, we need to answer the two related questions: (1) Is the empirical value of the signature significantly different from the one expected under the chosen benchmark? and (2) if so, is the deviation negative (hence, the signature is significantly smaller than expected) or positive (hence, the signature is significantly larger than expected)?\\

The first question can be answered upon calculating the two-sided p-value, reading

\begin{equation}\label{p2}
p_{ij}=2\times\min\left\{F(S_{ij}^*), 1-F(S_{ij}^*)\right\}
\end{equation}
with $F$ being the cumulative distribution function, defined as $F(S_{ij}^*)=\sum_{x\leq S_{ij}^*}P(S_{ij}=x)$: such a number evaluates the probability of observing a deviation from the expected value in either direction.

The second question can be answered upon calculating the sign of such a deviation, by determining if either

\begin{equation}
F(S_{ij}^*)<1-F(S_{ij}^*)
\end{equation}
or

\begin{equation}
F(S_{ij}^*)>1-F(S_{ij}^*)
\end{equation}
holds true. In the first case, $F(S_{ij}^*)<1/2$, the empirical value of the signature is smaller than the median of the distribution and the deviation is negative; in the second case, $F(S_{ij}^*)>1/2$, the empirical value of the signature is larger than the median of the distribution and the deviation is positive. Intuitively\footnote{The value $p_{th}$ will be determined later, at the third step of our algorithm: hereby, a merely discursive introduction to the proper validation step is provided.},

\begin{itemize}
\item[(1)] the two conditions $F(S_{ij}^*)<1/2$ and $p_{ij}\leq p_{th}$ indicate that nodes $i$ and $j$ have established a number of discordant motifs that is so large to induce a significantly negative signature - and, potentially, a negative link in the projection ($a_{ij}=-1$). In a sense, the two series are `in counterphase' most of the time, an expression that, in a more neuroscientific jargon, is intended to mean `found to be \emph{asynchronously} active a significantly large amount of the time';
\item[(2)] the two conditions $F(S_{ij}^*)>1/2$ and $p_{ij}\leq p_{th}$ indicate that nodes $i$ and $j$ have established a number of concordant motifs that is so large to induce a significantly positive signature - and, potentially, a positive link in the projection ($a_{ij}=+1$). In a sense, the two series are `in phase' most of the time, an expression that, in a more neuroscientific jargon, is intended to mean `found to be \emph{synchronously} active a significantly large amount of the time';
\item[(3)] the condition $p_{ij}>p_{th}$ individuates a value of the signature (i.e. of concordant/discordant motifs) that is \textit{compatible} with the one predicted by the chosen benchmark: stated otherwise, the empirical number of motifs could have been observed in configurations generated by the benchmark itself - hence, inducing a null link in the projection ($a_{ij}=0$).
\end{itemize}

Let us stress that the signs populating a projection carry a relative meaning, i.e. positive and negative signatures are so with respect to a benchmark: more explicitly, two series are revealed `in counterphase' if `not enough in phase' with respect to the benchmark. For a graphical representation of our validation procedure, see fig.~\ref{fig:2}.

\section{Validating the projection of a multivariate time series}\label{sec:V}

The second step of our method returns a symmetric matrix of p-values. Individuating the ones associated with the hypotheses to be actually rejected requires a procedure to deal with the comparison of multiple hypotheses at the same time. In very general terms, a threshold must be set: if the specific p-value is smaller than it, the associated event is interpreted as statistically significant and the corresponding nodes are connected in the projection.

One approach is that of adopting the `standard' threshold $\alpha=0.05$ and carrying out each test independently from the others but such a procedure leads to increase\footnote{A simple argument can, indeed, be provided. The probability that, solely by chance, at least one out of $|H|$ hypotheses is rejected, and the corresponding link validated, amounts to $\text{FWER}=1-(1-t)^{|H|}$: hence, $\text{FWER}\simeq1$ for (just) $|H|=100$ tests conducted at the significance level of $t=0.05$.} the number of incorrectly rejected null hypotheses (i.e. validated links). An alternative approach is that of adopting the Bonferroni correction~\cite{BonferroniCarloEmilio1936Tsdc,mantegna2011statistically,gualdi2016statistically} but such a procedure is known to severely increase\footnote{Indeed, the stricter condition $\text{FWER}=0.05$ leads to the threshold $p_{th}=t/|H|$, rapidly vanishing as $|H|$ grows: as a consequence, very sparse (if not empty) projections are obtained.} the number of incorrectly retained null hypotheses (i.e. discarded links).

In the present paper we apply the so-called False Discovery Rate (FDR) procedure~\cite{benjamini1995controlling}. FDR allows one to control for the expected number of `false discoveries' (i.e. incorrectly rejected null hypotheses or incorrectly validated links), regardless of the independence of the hypotheses tested. Whenever $|H|$ different hypotheses $H_1,\:H_2\dots$, characterised by $|H|$ different p-values, must be tested at the same time, FDR prescribes to first sort the $|H|$ p-values in increasing order

\begin{align}
\text{p-value}_1\le\dots\le\text{p-value}_{|H|}
\end{align}
and, then, identify the largest integer $\hat{i}$ satisfying the condition

\begin{align}
\text{p-value}_{\hat{i}}\le\frac{\hat{i}\alpha}{|H|}
\end{align}
with $\alpha$ representing the usual single-test significance level, hereby set to the value $\alpha=0.05$: since $H_i$ is the hypothesis that the distribution of dyadic motifs established by the $i$-th pair of nodes follows a binomial/Poisson-binomial, rejecting it amounts to connecting the corresponding nodes in the projection. Notice that $|H|=N(N-1)/2$.

Naturally, deciding which test is the best suited one for the problem at hand depends on the importance assigned to false positives and false negatives: as a rule of thumb, the Bonferroni correction can be deemed as appropriate when few tests are expected to be significant (i.e. when even a single false positive would be problematic); when, on the contrary, many tests are expected to be significant, using the Bonferroni correction may, in turn, produce a too large number of false negatives.

\section{Benchmarks for the analysis of multivariate time series}\label{sec:VI}

To evaluate the statistical significance of the similarity between any two series $i$ and $j$, a benchmark is required. As anticipated, a natural choice is that of adopting the class of null models known as ERGs. Within such a framework, the constrained maximisation of Shannon entropy

\begin{equation}
S=-\sum_{\mathbf{B}\in\mathbb{B}}P(\mathbf{B})\ln P(\mathbf{B})
\end{equation}
leads to assign the generic biadjacency matrix $\mathbf{B}$ the probability

\begin{equation}
P(\mathbf{B})=\frac{e^{-H(\bm{\theta},\bm{C}(\mathbf{B}))}}{Z(\bm{\theta})}
\end{equation}
whose value is determined by the vector $\bm{C}(\mathbf{B})$ of topological constraints via the Hamiltonian $H(\bm{\theta},\bm{C}(\mathbf{B}))=\bm{\theta}\times\bm{C}(\mathbf{B})=\sum_i\theta_iC_i(\mathbf{B})$. The use of linear constraints allows $P(\mathbf{B})$ to be written in a factorised form, i.e. as the product of pair-specific probability coefficients~\cite{gallo2025statistically}.

In order to determine the unknown parameters $\bm{\theta}$, the likelihood maximisation recipe can be adopted: given an observed biadjacency matrix $\mathbf{B}^*$, it translates into solving the system of equations

\begin{equation}
\langle C_i\rangle(\bm{\theta})=\sum_{\mathbf{B}\in\mathbb{B}}P(\mathbf{B})C_i(\mathbf{B})=C_i(\mathbf{B}^*),\quad\forall\:i
\end{equation}
which prescribes to equate each ensemble average to its observed counterpart.

Let us now generalise such a framework to accommodate models for studying binary multivariate time series. To this aim, we will follow~\cite{Livan2020,gallo2024testing}, which, in turn, builds upon the analytical approach introduced in~\cite{Park2004} and further developed in~\cite{Squartini2011}.

Let us, now, make a general observation: as each of the $N$ time series is defined by $T$ time steps, in corresponding of which either the value $-1$ or the value $+1$ can be assumed, the ensemble is constituted by all possible $|\mathbb{B}|=2^{N\times T}$ binary, signed biadjacency matrices.

\subsection{Signed Random Graph Model}

The first null model we consider is induced by the Hamiltonian

\begin{equation}
H(\mathbf{B})=\alpha\:B^+(\mathbf{B})+\gamma\:B^-(\mathbf{B})
\end{equation}
i.e. by the two global constraints $B^+(\mathbf{B})$ and $B^-(\mathbf{B})$. We will refer to it as Bipartite Signed Random Graph Model (bSRGM)~\cite{gallo2024testing}.

According to the bSRGM, each element of a multivariate time series is a random variable whose behaviour is described by the coefficients

\begin{align}
P(b_{it}=+1)&=\frac{e^{-\alpha}}{e^{-\alpha}+e^{-\gamma}}=p^+,\\
P(b_{it}=-1)&=\frac{e^{-\gamma}}{e^{-\alpha}+e^{-\gamma}}=p^-;
\end{align}
in other words, $b_{it}$ obeys a generalised Bernoulli distribution whose probability coefficients are determined by the (Lagrange multipliers of the) imposed constraints: each positive link appears with probability $p^+$ and each negative link appears with probability $p^-$.

In order to employ the bSRGM for studying real-world multivariate time series, the parameters that define it need to be properly tuned: more specifically, one needs to ensure

\begin{align}
\langle B^+\rangle_\text{bSRGM}&=B^+(\mathbf{B}^*),\\
\langle B^-\rangle_\text{bSRGM}&=B^-(\mathbf{B}^*)
\end{align}
with the symbol $\mathbf{B}^*$ indicating the empirical multivariate time series under analysis. To this aim, one can maximise the likelihood function $\mathcal{L}_\text{bSRGM}(\alpha,\gamma)=\ln P_\text{bSRGM}(\mathbf{B}^*|\alpha,\gamma)$, where $P_\text{bSRGM}(\mathbf{B}^*|\alpha,\gamma)$ is defined as in eq.~\ref{probgraph1}, with respect to the unknown parameters that define it~\cite{Garlaschelli2008}. Such a recipe leads us to find 

\begin{align}
p^+&=B^+(\mathbf{B}^*)/(N\times T),\\
p^-&=B^-(\mathbf{B}^*)/(N\times T)
\end{align}
with obvious meaning of the symbols (see also Appendix~\hyperlink{AppA}{A}).

\subsection{Signed Configuration Model}

The second null model we consider is induced by the Hamiltonian

\begin{align}
H(\mathbf{B})=&\sum_{i=1}^N[\alpha_ik_i^+(\mathbf B)+\gamma_ik_i^-(\mathbf B)]\nonumber\\
&+\sum_{t=1}^T[\delta_t\kappa_t^+(\mathbf B)+\eta_t\kappa_t^-(\mathbf B)];
\end{align}
i.e. by the $2(N+T)$ local constraints $\{k_i^+(\mathbf{B})\}_{i=1}^N$, $\{k_i^-(\mathbf{B})\}_{i=1}^N$, $\{\kappa_t^+(\mathbf{B})\}_{t=1}^T$ and $\{\kappa_t^-(\mathbf{B})\}_{t=1}^T$. We will refer to it as Bipartite Signed Configuration Model (bSCM)~\cite{gallo2024testing}.

According to the bSCM, each element of a multivariate time series is a random variable whose behaviour is described by the coefficients

\begin{align}
P(b_{it}=+1)&=\frac{e^{-(\alpha_i+\delta_t)}}{e^{-(\alpha_i+\delta_t)}+e^{-(\gamma_i+\eta_t)}}=p_{it}^+,\\
P(b_{it}=-1)&=\frac{e^{-(\gamma_i+\eta_t)}}{e^{-(\alpha_i+\delta_t)}+e^{-(\gamma_i+\eta_t)}}=p_{it}^-;
\end{align}
in other words, $b_{it}$ obeys a generalised Bernoulli distribution whose probability coefficients are determined by the (Lagrange multipliers of the) imposed constraints: the series $i$, at time $t$ assumes the value $+1$ with probability $p_{it}^+$ and the value $-1$ with probability $p_{it}^-$.

\begin{figure*}[t!]
\centering
\includegraphics[width=\textwidth]{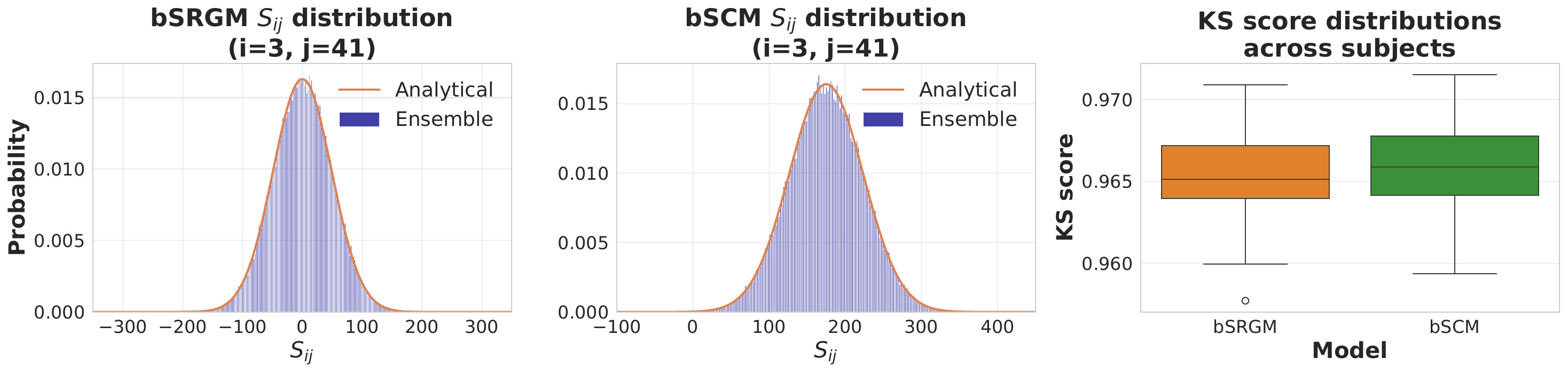}
\caption{\textbf{Consistency checks regarding the distribution of the signature for subject $\bm{\#100307}$.} Comparison between the distribution of the dyadic signature induced by the bSRGM/bSCM (left panel/middle panel) and their numerical counterparts, obtained by explicitly sampling $10^5$ realisations from the corresponding ensemble: in both cases, the KS test rejects the hypothesis that the two distributions coincide. The box-plots summing up the distributions of the KS-scores for the bSRGM and the bSCM, obtained by explicitly sampling $10^3$ realisations from the corresponding ensemble (right panel), show that, under both models, the median KS-score amounts to $\text{KS}_{5\%}\simeq0.965$, the $95\%$ of the values ranging between the $2.5$-th and the $97.5$-th percentiles reading $[q_{2.5},q_{97.5}]=[0.961,0.970]$. In words, our sampling procedure can be considered satisfactory enough to reproduce the analytical distributions defined by eq.~\ref{eq:bino} and eq.~\ref{eq:poibino}, even in case of statistical disagreement.}
\label{fig:3}
\end{figure*}

In order to tune the parameters defining the bSCM to ensure

\begin{align}
\langle k_i^+\rangle_\text{bSCM}&=k_i^+(\mathbf{B}^*),\quad\forall\:i,\\
\langle k_i^-\rangle_\text{bSCM}&=k_i^-(\mathbf{B}^*),\quad\forall\:i,\\
\langle\kappa_t^+\rangle_\text{bSCM}&=\kappa_t^+(\mathbf{B}^*),\quad\forall\:t,\\
\langle\kappa_t^-\rangle_\text{bSCM}&=\kappa_t^-(\mathbf{B}^*),\quad\forall\:t
\end{align}
let us maximise the likelihood function $\mathcal{L}_\text{bSCM}(\bm{\alpha},\bm{\gamma},\bm{\delta},\bm{\eta})=\ln P_\text{bSCM}(\mathbf{B}^*|\bm{\alpha},\bm{\gamma},\bm{\delta},\bm{\eta})$, where $P_\text{bSCM}(\mathbf{B}^*|\bm{\alpha},\bm{\gamma},\bm{\delta},\bm{\eta})$ is defined as in eq.~\ref{probgraph2}, with respect to the unknown parameters that define it~\cite{Garlaschelli2008} (see also Appendix~\hyperlink{AppA}{A}). Such a recipe leads us to find

\begin{align}
k_i^+(\mathbf{B}^*)&=\sum_{t=1}^T\frac{e^{-(\alpha_i+\delta_t)}}{e^{-(\alpha_i+\delta_t)}+e^{-(\gamma_i+\eta_t)}}=\langle k_i^+\rangle,\quad\forall\:i,\\
k_i^-(\mathbf{B}^*)&=\sum_{t=1}^T\frac{e^{-(\gamma_i+\eta_t)}}{e^{-(\alpha_i+\delta_t)}+e^{-(\gamma_i+\eta_t)}}=\langle k_i^-\rangle,\quad\forall\:i,\\
\kappa_t^+(\mathbf{B}^*)&=\sum_{i=1}^N\frac{e^{-(\alpha_i+\delta_t)}}{e^{-(\alpha_i+\delta_t)}+e^{-(\gamma_i+\eta_t)}}=\langle\kappa_t^+\rangle,\quad\forall\:t,\\
\kappa_t^-(\mathbf{B}^*)&=\sum_{i=1}^N\frac{e^{-(\gamma_i+\eta_t)}}{e^{-(\alpha_i+\delta_t)}+e^{-(\gamma_i+\eta_t)}}=\langle\kappa_t^-\rangle,\quad\forall\:t
\end{align}
a system that can be solved only numerically, along the guidelines provided in~\cite{vallarano2021} (see also Appendix~\hyperlink{AppB}{B}).

\section{Dataset description}\label{sec:VIII}

We have applied the methodology above to 3T resting-state fMRI data (rs-fMRI) data from $100$ unrelated subjects of the Young Adult Human Connectome Project (HCP)~\cite{VanEssen2013}.

Overall, this dataset includes $1200$ participants aged $22$ to $35$ of whom $1087$ underwent at least one rs-fMRI scan. Spontaneous, slowly fluctuating brain activity was measured during fMRI, using a multiband accelerated echo-planar imaging two-dimensional BOLD sequence with whole-brain coverage, while subjects maintained fixation on a central cross and were instructed to lie still and rest quietly - $\text{TR}=720$ ms, $\text{TE}=33.1$ ms, $\text{flip angle}=52$ deg, $\text{FOV}=208\times 180$ mm ($\text{RO}\times\text{PE}$), $\text{matrix}=104\times 90$ ($\text{RO}\times\text{PE}$), $\text{slice thickness}=2.0$ mm, $\text{number of slices}=72$, $\text{resolution}=2.0$ mm isotropic voxels, $\text{multiband factor}=8$, $\text{echo spacing}=0.58$ ms, $\text{BW}=2290$ Hz/Px, $\text{frames per run}=1200$, $\text{run duration\:(min:sec)}=14:33$. Within each session, oblique axial acquisitions alternated between phase encoding in a right-to-left direction in one run and phase encoding in a left-to-right direction in the other run.

\begin{figure*}[t!]
\centering
\includegraphics[width=\linewidth]{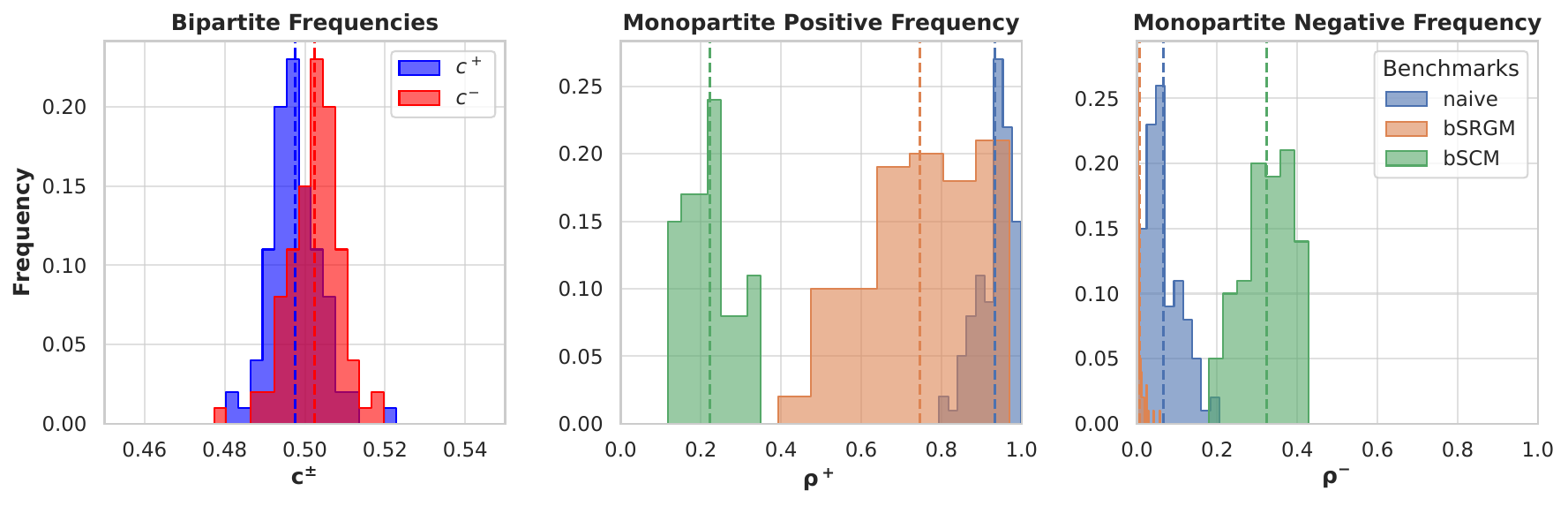}
\caption{\textbf{Distributions of the signed, bipartite, and monopartite connectance.} Left panel: distributions of the positive and negative bipartite connectance, respectively defined as $c^+=B^+/(N\times T)$ and $c^-=B^-/(N\times T)$, across our $100$ subjects. Middle and right panel: distributions of the positive and negative monopartite connectance, respectively defined as $\rho^+=2L^+/N(N-1)$ and $\rho^-=2L^-/N(N-1)$, across the three different projections of our $100$ subjects. Bin widths are computed according to the Freedman-Diaconis rule. Averages are plotted as vertical, dashed lines. How can a large number of bipartite, negative links co-exist with the large number of positive links characterising the na\"ive projection? Bipartite, negative links are evidently arranged into concordant motifs that, in turn, let the na\"ive and bSRGM-induced projections be populated by a majority of positive links. Employing the bSCM mitigates this situation, as the number of positive and negative links is, now, rebalanced.}
\label{fig:4}
\end{figure*}

\subsection{Pre-processing of time series}

fMRI recordings underwent a two-step pre-processing procedure. First, the data were motion-corrected, intensity-adjusted, and normalised~\cite{Glasser2013}. Afterwards, structured noise was corrected using the FIX denoising pipeline~\cite{Griffanti2014}. This clean-up included $24$ confound time series derived from the motion estimation (the $6$ rigid-body parameter time series, their backwards-looking temporal derivatives and all $12$ resulting regressors squared~\cite{Satterthwaite2013}). On the motion parameters, the temporal high-pass filtering was previously applied. They are, then, regressed out of the data aggressively, as they are not expected to contain variance of interest. To improve the length of BOLD time series, the right-to-left and left-to-right runs were concatenated after the mean across time removal. Finally, each brain was parcellated into $116$ regions of interest (ROIs) over which the signal was averaged. More specifically, the cortical regions were parcellated following the Schaefer $100$ atlas~\cite{Schaefer2018} and the subcortical ones following the Tian $16$ atlas~\cite{Tian2020}. As a final result, we obtained a collection of $100$ sets of time series, each one having dimensions $N\times T=116\times 2400$.

Let us, finally, notice that pre-processing procedures such as the global signal regression (GSR) have been criticised since biasing data to balance the number of positive and negative signs~\cite{Murphy2009}: for such a reason, we have opted for not adopting it, here.

\subsection{Pre-whitening of time series}

A documented property of rs-fMRI signals concerns the presence of a non-negligible amount of temporal autocorrelations: in other words, the sluggish nature of the hemodynamic response, which extends over timescales of $\simeq20$ s, lets the BOLD response at time $t$ to depend on its response at previous time steps~\cite{Friston1994,Glover1999,Woolrich2001}; since this may bias the number of co-occurrences, here we address such an issue by implementing a pre-whitening procedure aimed to removing it.

More quantitatively, we model the time series $X_{it}$ as

\begin{equation}
X_{it}=\sum_{k=1}^{p_i}\phi_{k}^{(i)}X_{it-k}+\varepsilon_{it},
\label{eq:ar_model}
\end{equation}
i.e. as an autoregressive process of order $p_i$, where $\phi_{1}^{(i)}\dots \phi_{p}^{(i)}$ are the coefficients and $\varepsilon_{it}$ is the residual, i.e. the component of the $i$-th signal at time $t$ that cannot be (linearly) predicted by employing the $p_i$ previous values: the autoregressive coefficients have been estimated via ordinary least squares separately for each of the $N=116$ ROIs\footnote{To this aim, we have employed the \texttt{AutoReg} estimator from the \texttt{statsmodels} Python library.}; orders have been evaluated separately as well, the optimum $\hat{p}_i$ having being selected via Bartlett's white noise test~\cite{Bartlett1955,Priestley1981,Durbin1967} (see also Appendix~\hyperlink{AppC}{C}). Once the coefficients and the order have been estimated, the pre-whitened residuals have been computed as

\begin{equation}
\varepsilon_{it}=X_{it}-\sum_{k=1}^{\hat{p}_i}\hat{\phi}_{k}^{(i)}X_{it-k},\quad t=\hat{p}_{i}+1\dots T.
\label{eq:residuals}
\end{equation}

Since different ROIs have different orders, the series of residuals have different lengths - for instance, series $i$ has a length of $T-\hat{p}_i$: as a consequence, the series of residuals must be `aligned'. To this aim, we have adopted a \emph{trimming strategy}, implementing the so-called \emph{trailing trim}, that prescribes to retain the last $T_\text{eff}=T-\max_i\hat{p}_i$ points of each series.

\begin{figure*}[t!]
\centering
\includegraphics[width=\linewidth]{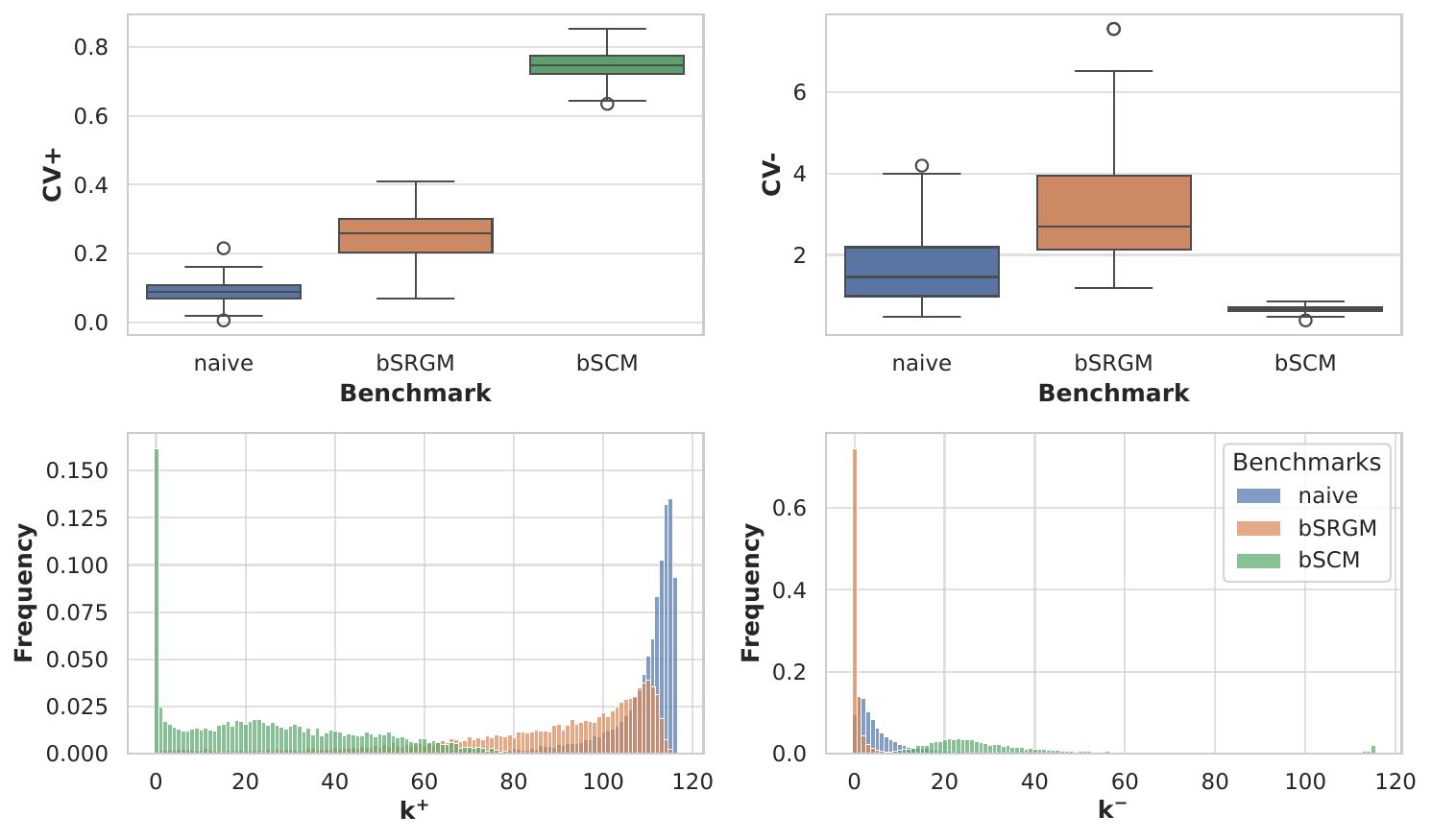}
\caption{\textbf{Distributions of CV$^+$, CV$^-$ and signed degrees.} Top panels: box-plots summing up the distributions of the coefficient of variation for positive degrees (CV$^+$, left) and negative degrees (CV$^-$, right) across the three different projections of our $100$ subjects. Each box-plot illustrates the distribution of the corresponding metric, indicating its central tendency and dispersion for each model: the positive variant of the CV suggests that the corresponding subgraphs are characterised by a quite homogeneous structure, while the negative one suggests that the opposite holds true; yet, these differences are levelled out when considering bSCM-induced projections. Bottom panels: distributions of the positive (left) and negative (right) degrees populating our projections, pooled across our subjects. While na\"ive and bSRGM-induced positive distributions appear as left-skewed, their negative counterparts appear as right-skewed; bSCM-induced distributions, instead, are much flatter in both cases, suggesting a larger heterogeneity of the degrees of this kind. The leftmost peak on $k^+$ and the rightmost peak on $k^-$ are due to subcortical regions.}
\label{fig:5}
\end{figure*}

Overall, thus, we are left with an $N\times T_\text{eff}$ rectangular matrix of residuals: the pre-whitened binarised series are, then, obtained by applying Iverson's brackets to each entry of such a standardised matrix, i.e.

\begin{equation}
w_{it}=\frac{\varepsilon_{it}-\underline{\varepsilon}_i}{\sigma_i[\varepsilon_i]},\quad\forall\:i,t
\end{equation}
where $\underline{\varepsilon}_i$ and $\sigma_i[\varepsilon_i]$ represent the (sample) average and standard deviation of series $\varepsilon_i=\{\varepsilon_{it}\}_{t=1\dots T}$; in formulas,

\begin{align}
b_{it}^-&=\left[w_{it}<0\right],\quad\forall\:i,t\\
b_{it}^+&=\left[w_{it}>0\right],\quad\forall\:i,t
\end{align}

\section{Results}\label{sec:IX}

Let us, now, employ our validation procedure to carry out a `binary-to-binary' analysis, i.e. obtain statistically validated binary projections of binary multivariate time series, defined by the `rule' that any two nodes sharing a significantly large number of concordant (discordant) motifs are connected by a positive (negative) edge.

\subsection{Consistency checks on the signature}

Before moving to analyse our data, let us clarify the nature and role of $S_{ij}$, verifying the extent to which $S_{ij}$ retains the sign of the Pearson correlation coefficient (see eq.~\ref{eq:1-Pearson Correlation}) between the time series $i$ and $j$, across all subjects. To this aim, we have proceeded in two ways: the first one, by identifying $i$ and $j$ with $w_i=\{w_{it}\}_{t=1\dots T}$ and $w_j=\{w_{jt}\}_{t=1\dots T}$; the second one, by identifying $i$ and $j$ with $b_i=\{b_{it}\}_{t=1\dots T}$ and $b_j=\{b_{jt}\}_{t=1\dots T}$. In the first case, the percentage of concordant, positive signs amounts to $\simeq97\%$, while the percentage of concordant, negative signs amounts to $\simeq78\%$; in the second case, the percentages above rise to $\simeq99\%$.

Let us, now, characterise our projections in a more quantitative way: to this aim, we will indicate the adjacency matrix of a generic projection with $\mathbf{A}$ and refer to the na\"ive ones (see eq.~\ref{eq:nai}), those validated via the bSRGM and those validated via the bSCM with $\mathbf{A}_\text{na\"ive}$, $\mathbf{A}_\text{bSRGM}$ and $\mathbf{A}_\text{bSCM}$, respectively.

\begin{figure*}[t!]
\centering
\includegraphics[width=\linewidth]{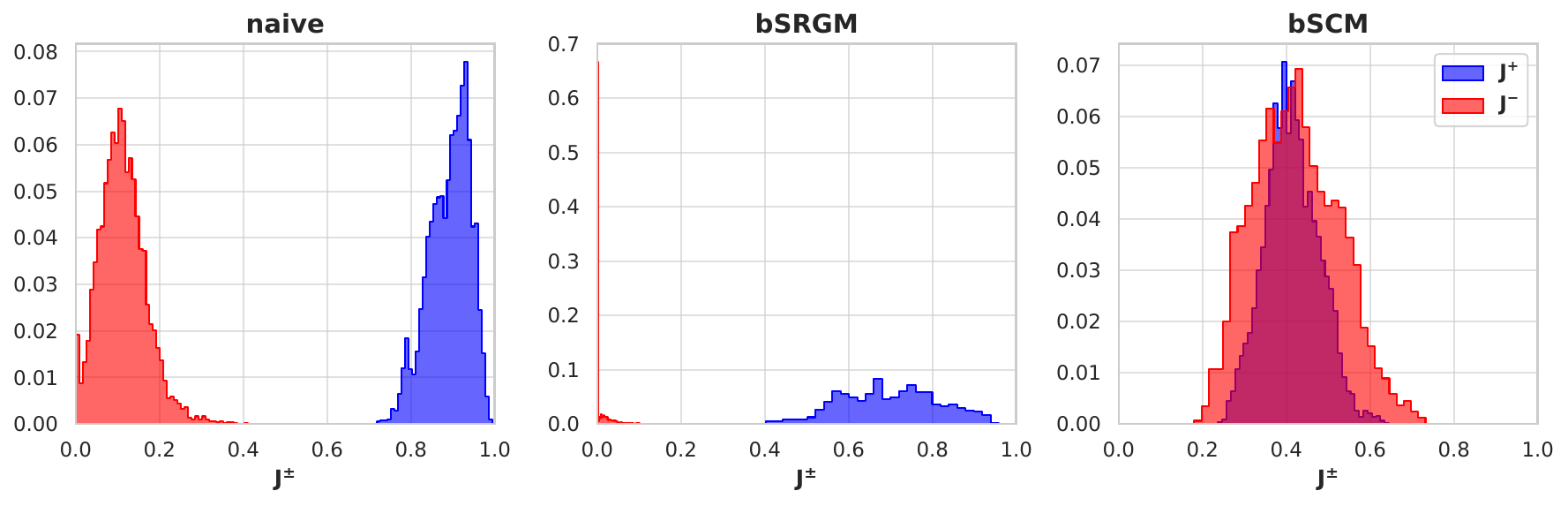}
\caption{\textbf{Distribution of the Jaccard similarity between pairs of projections.} The numerator of such an index represents the number of positive (negative) links occupying the same position in the adjacency matrix of the two subjects to compare and its denominator represents the cardinality of the union of the two sets of positive (negative) links. Bin widths are computed according to the Freedman-Diaconis rule. Since a value of $1$ indicates perfect overlap while a value of $0$ indicates no overlap, positive links yield a larger Jaccard similarity than negative links on both the na\"ive and bSRGM-induced projections, a result indicating the former ones as more spatially coherent than the latter ones. When considering bSCM-induced projections, instead, the spatial coherence of the two kinds of links appears very similar.}
\label{fig:6}
\end{figure*}

Let us, now, verify that the ensemble distribution of each pair-specific signature is consistent with its analytical counterpart, for all benchmarks considered here. Specifically, let us focus on each subject and test \emph{i)} if the distribution defined by eq.~\ref{eq:bino} is compatible with the ensemble distribution of $S_{ij}$, $\forall\:i<j$ induced by the bSRGM; \emph{ii)} if the distribution defined by eq.~\ref{eq:poibino} is compatible with the ensemble distribution of $S_{ij}$, $\forall\:i<j$ induced by the bSCM. Each subject can be associated with a KS-score, i.e. the percentage of times the Kolmogorov-Smirnov (KS) test (at the level of $5\%$) does not reject the hypothesis of compatibility: while a value of $1$ indicates perfect compatibility, a value of $0$ indicates perfect incompatibility. The results are summed up by the box-plots depicted in fig.~\ref{fig:3}: the median KS-score is $\text{KS}_{5\%}\simeq 0.965$ under both the bSRGM and the bSCM, the $95\%$ of these distributions ranging between the $2.5$-th and the $97.5$-th percentiles reading $[q_{2.5},q_{97.5}]=[0.961,0.970]$; in words, our sampling procedure satisfactorily reproduces the analytical distributions defined by eq.~\ref{eq:bino} and eq.~\ref{eq:poibino} even when the KS test rejects the hypothesis of compatibility (e.g. for subject $\#100307$).

\subsection{Patterns of brain (im)balance at the microscale}

\subsubsection{Connectance}

When considering the unsigned case, the connectance of the projection returned by the na\"ive approach is typically large~\cite{saracco2017}: this holds true here as well, for all the considered subjects (see fig.~\ref{fig:4}). It is worth noticing that many validated links are positive. This leads us to conclude that the time series associated with the related pairs of regions are concordant for the vast majority of the snapshots constituting the considered time window.

The positive connectance, defined as $\rho^+=2L^+/N(N-1)$ with $L^+=\sum_{i=1}^N\sum_{j(>i)}a_{ij}^+$, decreases as we employ stricter benchmarks: on average, in fact, $\rho^+_\text{na\"ive}=0.93$ while $\rho^+_\text{bSRGM}=0.75$ and $\rho^+_\text{bSCM}=0.22$. Such a result can be explained upon considering that concordant motifs represent the majority of motifs: filtering to a larger extent, thus, leads to a projection characterised by a smaller amount of positive links - precisely the ones induced by the aforementioned patterns. On the other hand, the negative connectance, defined as $\rho^-=2L^-/N(N-1)$ with $L^-=\sum_{i=1}^N\sum_{j(>i)}a_{ij}^-$, shows a (much) less trivial behaviour as, on average, $\rho^-_\text{na\"ive}=0.07$ while $\rho^-_\text{bSRGM}=0.01$ and $\rho^-_\text{bSCM}=0.32$.

Such a result can be explained upon considering that the global, bSRGM-based filter is defined by a binomial distribution centred on not-that-far right values of the signature: in fact, $c^-=B^-/(N\times T)\gtrsim c^+=B^+/(N\times T)$, as fig.~\ref{fig:4} confirms. Since the bipartite density of negative links is only slightly larger than the bipartite density of positive links, $\langle D_{ij}\rangle\gtrsim\langle C_{ij}\rangle$, i.e. $\langle S_{ij}\rangle\lesssim0$. As a consequence, several `na\"ively negative' signatures are no longer deemed as significant and the resulting negative links are cut; positive links, instead, are over-represented as the `validation threshold' becomes smaller: the global filter is, thus, stricter with negative than with positive links.

\begin{figure*}[t!]
\centering
\includegraphics[width=\linewidth]{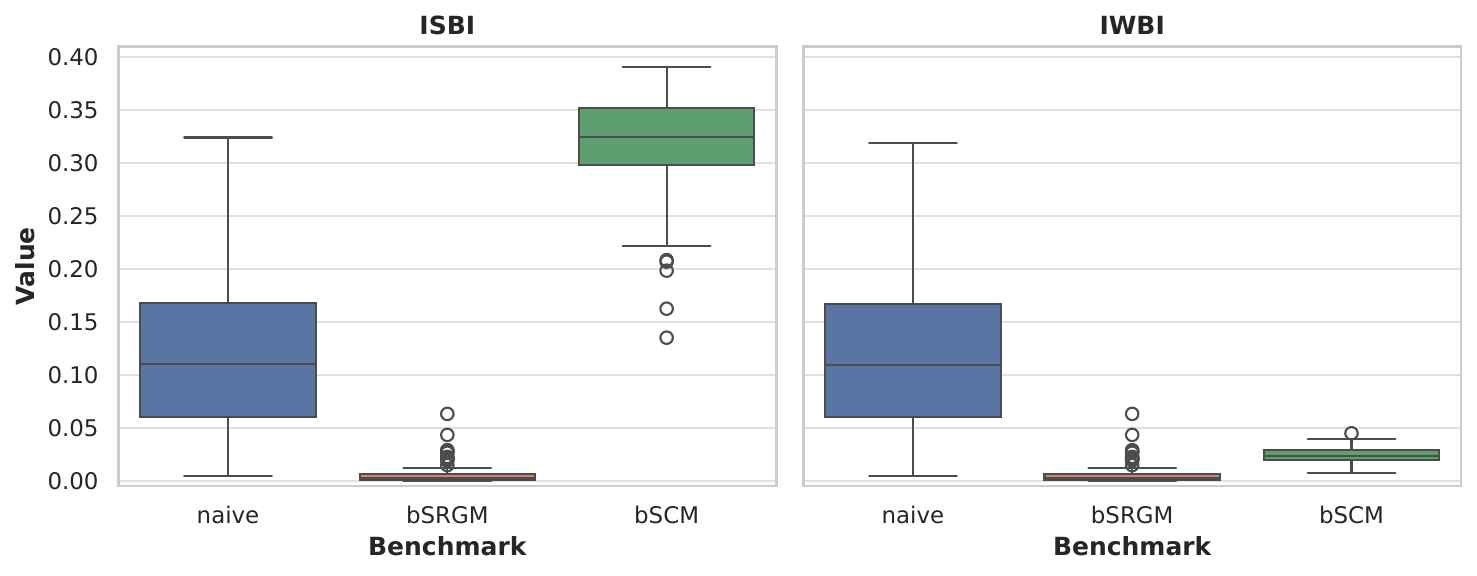}
\caption{\textbf{Distributions of the IBI values across subjects.} Box-plots summing up the distributions of the values of the strong (left) and weak (right) variants of the Index of Brain Imbalance (IBI) across the three different projections of our $100$ subjects. Each box-plot proxies the distribution of the corresponding metric, indicating its central tendency and dispersion: in both cases, the bSRGM-induced projections are characterised by the smallest median, while the na\"ive projections are peaked in correspondence of larger values, although showing more variability. The difference between the ISBI and the IWBI values characterising the bSCM-induced projections is due to the sensitivity of this benchmark to the abundance of the triad $(---)$, whose presence is enhanced by the predominantly negative relationships observed between subcortical regions as well as between limbic regions (see also fig.~\ref{fig:8}). Overall, however, each median IBI value is positive, a result confirming the non-null level of frustration characterising humans' brain networks.}
\label{fig:7}
\end{figure*}

The local, bSCM-based filter, instead, accounts for nodes heterogeneity: whenever the positive degree of a node is (much) larger than its negative degree, the distribution of its similarity with any other node moves to the (far) right, hence validating negative links to a much larger extent; the local filter is, thus, stricter with positive than with negative links, somehow rebalancing their number. In other words, when two regions are each strongly connected to most of the rest of the brain, yet share only a slight tendency towards `in phase' dynamics with one another, we assign a negative connection between them - loosely speaking, the bSCM work as an `enhancer' of negative signs, increasing the likelihood of observing asynchronies between pairs of regions.

\subsubsection{Tendency to make hubs}

A related index is the one named \emph{tendency to make hubs}, introduced in~\cite{Saberi2021} and quantifying the tendency of a given degree distribution - when dealing with signed graphs, both the positive and the negative degree distributions need to be considered - to host hubs. Hereby, we adopt a slightly different index, rooted in statistical theory and reducing the functional form of a given degree distribution to just one number: named \emph{coefficient of variation} (CV), it is defined as the ratio between the standard deviation and the expected value of the reference distribution, i.e.

\begin{equation}
\text{CV}^+=\frac{\text{std}[k^+]}{\overline{k^+}}=\frac{\text{std}[k^+]}{2L^+/N}
\end{equation}
and

\begin{equation}
\text{CV}^-=\frac{\text{std}[k^-]}{\overline{k^-}}=\frac{\text{std}[k^-]}{2L^-/N};
\end{equation}
while a CV smaller than $1$ indicates that the expected value is representative of the entire distribution (as its dispersion around the average is smaller than it), observing a CV larger than 1 indicates that the expected value is not representative of the entire distribution (as its dispersion around the average is larger than it) - from this perspective, a fat-tailed distribution is expected to obey $\text{CV}>1$.

As evident from fig.~\ref{fig:5}, none of our projections seems to be characterised by the presence of positive hubs ($\text{CV}^+_\text{na\"ive}=0.09$, $\text{CV}^+_\text{bSRGM}=0.25$, $\text{CV}^+_\text{bSCM}=0.75$); when considering the negative degrees, instead, na\"ive and bSRGM-induced distributions display a right-skewness that suggests the presence of nodes whose number of connections is (much) larger than that of the others ($\text{CV}^-_\text{na\"ive}=1.64$, $\text{CV}^-_\text{bSRGM}=3.16$): in a sense, thus, the subgraph induced by the positive links tends to be more homogeneous than the subgraph induced by the negative ones\footnote{As highlighted in~\cite{Luppi2024}, \emph{`[\dots] whereas cooperative (positive-valued) interactions in the effective connectivity tend to be strong, modular and relatively short-range, competitive interactions are weaker but more long-range, more diffuse and less clustered [\dots]'}.}. The bSCM, however, levels out such differences ($\text{CV}^-_\text{bSCM}=0.66$).

\begin{figure*}[t!]
\centering
\includegraphics[width=\linewidth]{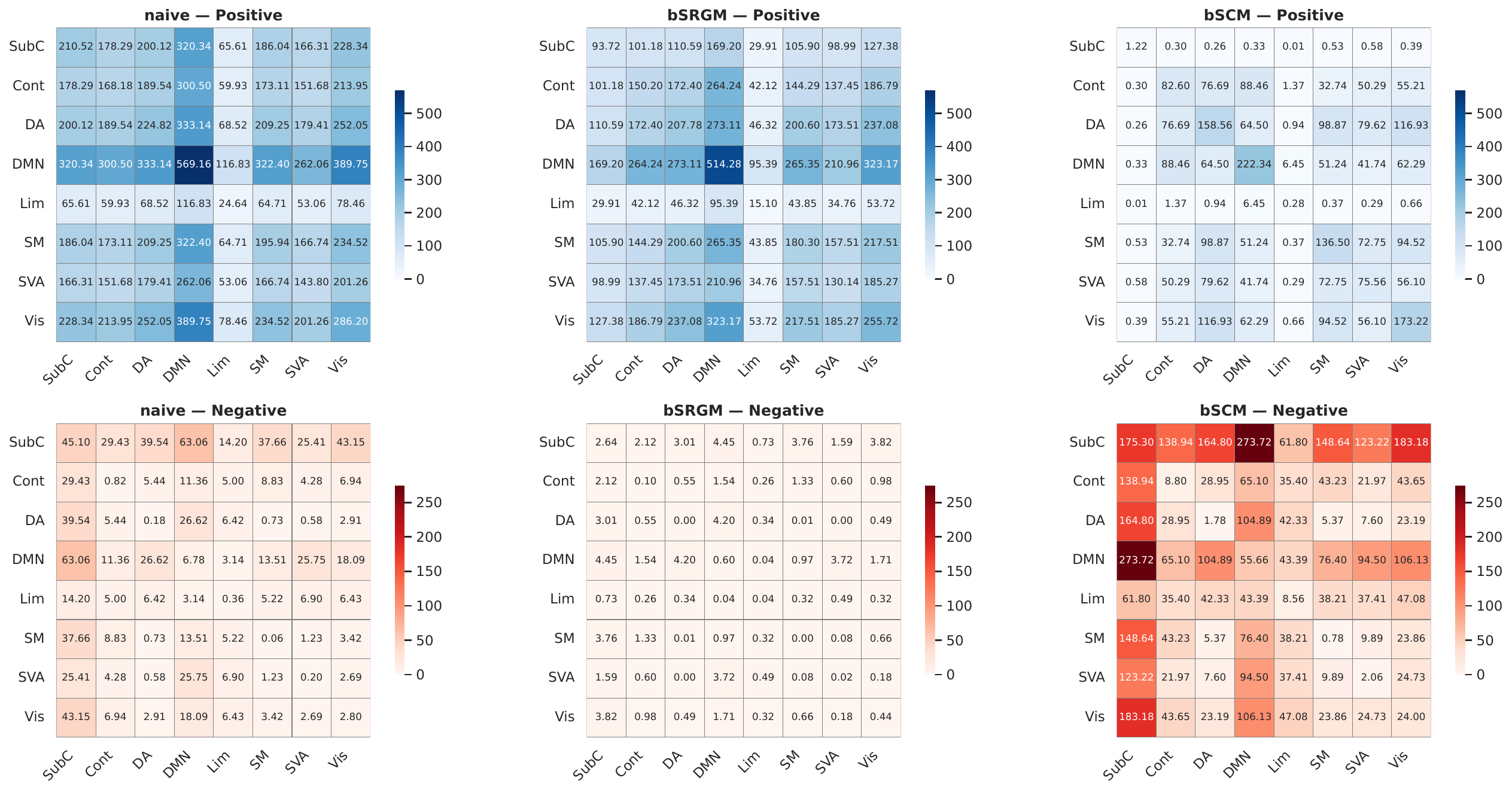}
\caption{\textbf{Heatmaps of the signed connections within and between Yeo's brain networks.} Values are averaged across subjects for each of our benchmarks. The intensity of each colour is proportional to the magnitude of the corresponding entry. Network labels stand for subcortical (SUBC), control (CONT), dorsal attention (DA), limbic (LIM), somatomotor (SM), salience/ventral attention (SVA), visual (VIS) networks, and default mode network (DMN). The neatest signal is observed for the bSCM-induced projections: while the largest number of positive links is found within the DMN, a large number of negative links is found between the DMN and every other network, as well as between the SUBC and every other network - the largest number of negative connections being located precisely between the DMN and the SUBC. Notice that, concerning the blocks lying on the main diagonal, $L^->L^+$ (only) for the SUBC and LIM networks (see also fig.~\ref{fig:7}).}
\label{fig:8}
\end{figure*}

We also explicitly show the distributions of the positive and negative degrees pooled across our subjects. As KS tests (at the level of $5\%$) reveal, practically all pairwise comparisons point out our three (na\"ive, bSRGM-induced, and bSCM-induced) positive degree distributions as significantly different; moreover, the appearance of positive hubs is enhanced when the global filter is used.

A related result concerns the degree distributions of cortical and subcortical regions: as evident from the figure in Appendix~\hyperlink{AppD}{D}, subcortical regions tend to have a larger negative degree than cortical regions.

\subsubsection{Spatial coherence of signed connections}

From what we have learned so far, the positive subgraphs are (overall) structurally homogeneous, while the negative ones are (overall) structurally heterogeneous. Let us, now, ask ourselves how similar positive and negative subgraphs are \emph{across subjects}.

In order to answer this question, we calculated the Jaccard similarity between pairs of (adjacency matrices representing) individual projections, per benchmark. As fig.~\ref{fig:6} shows, the positive variant of such an index (i.e. the percentage of positive links located in the same position) ranges between $0.4$ and $1$ while its negative variant (i.e. the percentage of negative links located in the same position) ranges between $0$ and $0.4$, on both the na\"ive and bSRGM-induced projections: in words, the spatial coherence of positive links is larger than the spatial coherence of negative links, which appear as those (better) characterising single subjects. This picture changes when bSCM-induced projections are considered: in such a case, in fact, the spatial coherence of the two kinds of links appears as very similar.

\begin{figure*}[t!]
\centering
\includegraphics[width=\linewidth]{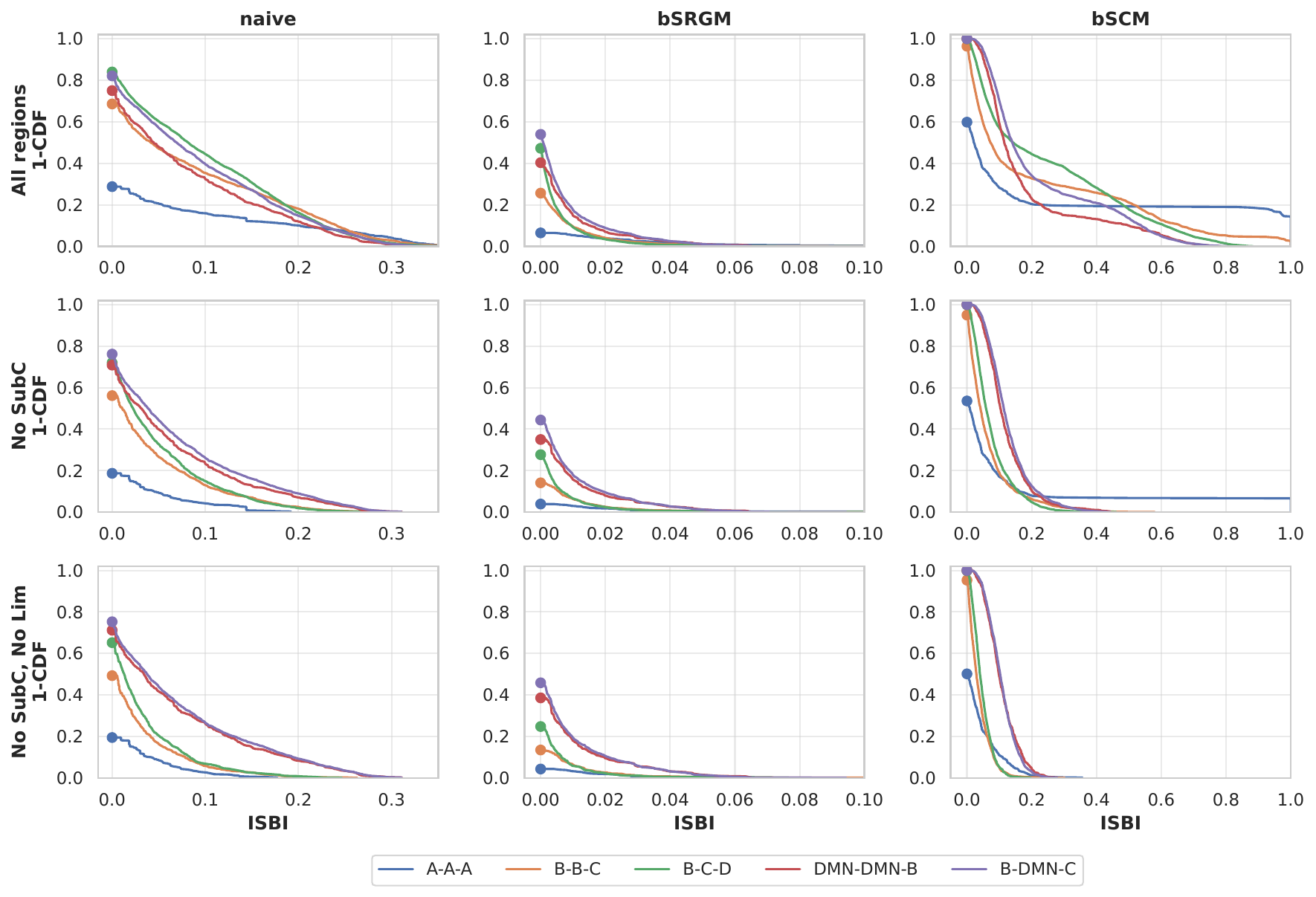}
\caption{\textbf{Distributions of the ISBI values associated with triplets of brain regions.} Cumulative distributions of the ISBI values across the three, different projections of our $100$ subjects: top panels refer to all regions, middle panels refer to all regions without subcortical ones, bottom panels refer to all regions without subcortical and limbic ones. The largest ISBI values characterise the triplets of subcortical regions: removing them leads to a different picture as the largest ISBI values, now, characterise the triplets within the LIM network; removing them as well reveals that the largest ISBI values characterise the triplets where the DMN is present once or twice - and, to a lesser extent, those within the DMN and CONT networks. The smallest ISBI values, instead, characterise those belonging to the same Yeo's network (especially the DA, SM and SVA networks).}
\label{fig:9}
\end{figure*}

\subsubsection{Assessing the brain (im)balance at the microscale}

Let us now inspect the arrangement of negative links within projections. To this aim, two choices are possible: (1) the first one amounts at considering the index named \emph{strong degree of balance}~\cite{gallo2024testing} and counting the triads characterised by an even number of negative links, i.e. either zero or two - in other words, the ones that are balanced according to the strong balance theory; (2) the second one amounts at considering the index named \emph{weak degree of balance}~\cite{gallo2024testing} and counting the triads characterised by zero, two or three negative links - in other words, the ones that are balanced according to the weak balance theory.

The indices above induce two variants of the Index of Brain Imbalance (IBI - practically, their complementary to $1$), i.e. the \emph{Index of Strong Brain Imbalance}

\begin{align}
\text{ISBI}&=\frac{T^{(+-+)}+T^{(---)}}{T^{(+++)}+T^{(-+-)}+T^{(+-+)}+T^{(---)}}
\end{align}
and the \emph{Index of Weak Brain Imbalance}

\begin{align}
\text{IWBI}&=\frac{T^{(+-+)}}{T^{(+++)}+T^{(-+-)}+T^{(+-+)}+T^{(---)}},
\end{align}
with

\begin{align}
T^{(+++)}&=\frac{\text{Tr}[(\mathbf{A}^+)^3]}{6},\\
T^{(-+-)}&=\frac{\text{Tr}[\mathbf{A}^-\mathbf{A}^+\mathbf{A}^-]}{2},\\
T^{(+-+)}&=\frac{\text{Tr}[\mathbf{A}^+\mathbf{A}^-\mathbf{A}^+]}{2},\\
T^{(---)}&=\frac{\text{Tr}[(\mathbf{A}^-)^3]}{6};
\end{align}
naturally, the closer their value to $1$, the larger the amount of unbalanced triads.

As fig.~\ref{fig:7} shows, the bSRGM-induced projections are the ones characterised by the largest amount of balanced triads - a result that is not surprising once considering the small amount of negative links populating them; the na\"ive projections, instead, are characterised by a number of balanced triads that amounts at five-to-six times the number of unbalanced triads. Lastly, and most interestingly, the behaviour of the bSCM-induced projections is sensitive to the one of the triad $(---)$, that can be quite abundant: such a behaviour characterises the subcortical and the limbic regions, partly reconciling our findings with those in~\cite{Luppi2024}\footnote{Speaking of cortical regions, there it is reported that \emph{`[\dots] it is less likely that competitively-interacting neighbours of a node will themselves be interacting competitively [\dots]'}.}. Overall, however, each median IBI value is positive\footnote{As confirmed by a one-sided Wilcoxon signed-rank test at the confidence level of $5\%$.}, a result confirming the non-null level of frustration characterising humans' brain networks: if, according to~\cite{Saberi2021}, the value $\text{IBI}=0$ is interpreted as the one characterising the energy ground state, the aforementioned results imply that humans' brain networks are found in some sort of excited state.\\

Let us, now, consider the brain parcellation carried out in~\cite{yeo2011organization}, where the authors propose a functional partition of brain regions by distinguishing communities of regions (hereby, \emph{Yeo's networks} or \emph{communities}) involved in sensory tasks (e.g. the visual and the somatomotor networks) from those involved in higher-order, cognitive tasks (e.g. the DMN): they are named \emph{control} (CONT), \emph{dorsal attention} (DA), \emph{limbic} (LIM), \emph{somatomotor} (SM), \emph{salience/ventral attention} (SVA), \emph{visual} (VIS) \emph{networks} and \emph{default mode network} (DMN). While the analysis carried out in~\cite{Demertzi2022} highlights the presence of negative correlations between the DMN and other networks, the one carried out in~\cite{Saberi2022} shows that, in healthy individuals, unbalanced triads predominantly occur between different Yeo's networks.

In order to verify the first of the aforementioned findings, we have constructed two $8\times 8$ matrices: each entry of the first (second) one encodes the number of positive (negative) links between the corresponding Yeo's networks - to which the \emph{subcortical regions} have been added - aggregated across individuals. As fig.~\ref{fig:8} shows, \emph{i)} the largest number of positive links is found within the DMN; \emph{ii)} a large number of negative links is found between the DMN and every other network, as well as between the SUBC and every other network. While the first result confirms the empirical `in phase' dynamics of the ROIs constituting the default mode network, the second one seems to (at least) partly reflect a filter-induced sign reversal; indeed, the subcortical regions are characterised by one of the highest degrees of independence from the remaining ROIs, comparable only to that of the limbic regions: as a consequence, the corresponding value of the signature falls below the validation threshold imposed by the local filter.

In order to verify the second finding above, we have inspected the distributions of the ISBI values associated with the admissible triplets of regions: as evident from fig.~\ref{fig:9}, \emph{i)} the SUBC triplets are those exhibiting the highest level of imbalance\footnote{As highlighted in~\cite{Saberi2022}, \emph{`Our results suggest that SUBC regions have a prominent role in frustration formation in the brain network at both nodal and connectional levels. So they can bring instability and alter properties that facilitate systemic level neural changes, providing adaptive characteristics to the brain'}.}; \emph{ii)} the cortical triplets belonging to LIM exhibit the highest level of imbalance, followed by those where the DMN is present once or, at most, twice; \emph{iii)} the cortical triplets belonging to the same Yeo's network are those exhibiting the lowest level of imbalance (the smallest values characterising the triplets involving nodes from the DA, SM and SVA networks) - with the exception of the regions belonging to LIM and, to a lesser extent, those belonging to DMN and CONT.

\subsection{Patterns of brain (im)balance at the mesoscale}

Following~\cite{gallo2024assessing}, one can adopt an `agnostic' attitude and explore the mesoscale organisation of a projection without aligning with any specific conceptual framework. A principled approach to achieve such a goal is that of minimising the quantity $\text{BIC}=\kappa\ln V-2\mathcal{L}$ where, now, $V=N(N-1)/2$. Since we aim at describing a projection mesoscale organisation, a natural choice is that of instantiating BIC with the Signed Stochastic Block Model (SSBM), defined by the likelihood function

\begin{align}
\mathcal{L}_\text{SSBM}=&\prod_{r=1}^k(p_{rr}^+)^{L_{rr}^+}(p_{rr}^-)^{L_{rr}^-}(1-p_{rr}^+-p_{rr}^-)^{\binom{N_r}{2}-L_{rr}}\nonumber\\
&\prod_{r=1}^k\prod_{\substack{s=1\\s>r}}^k(p_{rs}^+)^{L^+_{rs}}(p_{rs}^-)^{L^-_{rs}}(1-p_{rs}^+-p_{rs}^-)^{N_rN_s-L_{rs}}
\end{align}
and a number of parameters $\kappa_\text{SSBM}=k(k+1)$, with $k$ indicating the number of modules into which the projection is partitioned. Naturally, $N_r$ is the number of nodes constituting block $r$, $p_{rr}^+$ is induced by the empirical number of positive links within block $r$, i.e.

\begin{figure*}[t!]
\centering
\includegraphics[width=\linewidth]{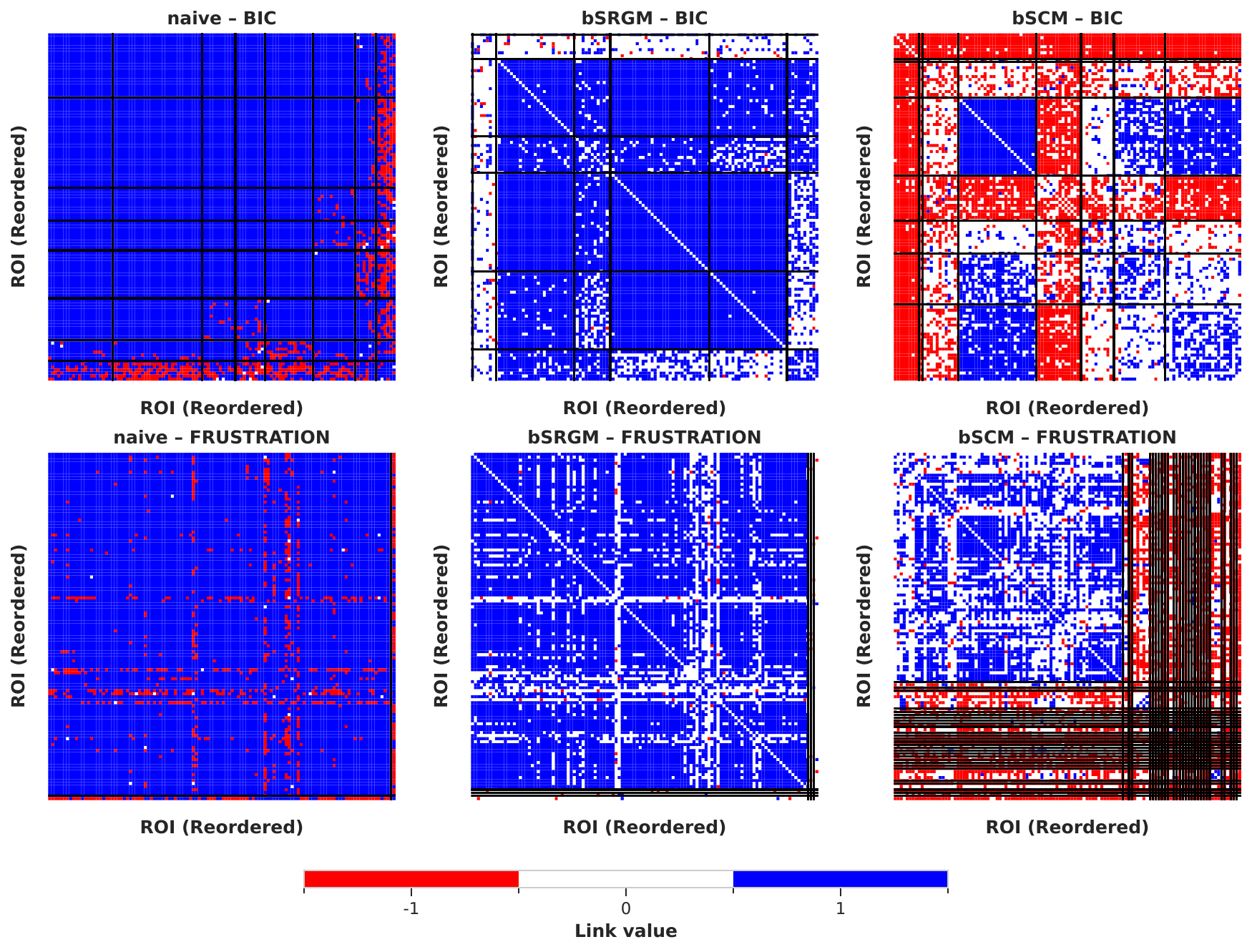}
\caption{\textbf{Signed communities partitioning the brain of subject $\bm{\#106016}$.} Top panels: heatmaps illustrating the communities partitioning the projections of subject $\#106016$, detected by minimising the Bayesian Information Criterion (BIC). Bottom panels: heatmaps illustrating the communities partitioning the projections of subject $\#106016$, detected by minimising the frustration ($F$). Although the number of detected communities varies, the overall message conveyed by the two methods can be summed up as follows: the brain seems to be constituted by groups of regions working `in phase' (i.e. in a \emph{synchronous} fashion) most of the time and `in counterphase' (i.e. in an \emph{asynchronous} fashion) with those of other groups; still, BIC minimisation admits regions from the same group working `in counterphase', as well as regions from different groups working `in phase' - while $F$ minimisation does not.}
\label{fig:10}
\end{figure*}

\begin{equation}
p_{rr}^+=\frac{2L_{rr}^+}{N_r(N_r-1)},
\end{equation}
$p_{rr}^-$ is induced by the empirical number of negative links within block $r$, i.e.

\begin{equation}
p_{rr}^-=\frac{2L_{rr}^-}{N_r(N_r-1)},
\end{equation}
$p_{rs}^+$ is induced by the empirical number of positive links between blocks $r$ and $s$, i.e.

\begin{equation}
p_{rs}^+=\frac{L_{rs}^+}{N_rN_s},
\end{equation}
and $p_{rs}^-$ is induced by the empirical number of negative links between blocks $r$ and $s$, i.e.

\begin{equation}
p_{rs}^-=\frac{L_{rs}^-}{N_rN_s}.
\end{equation}

\begin{figure*}[t!]
\centering
\includegraphics[width=\linewidth]{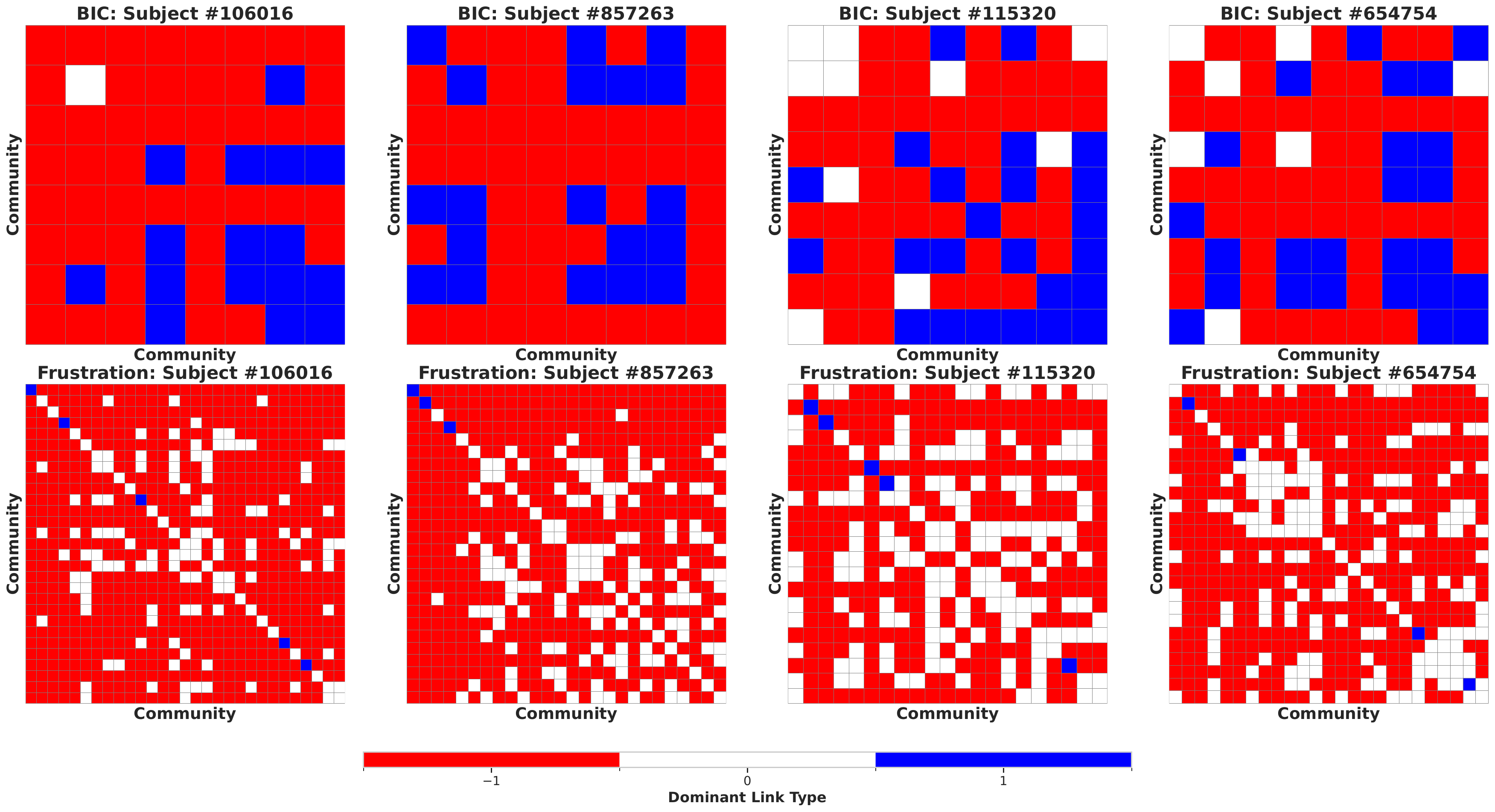}
\caption{\textbf{Heatmaps of the communities partitioning the bSCM-induced projections of four subjects.} Matrices are coloured according to eq.~\ref{eq:diagonal_balance} and eq.~\ref{eq:off_diagonal_balance}, i.e. blocks are blue (red) if the majority of links populating them is positive (negative); white blocks on the diagonal indicate singletons and white blocks off the diagonal indicate that $p_{rs}^+=p_{rs}^-$. Top panels: the communities partitioning the bSCM-induced projections of our subjects are detected by minimising the Bayesian Information Criterion (BIC); all subjects exhibit diagonal blocks with a majority of negative links and off-diagonal blocks with a majority of positive links, thereby satisfying the statistical variant of the Relaxed Balance Theory. Bottom panels: the communities partitioning the bSCM-induced projections of our subjects are detected by minimising the frustration ($F$); (left aside the white ones) all subjects admit blocks with a majority of negative links only off the main diagonal and blocks with a majority of positive links only on the main diagonal, hence aligning better with the statistical variant of the Traditional Balance Theory.}
\label{fig:11}
\end{figure*}

\subsubsection{A survey on individual projections}

Let us start by considering individual partitions, stressing that a large link density does not prevent BIC minimisation from detecting statistically significant mesoscopic structures: its sensitivity to both the number and the sign of the links, in fact, makes it capable of partitioning both denser and sparser configurations.\\

\noindent\textit{Na\"ive projections.} As fig.~\ref{fig:10} shows, BIC minimisation reveals the presence of $8$ communities: the vast majority of links \emph{within} communities - if not all, as evident upon looking at some of the modules - is positive, a result implying that the series describing the activity of these regions are concordant most of the time. For what concerns the links \emph{between} communities, instead, the picture becomes much more interesting: all communities in fig.~\ref{fig:10} are, in fact, connected by mostly negative links with the eighth, an evidence suggesting that such a block of regions is `in counterphase' with every other block most of the time; the first and second communities, instead, are connected by mostly positive links, i.e. even if these blocks of regions are separated, they still seem to work `in phase' (i.e. in a \emph{synchronous} fashion).\\

\noindent\textit{Validated projections.} Filtering our projections with a homogeneous benchmark makes the picture even clearer, as each block is now constituted by links of practically the same sign: more precisely, while positive links constitute the majority of links, (few) negative links are found only between modules. Filtering our projections with a heterogeneous benchmark, instead, returns a picture that is halfway between the na\"ive one and the bSRGM-induced one, as the blocks are sparser than the na\"ive ones, less homogeneous than those induced by the bSRGM, but richer in negative connections than both. The overall message conveyed by such an analysis, however, does not change: the brain seems to be constituted by groups of regions working `in phase' most of the time and `in counterphase' (i.e. in an \emph{asynchronous} fashion) with those of other groups; still, BIC minimisation admits regions working `in counterphase' even if belonging to the same group (see, for instance, the patterns highlighted in figs.~\ref{fig:7} and~\ref{fig:8}), as well as regions working `in phase' even if belonging to different groups.

\begin{figure*}[t!]
\centering
\includegraphics[width=\linewidth]{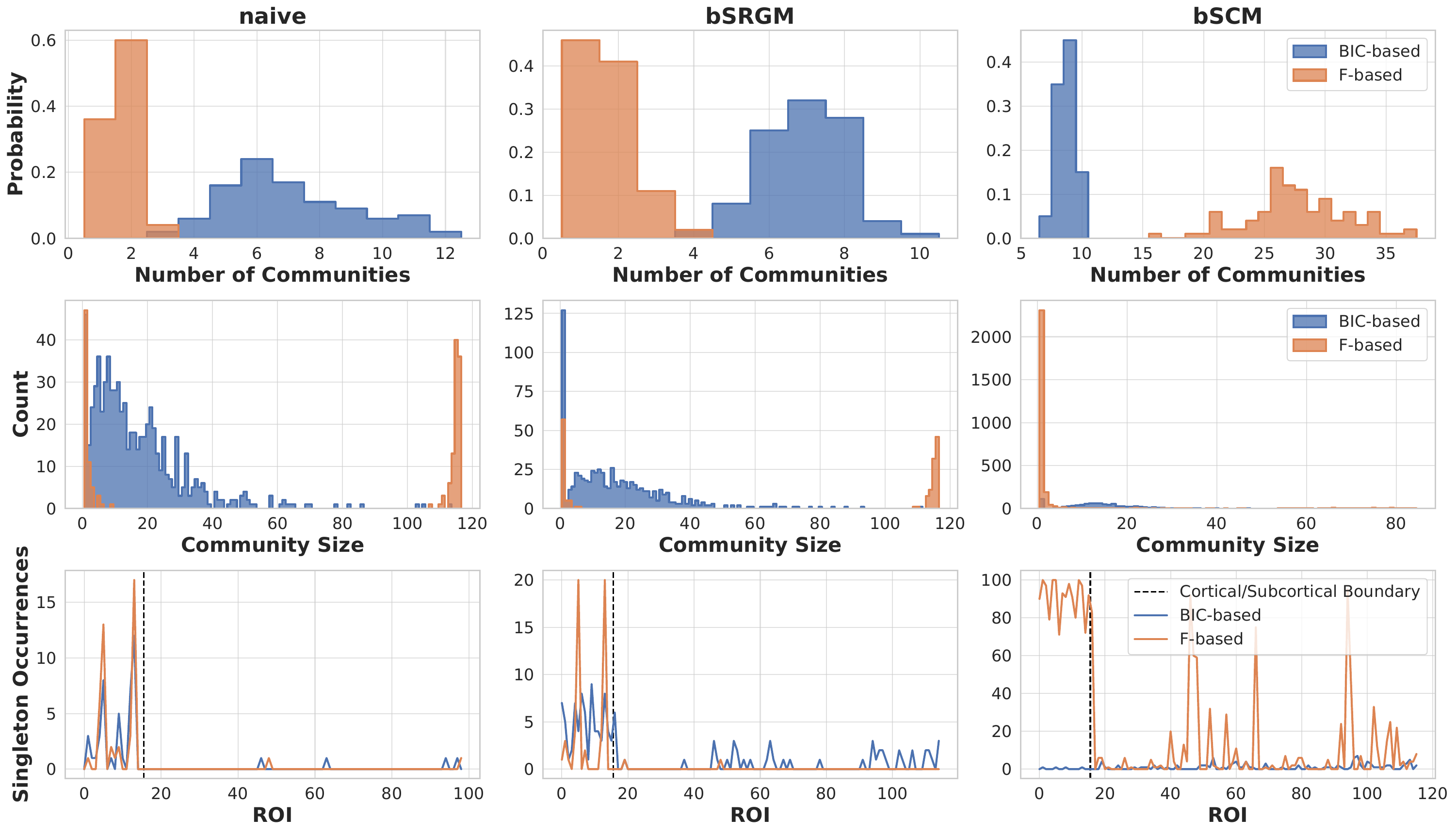}
\caption{\textbf{Basic statistics of the partitions across subjects.} Top panels: histograms of the number of communities detected on the projections returned by each benchmark and optimisation criterion. For what concerns na\"ive and bSRGM-induced projections, $F$ avoids creating frustrated patterns by individuating a smaller number of larger communities, while BIC singles out a larger number of smaller communities; for what concerns bSCM-induced projections, instead, $F$ individuates a larger number of smaller communities than BIC. Middle panels: histograms of the size of communities detected on the projections returned by each benchmark and optimisation criterion. These panels refine the picture provided by the previous ones, clarifying that $F$ tends to separate the singletons from the set of remaining nodes; BIC tends to individuate singletons as well, although the size of communities into which the set of remaining nodes is partitioned varies over a broader range. Bottom panels: number of times each brain region is individuated as a singleton. As evident from the plot, this is almost solely observed for subcortical regions, especially on $F$-induced partitions; the most `integrated' picture is, instead, returned by the bSCM-induced projections, partitioned by minimising BIC.}
\label{fig:12}
\end{figure*}

\subsubsection{Traditionally or relaxedly balanced?}

Motivated by the last observation, we now employ our benchmarks to probe the patterns of structural (im)balance at the mesoscopic level. We also explicitly acknowledge that discussing balance theory for brain networks may seem somewhat inappropriate; nonetheless, this is the only existing framework to interpret our results.\\

\noindent\textit{Traditional theory of balance.} According to the strong variant of the Traditional Balance Theory (TBT), the optimal partition of a network set of nodes consists of two groups, whose internal (external) connections are positive (negative); by contrast, the weak variant of the TBT allows for any number of groups. More formally, both versions of the TBT can be probed by finding the partition $\bm{\sigma}$ that minimises the \emph{frustration}, defined as

\begin{align}
F(\bm{\sigma})=L_\bullet^-+L_\circ^+,
\end{align}
i.e. as the number of negative links within modules (indicated with a filled dot) plus the number of positive links between modules (indicated with an empty dot). Let us also remark that minimising $F$ leads to results that are very similar to the ones obtained upon maximising the signed modularity~\cite{gallo2024assessing}.

The na\"ive, $F$-induced partition in fig.~\ref{fig:10} illustrates a first way the optimisation of $F$ works: a large number of positive links leads $F$ to collect them together, being more convenient to accommodate a small number of negative links within a larger, mostly positive, block than letting a large number of positive links be placed between blocks. The heterogeneous, $F$-induced partition in fig.~\ref{fig:10}, instead, illustrates a second way the optimisation of $F$ works: a large number of negative links leads $F$ to split them, even at the price of creating many singletons. In any case, the minimisation of $F$ further confirms that none of our projections obeys the traditional formulation of the Balance Theory.\\

\begin{figure*}[t!]
\centering
\includegraphics[width=\linewidth]{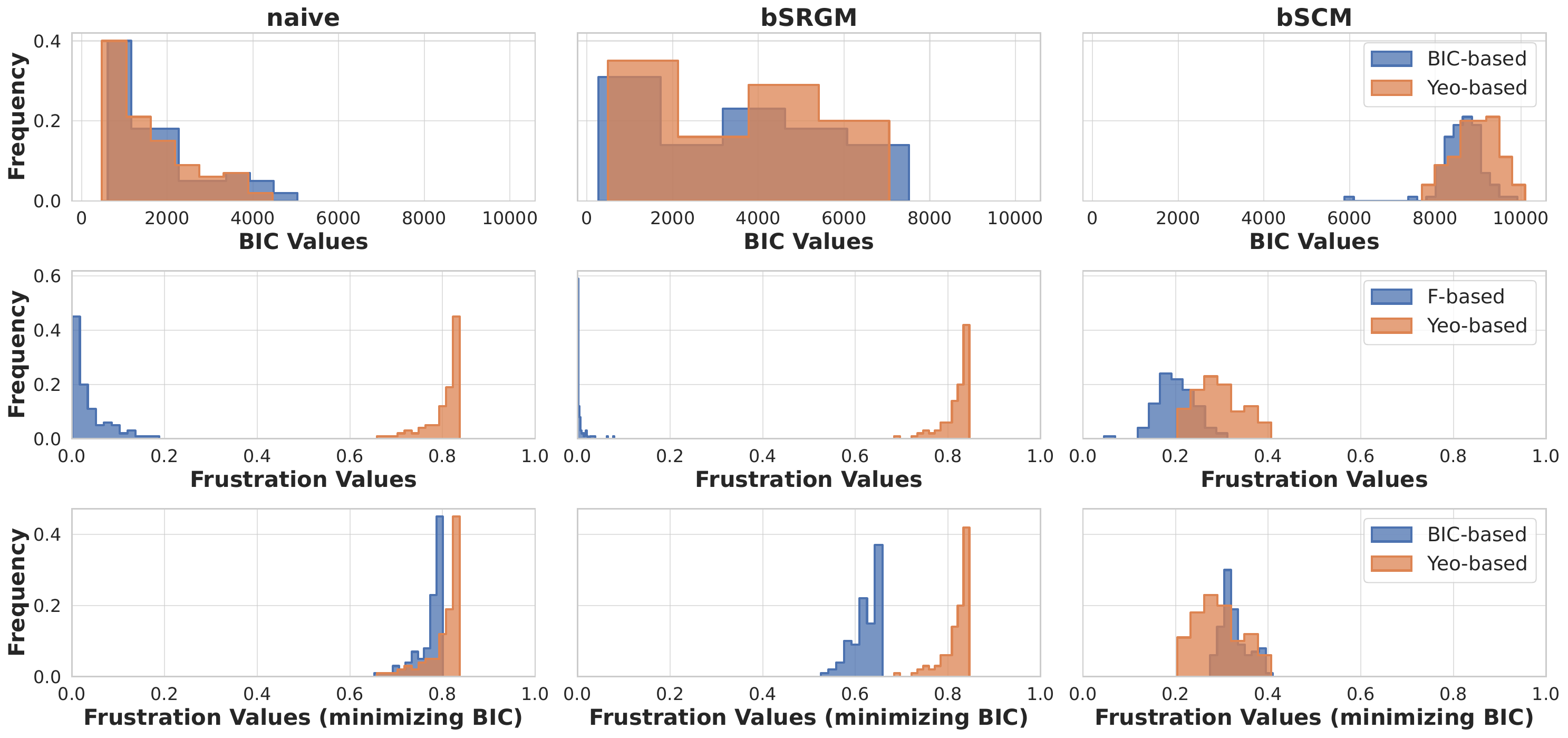}
\caption{\textbf{Distributions of the BIC and $F$ values across the partitions induced by their minimisation.} Top panels: histograms of the BIC values characterising the partitions determined by minimising it (in blue) and Yeo's partitions (in orange). As expected, the former are characterised by a slightly smaller value of such a score than Yeo's partitions. Middle panels: histograms of the $F$ values characterising the partitions determined by minimising it (in blue) and Yeo's partitions (in orange). In this second case, the former are characterised by a (much) smaller value of such a score than Yeo's partitions. Bottom panels: histograms of the $F$ values characterising the partitions determined by minimising BIC (in blue) and Yeo's partitions (in orange). In this third case, both are characterised by a quite large value of $F$. Bin widths are computed according to the Freedman-Diaconis rule. Taken together, our results indicate that \emph{i)} brain networks do not seek to minimise frustration - even when removing the subcortical regions; \emph{ii)} BIC minimisation embodies a principle returning a structure that captures (at least) part of the features of the Yeo's task-based one.}
\label{fig:13}
\end{figure*}

\noindent\textit{Statistical theory of balance.} Such a result motivates us to inspect the theory of balance best accommodating our observations. To answer such a question, let us first notice that the TBT can be formalised upon writing

\begin{equation}
p_{rr}^-=0,\quad\forall\:r
\end{equation}
and

\begin{equation}
p_{rs}^+=0,\quad\forall\:r<s
\end{equation}
with $p_{rr}^+$ indicating the probability that any two nodes belonging to the same block $r$ are connected by a positive link, $p_{rs}^+$ indicating the probability that any two nodes belonging to the different blocks $r$ and $s$ are connected by a positive link and analogously for their negative counterparts. As noticed in~\cite{gallo2024assessing}, however, the positions defining the TBT can be replaced with the milder ones reading

\begin{equation}\label{eq:diagonal_balance}
\text{sgn}[p_{rr}^+-p_{rr}^-]=+1,\quad\forall\:r
\end{equation}
which amounts to requiring $p_{rr}^+>p_{rr}^-$, $\forall\:r$ and

\begin{equation}\label{eq:off_diagonal_balance}
\text{sgn}[p_{rs}^+-p_{rs}^-]=-1,\quad\forall\:r<s
\end{equation}
which amounts to requiring $p_{rs}^+<p_{rs}^-$, $\forall\:r<s$: in words, the deterministic rules firstly defined by Cartwright, Harary and Davis are replaced by a set of probabilistic criteria individuating `a tendency' to obey, or not to obey, the TBT. More formally, a configuration satisfying eqs.~\ref{eq:diagonal_balance} and~\ref{eq:off_diagonal_balance} will be claimed to support the \emph{statistical variant} of the TBT - specifically, its strong variant if $k=2$ and its weak variant if $k>2$; if, instead, $p_{rr}^+\leq p_{rr}^-$ for some diagonal blocks or $p_{rs}^+\geq p_{rs}^-$ for some off-diagonal blocks, it will be claimed to support the \emph{statistical variant of the Relaxed Balance Theory} (RBT).\\

In order to evaluate the alignment of our projections with either statistical variants, let us take a look at the top panels of fig.~\ref{fig:10}, depicting the BIC-induced partitions: since $p_{rs}^+\leq p_{rs}^-$ for several, diagonal blocks (i.e. with $r=s$) and $p_{rs}^+\geq p_{rs}^-$ for several, off-diagonal blocks (i.e. with $r\neq s$), we are led to the conclusion that all projections obey the statistical variant of the RBT~\cite{gallo2024assessing} as a non-negligible number of negative (positive) links is found within (between) - and not only between (within) - clusters. As fig.~\ref{fig:11} further confirms, no subject obeys the statistical variant of the TBT, regardless of the benchmark employed, when considering the BIC-induced partitions.\\

Let us, now, take a look at the bottom panels of fig.~\ref{fig:10}, depicting the $F$-induced partitions: although eqs.~\ref{eq:diagonal_balance} and~\ref{eq:off_diagonal_balance} are not strictly obeyed, our subject, now, admits blocks with a majority of negative links only off the main diagonal and blocks with a majority of positive links only on the main diagonal; we are, thus, led to the conclusion that all projections align better with the statistical variant of the TBT. As fig.~\ref{fig:11} further confirms, all subjects align better with the statistical variant of the TBT, regardless of the benchmark employed, when considering the $F$-induced partitions.

\begin{figure*}[t!]
\centering
\includegraphics[width=\linewidth]{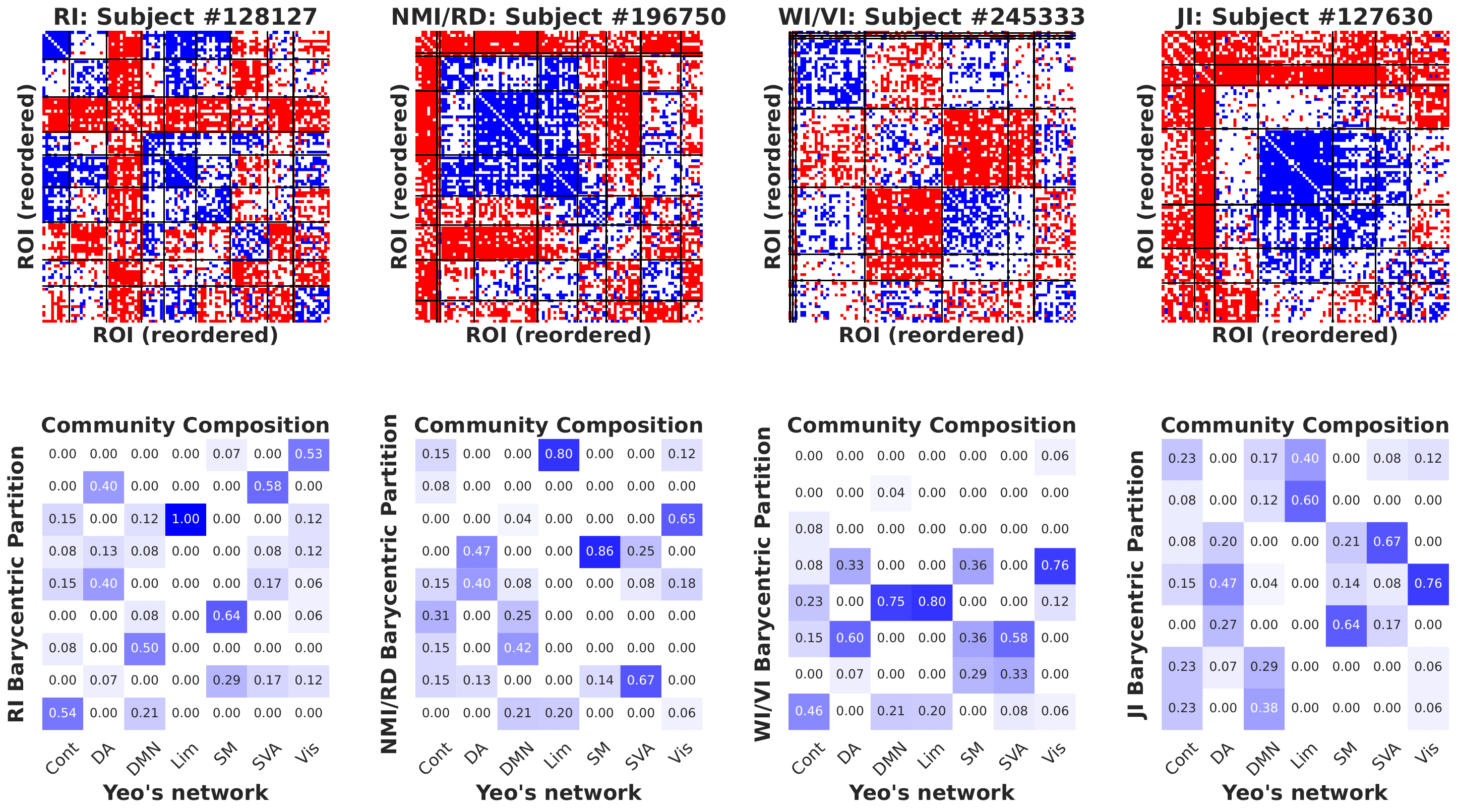}
\caption{\textbf{Brain centroids individuated by the considered distances between partitions.} Top panels: brain centroids individuated by adopting the Rand Index (RI), the Normalised Mutual Information (NMI), the Rajski distance (RD), the Wallace Index (WI), the Jaccard Index (JI), and the Variation of Information (VI). Generally speaking, while RI, NMI, and RD individuate projections with a larger number of predominantly positive communities, WI, JI, and VI individuate projections with a smaller number of predominantly positive communities. Such a result may be explained by noticing that indices like WI and JI are solely defined in terms of true positives: as a consequence, identifying a smaller number of larger communities may enhance the detection of such events. Indices like RI, on the other hand, account for true negatives as well, whence the larger number of smaller communities. Bottom panels: different metrics induce a different overlap between our structure-based partitions and Yeo's task-based one. Overall, the communities identified by WI, JI, VI, and RD are able to capture up to $86\%$ of certain Yeo's networks, while RI recovers the LIM network in its entirety.}
\label{fig:14}
\end{figure*}

\subsection{Towards a representative brain?}

Let us now move from individual analyses to the aggregated one, the aim being to understand the extent to which the previous results can be generalised.

\subsubsection{Distribution of the number of communities}

First, let us plot the distribution of the number of communities per benchmark and optimisation criterion. The results are shown in fig.~\ref{fig:12}: for what concerns na\"ive and bSRGM-induced projections, the number of communities detected by minimising the frustration is practically always smaller than the number of communities detected by minimising BIC - the $F$-induced mode being $2$ and the BIC-induced mode being $6$-$7$; moreover, the $F$-induced distributions are narrower than the BIC-induced ones, a result hinting at a more pronounced inter-subject variability, in the second case. Concerning the bSCM-induced projections, instead, $F$ identifies a larger and more dispersed number of smaller communities than BIC.

\subsubsection{The role of singletons}

The histograms of the size of communities detected on the same projections refine the aforementioned picture, clarifying that $F$ tends to individuate simpler structures than BIC, such as fewer modules accompanied by several singletons detached from the rest of the nodes - although BIC may individuate singletons as well, the size of communities into which the rest of the nodes is partitioned varies over a broader range.

Interestingly, singletons often coincide with the so-called \emph{subcortical regions}, i.e. regions serving as hubs `integrating' the \emph{cortical regions}, among the other things: more quantitatively, the fraction of subcortical singletons amounts to \emph{i)} $\simeq0.96$ according to $F$ and $\simeq0.91$ according to BIC, on the na\"ive projections; \emph{ii)} $\simeq0.96$ according to $F$ and $\simeq0.60$ according to BIC, on the bSRGM-induced projections; \emph{iii)} $\simeq0.63$ according to $F$ and $\simeq0.04$ according to BIC, on the bSCM-induced projections. In words, carrying out a BIC-based community detection on bSCM-induced projections rarely leads to isolate subcortical regions, a result indicating that accounting for local properties such as the series- and the time-specific degrees leads to recover projections where all regions are (practically always) integrated.

\subsubsection{Distribution of BIC and $F$ values}

A related analysis concerns the distributions of the BIC and $F$ values across the partitions determined by minimising each of them. As fig.~\ref{fig:13} confirms, the partitions retrieved by minimising BIC are characterised by a smaller value of such a score than Yeo's partitions; besides, both kinds of partitions are characterised by a similar, and quite large, value of $F$. Inspecting the distribution of the values of $F$, instead, reveals that the partitions retrieved by minimising it are much less frustrated than Yeo's partitions. Taken together, our results suggest that \emph{i)} brain networks do not seek to minimise frustration (even when removing subcortical regions - as in this case, to carry out a fair comparison with Yeo's classification); \emph{ii)} BIC minimisation is capable of capturing a structure that shares several similarities with the one defined by Yeo's task-based networks.

\begin{figure*}[t!]
\centering
\includegraphics[width=\linewidth]{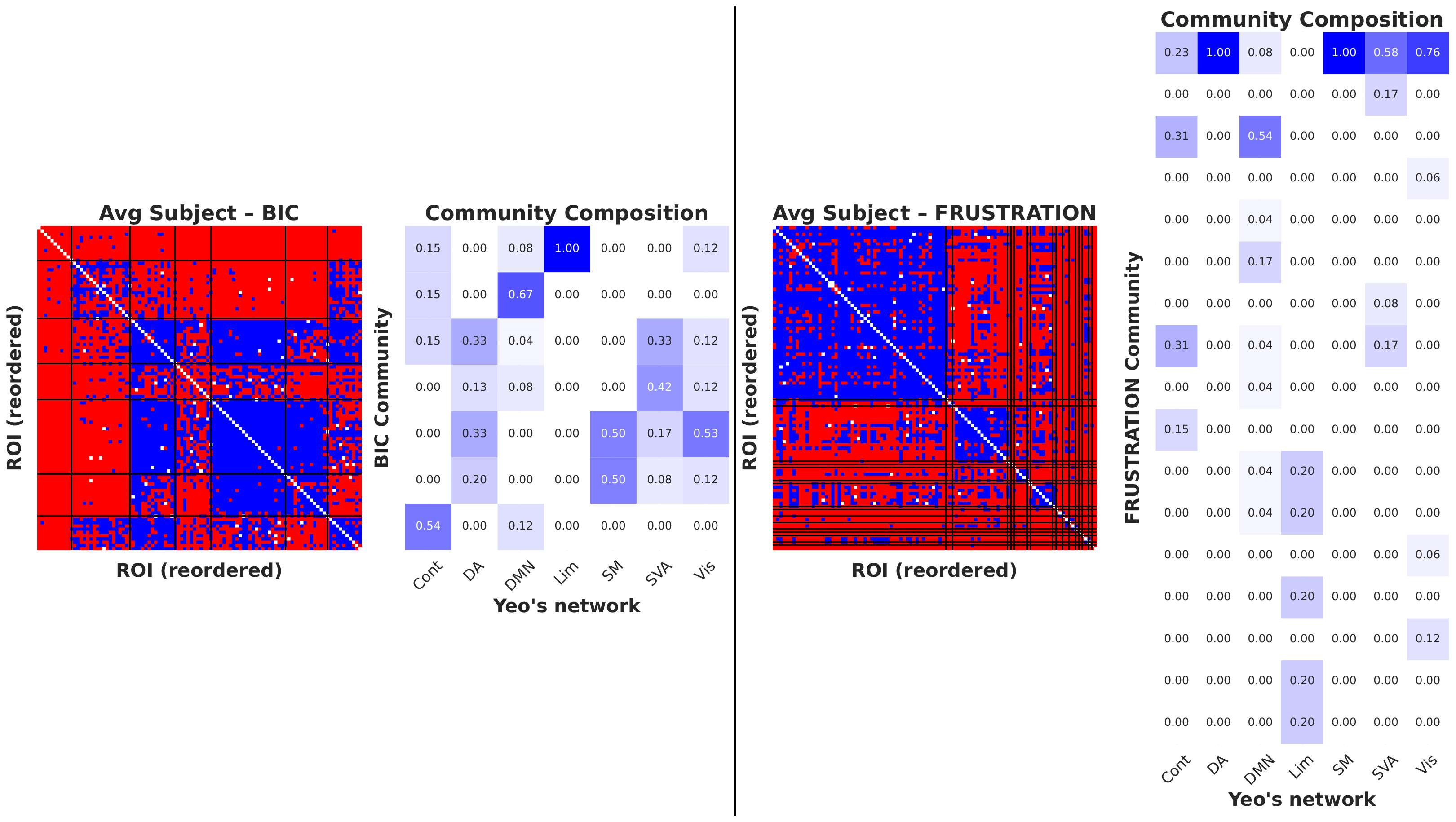}
\caption{\textbf{Average brain, partitioned according to BIC minimisation and $F$ minimisation.} As the presence of a vast majority of positive links causes most of the negative ones populating the na\"ive and bSRGM-induced projections to be deleted, only the bSCM-induced projections have been considered: while the average brain partitioned according to $F$ minimisation is characterised by $2$ predominantly positive communities that gather the vast majority of Yeo's networks together and $13$ predominantly negative modules that fragment the remaining ones, the average brain partitioned according to BIC minimisation is characterised by modules sharing similarities with the ones induced by NMI and those induced by RI and NMI, and capable to recover the LIM network in its entirety.}
\label{fig:15}
\end{figure*}

\subsubsection{The barycentric partition}

Let us, now, ask ourselves if a criterion to condense the information embodied by each individual partition can be devised. To this aim, one could try to single out the `most representative' brain by individuating the `most representative' mesoscale structure: finding an answer to this question ultimately amounts to finding a method to compare any two partitions, say $X=\{x_1\dots x_n\}$ and $Y=\{y_1\dots y_m\}$.\\

\noindent\textit{Rand index (RI).} One of the simplest metrics is the RI, reading

\begin{equation}
\text{RI}=\frac{\text{TP}+\text{TN}}{\text{TP}+\text{FP}+\text{TN}+\text{FN}}=\frac{\text{TP}+\text{TN}}{N(N-1)/2}
\end{equation}
and computing the ratio between the sum of \emph{true positives} (TP; the number of nodes that are `together' in both partitions, i.e. in the same subset in $X$ and in the same subset in $Y$) and \emph{true negatives} (TN; the number of nodes that are `separated' in both partitions, i.e. in different subsets in $X$ and in different subsets in $Y$) and the total number of pairs of nodes. Naturally, $0\leq\text{RI}\leq1$.

Other variants are the Wallace Index (WI), defined as $\text{WI}=\text{TP}/(\text{TP}+\text{FP})$, and the Jaccard Index (JI), defined as $\text{JI}=\text{TP}/(\text{TP}+\text{FP}+\text{FN})$: notice, however, that both WI and JI measure the agreement between two partitions by solely considering the number of TP.\\

\noindent\textit{Normalised mutual information (NMI).} A second metric is the NMI, reading

\begin{align}
\text{NMI}(X,Y)&=\frac{2\text{MI}(X,Y)}{S(X)+S(Y)};
\end{align}
here,

\begin{equation}
\text{MI}(X,Y)=\sum_{i=1}^N\sum_{j=1}^Nh_{ij}\ln\left(\frac{h_{ij}}{f_ig_j}\right),
\end{equation}
with $h_{ij}=|X_i\cap Y_j|/N$ being the fraction of nodes common to subsets $X_i$ and $Y_j$, $S(X)=-\sum_{i=1}^Nf_i\ln f_i$ being the Shannon entropy induced by the partition $X$ and $f_i=|X_i|/N$ being the fraction of nodes within the cluster $X_i$, $S(Y)=-\sum_{j=1}^Ng_j\ln g_j$ being the Shannon entropy induced by the partition $Y$ and $g_j=|Y_j|/N$ being the fraction of nodes within the cluster $Y_j$. Naturally, $0\leq\text{NMI}\leq1$.\\

\noindent\textit{Rajski distance (RD).} A third metric is the \emph{variation of information} (VI), reading

\begin{align}
\text{VI}(X,Y)&=S(X,Y)-\text{MI}(X,Y)\nonumber\\
&=2S(X,Y)-S(X)-S(Y)\nonumber\\
&=S(X)+S(Y)-2\text{MI}(X,Y)
\end{align}
 or, more explicitly,

\begin{equation}
\text{VI}(X,Y)=-\sum_{i=1}^N\sum_{j=1}^Nh_{ij}\left[\ln\left(\frac{h_{ij}}{f_i}\right)+\ln\left(\frac{h_{ij}}{g_j}\right)\right];
\end{equation}
such an expression can be normalised by dividing it by the joint entropy $S(X,Y)=-\sum_{i=1}^N\sum_{j=1}^Nh_{ij}\ln h_{ij}$, i.e.

\begin{equation}
\text{RD}(X,Y)=\frac{\text{VI}(X,Y)}{S(X,Y)}=1-\frac{\text{MI}(X,Y)}{S(X,Y)};
\end{equation}
naturally, $0\leq\text{RD}\leq1$.\\

\noindent\textit{Finding the barycentre.} Any of the metrics above can be employed to assess how different the partitions characterising the brains of any two subjects, say $u$ and $u'$, are. Once such a matrix of distances, say $\mathbf{D}=\{D_{uu'}\}_{u,u'=1}^U$, with $U=100$, has been obtained, a second method is needed to condense such a wealth of information. Our proposal is that of calculating the vector $\underline{\mathbf{D}}=\{\underline{D}_u\}$, where

\begin{equation}
\underline{D}_u=\frac{\sum_{u'=1}^U D_{uu'}}{U},
\end{equation}
and take the partition lying at distance

\begin{equation}
d=\min\left\{\underline{D}_u\right\}_{u=1}^U
\end{equation}
from the others, i.e. at the minimum, average one. Such a partition will be referred to as the \emph{barycentric partition} or \emph{brain centroid}. Let us explicitly notice that while VI and RD are proper distances over the space of possible partitions of a system, hence inducing the positions $D_{uu'}^\text{VI}=\text{VI}_{uu'}$ and $D_{uu'}^\text{RD}=\text{RD}_{uu'}$, RI and NMI quantify the similarity of any two partitions\footnote{In fact, both RI and NMI could be directly maximised.}, hence inducing the positions $D_{uu'}^\text{RI}=1-\text{RI}_{uu'}$ and $D_{uu'}^\text{NMI}=1-\text{NMI}_{uu'}$.

The observations about the importance of considering filtered projections and the prominent role played by heterogeneous benchmarks motivate us to focus on the partitions obtained by minimising BIC on the projections induced by the bSCM.\\

Upon looking at fig.~\ref{fig:14}, we realise that \emph{i)} RI individuates a barycentric projection with $7$ predominantly positive communities; \emph{ii)} NMI and RD individuate the same barycentric projection with $5$ predominantly positive communities; \emph{iii)} WI and VI individuate a barycentric projection with $4$ predominantly positive communities; \emph{iv)} JI individuates a barycentric projection with $5$, major communities defined by predominantly positive connections. The explanation of such a result may lie in the evidence that indices like WI and JI are solely defined in terms of true positives: as a consequence, identifying a smaller number of larger communities may enhance the detection of such events - although negative connections become under-represented; indices like RI, on the other hand, account for true negatives as well, whence the larger number of smaller communities - connected by a vast majority of negative links.

Let us, now, compare the community structure of the different brain centroids with the partition defined by Yeo. As fig.~\ref{fig:14} shows, while all metrics tend to individuate communities capturing large portions of certain Yeo's communities, each one induces a different overlap between our structure-based partitions and Yeo's task-based one.\\

A simpler recipe to identify the representative brain could be that of averaging the individual projections in a benchmark-specific fashion; as the presence of a vast majority of positive links causes most of the negative ones populating the na\"ive and bSRGM-induced projections to be deleted, here we limit ourselves to consider the bSCM-induced projections. As fig.~\ref{fig:15} shows, BIC minimisation leads to single out $7$ communities, while $F$ minimisation leads to single out $17$ communities: in the second case, these modules appear as quite uninformative as $15$ are very small groups of nodes fragmenting a single Yeo's network in a quite unrealistic fashion, while the first two ones gather all the remaining regions together; in the first case, instead, these modules share similarities with the ones induced by RI and those induced by NMI, and are capable to recover the LIM network in its entirety.

\section{Discussion}
 
The role played by negative connections within and between brain regions represents a topic that has recently gained attention within the neuroscientific literature~\cite{Saberi2021,Saberi2022,Luppi2024}; still, they remain difficult to interpret because their estimation depends on pre-processing choices, correlation-based construction pipelines and thresholding procedures. Here, we propose a statistically validated route to signed functional projections that does not require (constructing and thresholding) a correlation matrix: instead, signed links are inferred by counting (concordant and discordant) temporal motifs and validating their abundance against two bechmarks, i.e. the bSRGM (which controls for positive and negative link densities) and the bSCM (which additionally preserves local heterogeneities).

Such a validation scheme allows us to address a number of neuroscientific questions while overcoming specific pitfalls characterising traditional analyses. Crucially, however, our method goes beyond simply confirming known patterns: from a methodological perspective, the entire analysis represents a novel approach to network construction from time series, circumventing the intermediate, and problematic, step of handling a correlation matrix. Standard correlation-based analyses, in fact, face well-documented challenges: arbitrary threshold selection, sensitivity to GSR and the difficulty of distinguishing genuine anti-correlations from preprocessing artifacts~\cite{Murphy2009,Saad2012,van2017}. Our validation scheme sidesteps these issues by contrasting the occurrence of (concordant and discordant) temporal motifs against properly defined benchmarks (bSRGM and bSCM) which, differently from traditional approaches, can account for the heterogeneity of each region. This approach provides several advantages that manifest in our results.

\subsubsection{Negative links reveal brain (im)balance}
 
While previous studies have noticed the presence of frustrated triads in brain networks, our IBI metrics provide a systematic quantification of this phenomenon: the evidence that the median IBI value is positive, regardless of the benchmark employed, allows us conclude that brain networks are populated by a non-negligible number of $(+-+)$ and $(---)$ triads. 

However, the contrasting IBI values of bSRGM- and bSCM-induced projections illuminate a fundamental aspect of brain organisation: the global filter (bSRGM), which levels out differences between ROIs, produces densely positive subgraphs with few negative connections; the local filter (bSCM), on the contrary, reveals a more balanced distribution of positive and negative links - a result demonstrating that not all regions participate equally in concordant and discordant dynamics.

As shown in Appendix~\hyperlink{AppE}{E}, our framework generalises the one defined by Pearson's test of hypothesis, whose outcome overlaps to a very large extent with the one of the bSRGM: left aside the particularly large value of the Jaccard similarity (amounting to 1, in some cases), such a result indicates that Pearson's test of hypothesis is a homogeneous one, i.e. ignores the peculiarities of the two ROIs whose empirical Pearson correlation coefficient is being tested; moreover, it returns projections with a majority of positive signs. From this perspective, the bSCM-induced projection scheme generalises it, by properly accounting for the role played by any two series asynchrony.

Interestingly, the negative connections constituting the frustrated triads are mostly found \emph{i)} within groups of SUBC and LIM regions and \emph{ii)} between groups of nodes constituting the DMN and those constituting task-positive regions: these patterns are consistent with the established architecture of rs-networks~\cite{Greicius2003functional,Fox2005Human,Hampson2010Functional,Demertzi2022}, providing strong evidence that our method captures a genuine structure rather than a methodological artifact; similarly, the prominent role of subcortical regions in network frustration aligns with recent observations about their integrative function and contribution to network metastability~\cite{Saberi2022,Luppi2024}. These consistencies validate our approach: the statistically significant signed links emerging from our benchmarks correspond to (functionally) meaningful neural connections.

\subsubsection{Mesoscopic organisation beyond traditional balance}
 
At the mesoscale, our results indicate that the signed organisation of rs-networks is not in accordance with the TBT: if traditional balance were the dominant organising principle, one would expect predominantly positive links within communities and predominantly negative links between communities; here, instead, the observed partitions obtained through BIC minimisation reveal that both negative (positive) connections can populate diagonal (off-diagonal)/within (between) blocks at the mesoscopic level.

While minimising frustration leads to partitions that make the network as close as possible to a traditionally balanced state, BIC minimisation identifies the block structure best describing the observed projection while penalising model complexity; the two criteria, therefore, answer different questions and their disagreement suggests that the brain mesoscale is not primarily shaped by frustration reduction: rather, the observed structure appears to preserve a mixture of cooperative and competitive interactions, both within and between functional systems.
 
Earlier studies reported that empirical brain networks are more balanced than suitably randomised configurations, thus suggesting that functional brain organisation may be biased towards relatively stable signed configurations~\cite{Saberi2021}; more recently, Saberi et al.~\cite{Saberi2025} provided dynamic evidence in favour of TBT by showing that balanced triads display longer lifetimes and larger peak energies than imbalanced ones - with a separation of triadic states consistent with the strong formulation of the TBT.

These findings may appear in tension with ours. However, they mainly assess \emph{triadic} balance in correlation-based functional networks and in shorter time windows: our analysis, instead, focuses on statistically validated signed projections and on their \emph{mesoscopic} block organisation. In this setting, the resulting partitions admit negative links within modules and positive links between modules, a result suggesting that the brain may operate in a functionally `excited' state rather than seeking a `ground' state of minimal frustration - the (statistical variant of the) RBT may, thus, represent a better framework than the TBT to interpret our results.

\subsubsection{Inter-subject variability and representative partitions}
 
The analysis of Jaccard similarity to assess inter-subject robustness of positive and negative connections reveals an interesting asymmetry on na\"ive and bSRGM-induced projections: positive links show a larger spatial coherence, while negative links are more variable. This pattern suggests that positive connections, mirroring a concordant dynamics, follow a relatively stereotyped architecture across individuals, whereas negative connections, mirroring a discordant, or competitive, dynamics, may be more subject-specific or state-dependent. By accounting for local heterogeneity, the bSCM filter reconciles this asymmetry, a finding that may have implications in fingerprinting analysis and personalised neuroscience and that standard (i.e. correlation-based) analyses, which often threshold away negative connections, lose.
 
Given the high level of inter-subject variability, the `most representative' brain network remains challenging to define, as different distance metrics yield different barycentric partitions: while those emphasising true positives (WI, JI) favour a smaller number of larger communities, the ones incorporating true negatives (RI) identify a larger number of smaller communities. This sensitivity highlights a fundamental tension in network neuroscience, i.e. whether to prioritise the detection of strong and consistent connections (potentially missing sparse but important links) or to adopt a more fine-grained view (potentially fragmenting coherent functional systems).
 
Our averaged projections suggest a practical compromise: BIC minimisation on bSCM-induced projections identifies communities that capture up to $80\%$ (in one case, 100\%) of certain Yeo's networks, while maintaining an interpretable structure that does not discard negative interactions: this approach may prove valuable for defining reference networks in clinical applications or for generating prior hypotheses in connectivity studies.

\subsubsection{Robustness and technical considerations}
 
An important concern with any fMRI-based analysis is the potential confound of autocorrelations induced by the hemodynamic response. We have verified that our results are robust against standard cleaning procedures: although a certain degree of autocorrelation is present, it primarily affects signals at small time-lags - typically of the same order of magnitude of the hemodynamic response itself; running an autoregressive model of order up to $7$ is sufficient to remove these effects without substantially altering the validated network structure.
 
The choice of temporal resolution and window length represents another important consideration: our analysis used the full rs-scan duration but the framework naturally extends to sliding-window approaches for capturing dynamic FC. The key requirement is that the temporal window be sufficiently long to accumulate enough concordant and discordant motifs for statistical testing.
 
\subsubsection{Limitations and caveats}

While our method offers several advantages, certain limitations warrant discussion. First, the binary nature of our current framework, classifying temporal snapshots as simply concordant or discordant, discards information about the magnitude of co-fluctuations, which typically drives FC and encodes subject-specific information~\cite{Esfahlani2020}: future extensions incorporating weighted edges (as mentioned in Conclusions) would address this limitation; second, our analysis focuses on triadic motifs built upon pairwise relationships, while higher-order interactions beyond triads may play important roles in shaping and characterising (dis)concordant neural dynamics; third, the choice between bSRGM and bSCM represents a fundamental modelling decision: while the bSCM preserves heterogeneity, the bSRGM provides a simpler null model that may be preferable when strong prior hypotheses about degree distributions are lacking; fourth, our benchmarks treat time steps as exchangeable, hence the framework is insensitive to the temporal ordering of the original data: the resulting projection is identical for any permutation of time points that preserves the simultaneity structure across brain regions - although this renders the framework unsuitable for causal analyses, it is appropriate when focusing on (simultaneous) co-occurrences of ROIs; finally, we considered a single dataset (the HCP-Young Adults~\cite{VanEssen2013}) with a single parcellation (Schaefer+Tian~\cite{Schaefer2018,Tian2020}): additional results on different datasets and parcellations are needed to establish their robustness.

The interpretation of `frustration' in neural contexts also deserves caution: unlike in social networks, where frustration may represent genuine conflict, in neural networks it may reflect more subtle physiological or functional mechanisms; the positive IBI values we observe may, thus, represent a more subtle signature of a complex, multi-scale system that is maintained in an `excited' state through a balance of cooperative and competitive interactions.

\subsubsection{Physiological mechanisms}

For the physiological mechanisms underlying the emergence of negative connections, including potential contributions from neural inhibition, neurovascular coupling and network-level oscillatory dynamics, we redirect the interested reader to more technical contributions~\cite{Harita2024}; the validation framework presented here is agnostic to these mechanisms, focusing on the statistical significance and topological organisation of the observed patterns.

\section{Conclusions}\label{sec:X}

A critical, yet often overlooked, aspect of functional brain data is the role played by negative links: frequently ignored due to the interpretative challenges brought along, emerging evidence indicates negative links to be fundamental in shaping resting-state networks - especially the properties affecting self-stabilisation.

As excluding negative links, thus, means underestimating the brain ability to function properly, we have proposed a validation technique allowing statistically significant signed links to emerge out of the noise accompanying raw data; two, different benchmarks have been employed, filtering multivariate time series to a different extent and letting diverse patterns emerge out of the wealth of empirical observations: while the bSRGM levels out the differences between time series, the bSCM preserves their heterogeneity, thus inducing sparser projections that retain more information about the original structure.

In particular, our analysis points out that brain networks are, indeed, frustrated, the patterns induced by signed connections overlapping (in some cases, to a quite large extent) with Yeo's task-based partitions and returning the picture of a system aligning with the statistical variant of the Relaxed Balance Theory rather than seeking to minimise frustration. Negative links are not evenly distributed, being concentrated between specific ROIs, especially those belonging to Yeo's SUBC and LIM networks.

A future research direction is that of employing our algorithm to obtain statistically validated \emph{binary} and \emph{weighted} projections of multivariate \emph{weighted} time series. Such an extension would preserve the interpretability of the present framework, while enabling a more nuanced characterisation of cooperative and competitive dynamics in functional brain networks.

\section{Acknowledgements}
This work is supported by the PNRR-M4C2-Investimento 1.3, Partenariato Esteso PE00000013-``FAIR-Future Artificial Intelligence Research''-Spoke 1 ``Human-centered AI'', funded by the European Commission under the NextGeneration EU programme. This work has also been supported by the European Union under the scheme HORIZON-INFRA-2021-DEV-02-01 -- Preparatory phase of new ESFRI research infrastructure projects, Grant Agreement No.~101079043, ``SoBigData RI PPP: SoBigData RI Preparatory Phase Project'' by ``Reconstruction, Resilience and Recovery of Socio-Economic Networks'' RECON-NET EP\_FAIR\_-005PE0000013 ``FAIR,'' Funded by the European Union NextGeneration EU, PNRR Mission 4 Component 2, Investment 3.1; and by the European Community programme under the funding schemes: ERC-2018-ADG G.A.~834756 ``XAI: Science and technology for the explanation of AI decision making''.

\section{Data availability}

Data concerning rs-fMRI from the HCP can be found at the address \url{https://www.humanconnectome.org/study/hcp-young-adult}; for a description, see~\url{https://www.humanconnectome.org/storage/app/media/documentation/s1200/HCP_S1200_Release_Reference_Manual.pdf}.

\section{Code availability}

The Python package named \texttt{SIBERIA} (\textit{SIgned BEnchmarks foR tIme series Analysis}), implementing our benchmarks for the analysis of multivariate time series, is available on PyPI and at the URL \url{https://github.com/MarsMDK/SIBERIA}.

\section{Author contributions}

Study conception and design: MIB, MDV, TG, TS. Literature review: EA, MDV, ST. Data collection: EA. Analysis and interpretation of results: EA, MDV, TS. Draft manuscript preparation: EA, MDV, TS. Draft manuscript revision: EA, MIB, MDV, ST, TG, TS.

\section{Competing Interests}

The authors declare no competing interests.

\bibliography{bibmain.bib}

\clearpage

\onecolumngrid

\appendix

\hypertarget{AppA}{}
\section{Probabilistic models for binary undirected bipartite signed networks}
\label{AppA}

The generalisation of the ERG formalism for the analysis of binary undirected bipartite signed graphs rests upon the constrained maximisation of Shannon entropy, i.e.

\begin{equation}
\mathscr{L}=S[P]-\sum_{i=0}^M\theta_i[P(\mathbf{B})C_i(\mathbf{B})-\langle C_i\rangle]
\end{equation}
where

\begin{equation}
S=-\sum_{\mathbf{B}\in\mathbb{B}}P(\mathbf{B})\ln P(\mathbf{B}),
\end{equation}
$C_0=\langle C_0\rangle=1$ sums up the normalisation condition and the remaining $M-1$ constraints represent proper, topological properties. Such an optimisation procedure defines the expression

\begin{equation}
P(\mathbf{B})=\frac{e^{-H(\mathbf{B})}}{Z}=\frac{e^{-H(\mathbf{B})}}{\sum_{\mathbf{B}\in\mathbb{B}}e^{-H(\mathbf{B})}}=\frac{e^{-\sum_{i=1}^M\theta_iC_i(\mathbf{B})}}{\sum_{\mathbf{B}\in\mathbb{B}}e^{-\sum_{i=1}^M\theta_iC_i(\mathbf{B})}}
\end{equation}
that can be made explicit only after having chosen a specific set of constraints.

\subsection*{1. Signed Random Graph Model}

Let us consider the properties $L^+(\mathbf{B})$ and $L^-(\mathbf{B})$. The Hamiltonian describing such a problem reads

\begin{equation}
H(\mathbf{B})=\alpha\:B^+(\mathbf{B})+\gamma\:B^-(\mathbf{B})=\sum_{i=1}^N\sum_{t=1}^T(\alpha\:b_{it}^++\gamma\:b_{it}^-)
\end{equation}
and induces a partition function reading

\begin{align}
Z&=\sum_{\mathbf{B}\in\mathbb{B}}e^{-H(\mathbf{B})}\nonumber\\
&=\sum_{\mathbf{B}\in\mathbb{B}}e^{-\alpha\:B^+(\mathbf{B})-\gamma\:B^-(\mathbf{B})}\nonumber\\
&=\sum_{\mathbf{B}\in\mathbb{B}}e^{-\sum_{i=1}^N\sum_{t=1}^T(\alpha\:b_{it}^++\gamma\:b_{it}^-)}\nonumber\\
&=\sum_{\mathbf{B}\in\mathbb{B}}\prod_{i=1}^N\prod_{t=1}^Te^{-(\alpha\:b_{it}^++\gamma\:b_{it}^-)}\nonumber\\
&=\prod_{i=1}^N\prod_{t=1}^T\left[\sum_{b_{it}=-1,1}e^{-(\alpha\:b_{it}^++\gamma\:b_{it}^-)}\right]\nonumber\\
&=\prod_{i=1}^N\prod_{t=1}^T(e^{-\alpha}+e^{-\gamma})\nonumber\\
&=(e^{-\alpha}+e^{-\gamma})^{N\times T}.
\end{align}

The expression above leads us to find

\begin{align}\label{probgraph1}
P_\text{bSRGM}(\mathbf{B})=\frac{e^{-\alpha\:B^+(\mathbf{B})-\gamma\:B^-(\mathbf{B})}}{(e^{-\alpha}+e^{-\gamma})^{N\times T}}=\frac{x^{B^+(\mathbf{B})}y^{B^-(\mathbf{B})}}{(x+y)^{N\times T}}=(p^-)^{B^-(\mathbf{B})}(p^+)^{B^+(\mathbf{B})}
\end{align}
having posed $p^-=\frac{e^{-\gamma}}{e^{-\alpha}+e^{-\gamma}}=\frac{y}{x+y}$ and $p^+=\frac{e^{-\alpha}}{e^{-\beta}+e^{-\gamma}}=\frac{x}{x+y}$, where $p^+$ is the probability that a generic series assumes a positive value at time $t$ and $p^-$ is the probability that a generic series assumes a negative value at time $t$.\\

In order to determine the parameters that define the bSRGM, let us maximise the likelihood function

\begin{equation}
\mathcal{L}_\text{bSRGM}(x,y)=\ln P_\text{bSRGM}(\mathbf{B}^*|x,y)=B^+(\mathbf{B}^*)\ln x+B^-(\mathbf{B}^*)\ln y-(N\times T)\ln(x+y)
\end{equation}
with respect to $x$ and $y$. Upon doing so, we obtain the pair of equations

\begin{align}
\frac{\partial\mathcal{L}_\text{bSRGM}(x,y)}{\partial x}&=\frac{B^+(\mathbf{B}^*)}{x}-\frac{N\times T}{x+y},\\
\frac{\partial\mathcal{L}_\text{bSRGM}(x,y)}{\partial y}&=\frac{B^-(\mathbf{B}^*)}{y}-\frac{N\times T}{x+y};
\end{align}
equating them to zero leads us to find

\begin{align}
p^+&=B^+(\mathbf{B}^*)/(N\times T),\\
p^-&=B^-(\mathbf{B}^*)/(N\times T).
\end{align}

\subsection*{2. Signed Configuration Model}

Let us consider the properties $\{k_i^+(\mathbf{B})\}_{i=1}^N$, $\{k_i^-(\mathbf{B})\}_{i=1}^N$, $\{\kappa_t^+(\mathbf{B})\}_{t=1}^T$ and $\{\kappa_t^-(\mathbf{B})\}_{t=1}^T$. The Hamiltonian describing such a problem reads

\begin{align}
H(\mathbf{B})=\sum_{i=1}^N[\alpha_ik_i^+(\mathbf B)+\gamma_ik_i^-(\mathbf B)]+\sum_{t=1}^T[\delta_t\kappa_t^+(\mathbf B)+\eta_t\kappa_t^-(\mathbf B)]
\end{align}
and induces a partition function reading 

\begin{align}
Z&=\sum_{\mathbf{B}\in\mathbb{B}}e^{-H(\mathbf B)}\nonumber\\
&=\sum_{\mathbf{B}\in\mathbb{B}}e^{-\sum_{i=1}^N[\alpha_ik_i^+(\mathbf B)+\gamma_ik_i^-(\mathbf B)]-\sum_{t=1}^T[\delta_t\kappa_t^+(\mathbf B)+\eta_t\kappa_t^-(\mathbf B)]}\nonumber\\
&=\sum_{\mathbf{B}\in\mathbb{B}}e^{-\sum_{i=1}^N\sum_{t=1}^T[(\alpha_i+\delta_t)b_{it}^++(\gamma_i+\eta_t)b_{it}^-]}\nonumber\\
&=\sum_{\mathbf{B}\in\mathbb{B}}\prod_{i=1}^N\prod_{t=1}^Te^{-(\alpha_i+\delta_t)b_{it}^+-(\gamma_i+\eta_t)b_{it}^-}\nonumber\\
&=\prod_{i=1}^N\prod_{t=1}^T\left[\sum_{b_{it}=-1,1}e^{-(\alpha_i+\delta_t)b_{it}^+-(\gamma_i+\eta_t)b_{it}^-}\right]\nonumber\\
&=\prod_{i=1}^N\prod_{t=1}^T\left[e^{-(\alpha_i+\delta_t)}+e^{-(\gamma_i+\eta_t)}\right].
\end{align}

The expression above leads us to find

\begin{align}\label{probgraph2}
P_\text{bSCM}(\mathbf{B})&=\frac{e^{-\sum_{i=1}^N[\alpha_ik_i^+(\mathbf B)+\gamma_ik_i^-(\mathbf B)]-\sum_{t=1}^T[\delta_t\kappa_t^+(\mathbf B)+\eta_t\kappa_t^-(\mathbf B)]}}{\prod_{i=1}^N\prod_{t=1}^T\left[e^{-(\alpha_i+\delta_t)}+e^{-(\gamma_i+\eta_t)}\right]}\nonumber\\
&=\prod_{i=1}^N\prod_{t=1}^T\frac{x_i^{k_i^+(\mathbf{B})}y_i^{k_i^-(\mathbf{B})}z_t^{\kappa_t^+(\mathbf{B})}w_t^{\kappa_t^-(\mathbf{B})}}{x_iz_t+y_iw_t}\nonumber\\
&=\prod_{i=1}^N\prod_{t=1}^T(p_{it}^-)^{b_{it}^-}(p_{it}^+)^{b_{it}^+}
\end{align}
having posed $p_{it}^-=\frac{e^{-(\gamma_i+\eta_t)}}{e^{-(\alpha_i+\delta_t)}+e^{-(\gamma_i+\eta_t)}}=\frac{y_iw_t}{x_iz_t+y_iw_t}$ and $p_{it}^+=\frac{e^{-(\alpha_i+\eta_t)}}{e^{-(\alpha_i+\delta_t)}+e^{-(\gamma_i+\eta_t)}}=\frac{x_iz_t}{x_iz_t+y_iw_t}$, where $p_{it}^+$ is the probability that series $i$ assumes a positive value at time $t$ and $p_{it}^-$ is the probability that series $i$ assumes a negative value at time $t$.\\

In order to determine the parameters that define the bSCM, let us maximise the likelihood function

\begin{align}
&\mathcal{L}_\text{bSCM}(\bm{x},\bm{y},\bm{z},\bm{w})=\ln P_\text{bSCM}(\mathbf{B}^*|\bm{x},\bm{y},\bm{z},\bm{w})\nonumber\\
&=\sum_{i=1}^Nk_i^+(\mathbf{B}^*)\ln x_i+\sum_{i=1}^Nk_i^-(\mathbf{B}^*)\ln y_i+\sum_{t=1}^T\kappa_t^+(\mathbf{B}^*)\ln z_t+\sum_{t=1}^T\kappa_t^-(\mathbf{B}^*)\ln w_t-\sum_{i=1}^N\sum_{t=1}^T\ln(x_iz_t+y_iw_t)
\end{align}
with respect to $x_i$, $y_i$, $z_t$ and $w_t$, $\forall\:i,t$. Upon doing so, we obtain the system of equations

\begin{align}
\frac{\partial\mathcal{L}_\text{bSCM}(\bm{x},\bm{y},\bm{z},\bm{w})}{\partial x_i}&=\frac{k_i^+(\mathbf{B}^*)}{x_i}-\sum_{t=1}^T\frac{z_t}{x_iz_t+y_iw_t},\quad\forall\:i,\\
\frac{\partial\mathcal{L}_\text{bSCM}(\bm{x},\bm{y},\bm{z},\bm{w})}{\partial y_i}&=\frac{k_i^-(\mathbf{B}^*)}{y_i}-\sum_{t=1}^T\frac{w_t}{x_iz_t+y_iw_t},\quad\forall\:i,\\
\frac{\partial\mathcal{L}_\text{bSCM}(\bm{x},\bm{y},\bm{z},\bm{w})}{\partial z_\alpha}&=\frac{\kappa_t^+(\mathbf{B}^*)}{z_t}-\sum_{i=1}^N\frac{x_i}{x_iz_t+y_iw_t},\quad\forall\:t,\\
\frac{\partial\mathcal{L}_\text{bSCM}(\bm{x},\bm{y},\bm{z},\bm{w})}{\partial w_\alpha}&=\frac{\kappa_t^-(\mathbf{B}^*)}{w_t}-\sum_{i=1}^N\frac{y_i}{x_iz_t+y_iw_t},\quad\forall\:t;
\end{align}
equating them to zero leads us to find

\begin{align}
k_i^+(\mathbf{B}^*)&=\sum_{t=1}^T\frac{x_iz_t}{x_iz_t+y_iw_t}=\sum_{t=1}^Tp_{it}^+=\langle k_i^+\rangle,\quad\forall\:i,\\
k_i^-(\mathbf{B}^*)&=\sum_{t=1}^T\frac{y_iw_t}{x_iz_t+y_iw_t}=\sum_{t=1}^Tp_{it}^-=\langle k_i^-\rangle,\quad\forall\:i,\\
\kappa_t^+(\mathbf{B}^*)&=\sum_{i=1}^N\frac{x_iz_t}{x_iz_t+y_iw_t}=\sum_{i=1}^Np_{it}^+=\langle \kappa_t^+\rangle,\quad\forall\:t,\\
\kappa_t^-(\mathbf{B}^*)&=\sum_{i=1}^N\frac{y_iw_t}{x_iz_t+y_iw_t}=\sum_{i=1}^Np_{it}^-=\langle \kappa_t^-\rangle,\quad\forall\:t.
\end{align}

The system above can be solved only numerically (see also Appendix~\ref{AppB}).

\clearpage

\hypertarget{AppB}{}
\section{Numerical optimisation of likelihood functions}
\label{AppB}

In order to numerically solve the system of equations defining the bSCM, we can follow the guidelines provided in~\cite{vallarano2021}: more specifically, we will adapt the iterative recipe provided there to our setting. First, let us notice that, upon posing $x_i=e^{-\alpha_i}$, $y_i=e^{-\gamma_i}$, $\forall\:i$ and $z_t=e^{-\delta_t}$, $w_t=e^{-\eta_t}$, $\forall\:t$, the system of equations defining the bSCM can be re-written as

\begin{align}
x_i&=\frac{k_i^+(\mathbf{B^*})}{\sum_{t=1}^T\dfrac{z_t}{x_iz_t+y_iw_t}}\:\Longrightarrow\: x_i^{(n)}=\frac{k_i^+(\mathbf{B^*})}{\sum_{t=1}^T\dfrac{z_t^{(n-1)}}{x_i^{(n-1)}z_t^{(n-1)}+y_i^{(n-1)}w_t^{(n-1)}}},\quad\forall\:i,\\
y_i&=\frac{k_i^-(\mathbf{B^*})}{\sum_{t=1}^T\dfrac{w_t}{x_iz_t+y_iw_t}}\:\Longrightarrow\: y_i^{(n)}=\frac{k_i^-(\mathbf{B^*})}{\sum_{t=1}^T\dfrac{w_t^{(n-1)}}{x_i^{(n-1)}z_t^{(n-1)}+y_i^{(n-1)}w_t^{(n-1)}}},\quad\forall\:i,\\
z_t&=\frac{\kappa_t^+(\mathbf{B^*})}{\sum_{i=1}^N\dfrac{x_i}{x_iz_t+y_iw_t}}\:\Longrightarrow\: z_t^{(n)}=\frac{\kappa_t^+(\mathbf{B^*})}{\sum_{i=1}^N\dfrac{x_i^{(n-1)}}{x_i^{(n-1)}z_t^{(n-1)}+y_i^{(n-1)}w_t^{(n-1)}}},\quad\forall\:t,\\
w_t&=\frac{\kappa_t^-(\mathbf{B^*})}{\sum_{i=1}^N\dfrac{y_i}{x_iz_t+y_iw_t}}\:\Longrightarrow\: w_t^{(n)}=\frac{\kappa_t^-(\mathbf{B^*})}{\sum_{i=1}^N\dfrac{y_i^{(n-1)}}{x_i^{(n-1)}z_t^{(n-1)}+y_i^{(n-1)}w_t^{(n-1)}}},\quad\forall\:t;
\end{align}
in order for our iterative recipe to converge, an appropriate vector of initial conditions needs to be chosen; here, we have opted for the following one: $x_i=k_i^+(\mathbf{B}^*)/\sqrt{L^+(\mathbf{B}^*)}$, $\forall\:i$, $y_i=k_i^-(\mathbf{B}^*)/\sqrt{L^-(\mathbf{B}^*)}$, $\forall\:i$, $z_t=\kappa_t^+(\mathbf{B}^*)/\sqrt{L^+(\mathbf{B}^*)}$, $\forall\:t$ and $w_t=\kappa_t^-(\mathbf{B}^*)/\sqrt{L^-(\mathbf{B}^*)}$, $\forall\:t$.

Moreover, we have adopted two, different stopping criteria: the first one is a condition on the Euclidean norm of the vector of differences between the values of the parameters at subsequent iterations, i.e. $||\Delta\vec\theta||_2=\sqrt{\sum_{i=1}^N(\Delta\theta_i)^2}\le10^{-8}$; the second one is a condition on the maximum number of iterations of our iterative algorithm, set to $10^3$.

The accuracy of our method in estimating the constraints has been evaluated by computing the \textit{maximum absolute error} (MAE), defined as 

\begin{align}
\text{bMAE}&=\max_{i,t}\left\{|k_i^+(\mathbf{B^*})-\langle k_i^+\rangle|,\:|k_i^-(\mathbf{B^*})-\langle k_i^-\rangle|,\:|\kappa_t^+(\mathbf{B^*})-\langle \kappa_t^+\rangle|,\:|\kappa_t^-(\mathbf{B^*})-\langle \kappa_t^-\rangle|\right\}
\end{align}
(i.e. as the infinite norm of the difference between the vector of the empirical values of the constraints and the vector of their expected values) and the \textit{maximum relative error} (MRE), defined as

\begin{align}
\text{bMRE}&=\max_{i,t}\left\{\frac{|k_i^+(\mathbf{B^*})-\langle k_i^+\rangle|}{k_i^+(\mathbf{B^*})},\:\frac{|k_i^-(\mathbf{B^*})-\langle k_i^-\rangle|}{k_i^-(\mathbf{B^*})},\:\frac{|\kappa_t^+(\mathbf{B^*})-\langle\kappa_t^+\rangle|}{\kappa_t^+(\mathbf{B^*})},\:\frac{|\kappa_t^-(\mathbf{B^*})-\langle\kappa_t^-\rangle|}{\kappa_t^-(\mathbf{B^*})}\right\}
\end{align}
(i.e. as the infinite norm of the relative difference between the vector of the empirical values of the constraints and the vector of their expected values).

Table \ref{tab:iterative}, providing information on the basic statistics of $5$, illustrative time series coming from the HCP-Young Adult dataset shows the time employed by our algorithm to converge as well as its accuracy in reproducing the constraints defining the bSCM: our method is, overall, fast and accurate, as the absolute error is steadily around $10^{-3}$ and the relative error is steadily around $10^{-6}$, the time employed to achieve such an accuracy being always less than a minute.

\begin{table}[t!]
\centering
\renewcommand{\arraystretch}{1.2}
\begin{tabular}{l|c|c|c|c|c|c|c|c|c}
\hline
\multicolumn{7}{c|}{\textbf{Basic statistics of time series}} & \multicolumn{3}{c|}{\textbf{bSCM}}\\
\hline
\hline
& $N$ & $T$ & $B^+$ & $B^-$ & $c^+$ & $c^-$ & MAE & MRE & Time (s)\\
\hline
\hline
HCP$_{100408}$ & 116 & 2.400 & 137.804 & 140.596 & $\simeq 0.495$ & $\simeq 0.505$ & $\simeq 1.30 \times 10^{-3}$ & $\simeq 1.15 \times 10^{-6}$ & $\simeq 37.8$ \\
\hline
HCP$_{106016}$ & 116 & 2.400 & 142.407 & 135.993 & $\simeq 0.512$ & $\simeq 0.488$ & $\simeq 1.10 \times 10^{-3}$ & $\simeq 9.94 \times 10^{-7}$ & $\simeq 28.8$ \\
\hline
HCP$_{125525}$ & 116 & 2.400 & 139.054 & 139.346 & $\simeq 0.499$ & $\simeq 0.501$ & $\simeq 1.26 \times 10^{-3}$ & $\simeq 1.17 \times 10^{-6}$ & $\simeq 37.8$ \\
\hline
HCP$_{130316}$ & 116 & 2.400 & 137.196 & 141.204 & $\simeq 0.493$ & $\simeq 0.507$ & $\simeq 1.97 \times 10^{-3}$ & $\simeq 1.89 \times 10^{-6}$ & $\simeq 37.7$ \\
\hline
HCP$_{857263}$ & 116 & 2.400 & 136.376 & 142.024 & $\simeq 0.490$ & $\simeq 0.510$ & $\simeq 1.10 \times 10^{-3}$ & $\simeq 9.90 \times 10^{-7}$ & $\simeq 33.3$ \\
\hline
\end{tabular}
\renewcommand{\arraystretch}{1.}
\caption{\label{tab:iterative} Performance of our fixed-point algorithm to solve the system of equations defining the bSCM on $5$, illustrative time series coming from the HCP-Young Adult dataset (each pedex refers to a subject): $N$ is the number of brain regions, $T$ is the duration of the related time series, $B^+$ is the overall number of positive values, $B^-$ is the overall number of negative values, $c^+=B^+/(N\times T)$ is the density of positive values and $c^-=B^-/(N\times T)$ is the density of negative values. Our fixed-point algorithm leads to MAE values around $10^{-3}$ and MRE values of the order of $10^{-6}$ in less than forty seconds.}
\end{table}

\clearpage

\hypertarget{AppC}{}
\section{Temporal autocorrelation and pre-whitening of BOLD time series}
\label{AppC}

\begin{figure*}[t!]
\includegraphics[width=\linewidth]{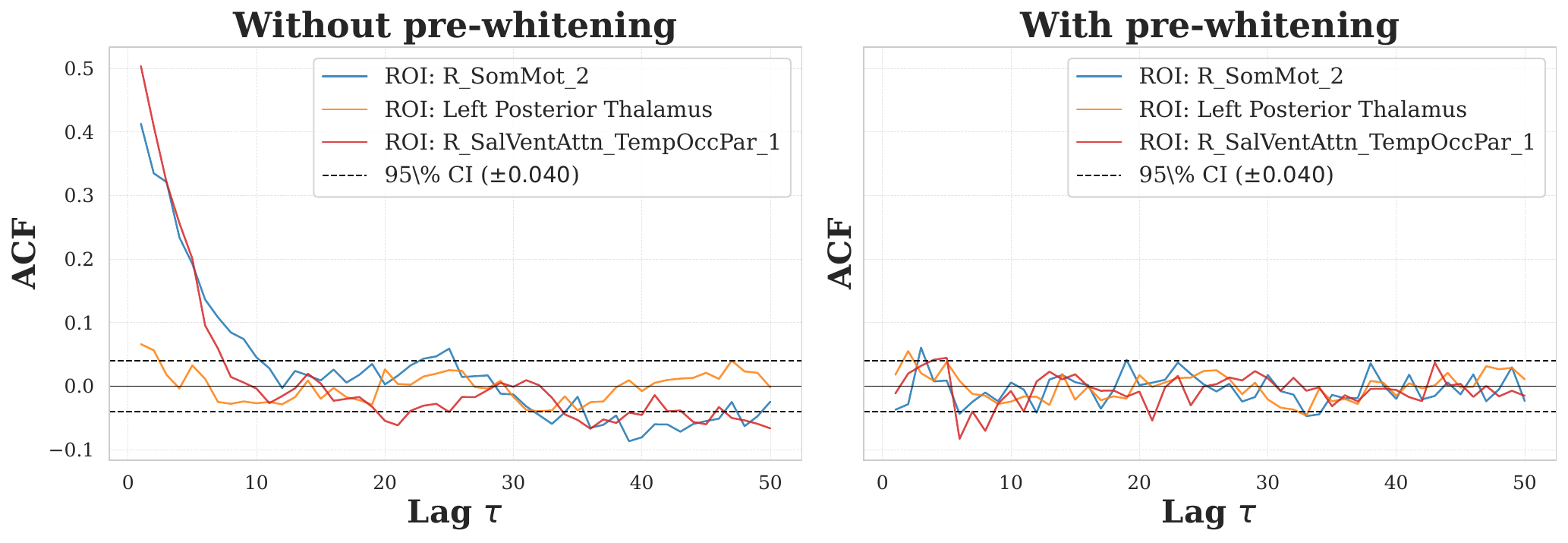}
\caption{{\bf Autocorrelation function of subject $\bm{\#106016}$, before and after pre-whitening.} Left panel: autocorrelation function $r_{ii}^\tau$ for three ROIs of the subject $\#106016$, as a function of lag $\tau$; without pre-whitening, two of the three series exhibit positive autocorrelations at short lags - a signature of the hemodynamic response. Right panel: autocorrelation function $r_{ii}^\tau$ for the pre-whitened binarised series; the autocorrelation function, now, lies within the $95\%$ confidence interval induced by white noise and individuated by the values $\pm1.96/\sqrt{T}$.}
\label{appC:autocorr_fig}
\end{figure*}

A documented property of rs-fMRI signals concerns the presence of a non-negligible amount of temporal autocorrelations. Such a presence can be explicitly inspected by computing the autocorrelation function of the time series $X_{it}$:

\begin{equation}\label{AppC:autocorrelation_function}
r_{ii}^\tau=\frac{\sum_{t=1}^{T}(X_{it}-\underline{X}_i)\times(X_{it+\tau}-\underline{X}_i)}{\sum_{t=1}^{T}(X_{it}-\underline{X}_i)^2};
\end{equation}
as shown if fig.~\ref{appC:autocorr_fig}, the selected series exhibits a large autocorrelation at small lags, decaying to zero over a timescale that is consistent with that of the hemodynamic response~\cite{Friston1994,Glover1999,Woolrich2001}. The pre-whitening procedure described below is intended to remove it from each time series $X_{it}$.

\subsection*{1. Bartlett's white noise test}

For a given order $p_i$, Bartlett's white noise test~\cite{Bartlett1955,Priestley1981,Durbin1967}, prescribes to compute the \emph{periodogram} of residuals, i.e.

\begin{equation}
I_i(\omega_k)=\left|\sum_{t=p_i+1}^T\varepsilon_{it} e^{-2\pi i\omega_k t}\right|^2,\quad\omega_k=\frac{k}{T-p_i},\quad k=1\dots m
\end{equation}
where $m=\lfloor(T-p)/2\rfloor$. Under the null hypothesis of \emph{white noise}, i.e. $\varepsilon_{it}\sim\mathcal{N}(0,\sigma^2)$, $\forall\:i,t$, one finds that

\begin{equation}
I_i(\omega_k)=\left(\sum_{t=p_i+1}^T\varepsilon_{it} e^{-2\pi i\omega_k t}\right)\left(\sum_{s=p_i+1}^T\varepsilon_{is} e^{2\pi i\omega_k s}\right)=\sum_{t=p_i+1}^T\sum_{s=p_i+1}^T\varepsilon_{it} \varepsilon_{is}e^{-2\pi i\omega_k(t-s)},
\end{equation}
an expression whose expected value reads $(T-p_i)\sigma^2$; as a consequence, the expected value of the normalised cumulative periodogram

\begin{equation}
F_i(\omega_k)=\frac{\sum_{j=1}^k I_i(\omega_j)}{\sum_{j=1}^m I_i(\omega_j)},\quad k=1\dots m
\end{equation}
reads $k/m$, i.e. grows linearly with $k$: in other words, the normalised periodogram obeys a uniform distribution on $[0,1]$~\cite{Priestley1981}. Such a compatibility - any deviation from which would indicate residual autocorrelation - is assessed via a KS test~\cite{Durbin1967} at the confidence level of $5\%$ and the optimal order is selected as the minimum for which the KS test fails to reject the null hypothesis of white noise - to this aim, the upper bound $P=30$ has been pre-specified.

\subsection*{2. Correcting for multiple hypothesis testing}

Since the Bartlett's white noise test is carried out $P$ times per series, a procedure accounting for the fact that multiple hypotheses are tested at the same time is needed. Here, we adopt the Benjamini-Yekutieli procedure~\cite{BenjaminiYekutieli2001}, which controls the False Discovery Rate under arbitrary dependence of p-values. Operatively, the $P$ p-values are sorted in increasing order and the selected one is the minimum satisfying the condition

\begin{equation}
\text{p-value}_j\geq\frac{j\alpha}{P\times c(P)},
\end{equation}
with $j=1\dots P$ indexing the p-values associated with each ROI, $c(P)\simeq\ln P$ being the Euler-Mascheroni constant, $\alpha=0.05$ and $P=30$ - a choice that has been further validated a posteriori: across all ROIs of our subjects no value $\hat{p}_i=P$ has been observed, a result confirming that the results are independent of such a specific value.\\

\begin{figure*}[t!]
\includegraphics[width=\linewidth]{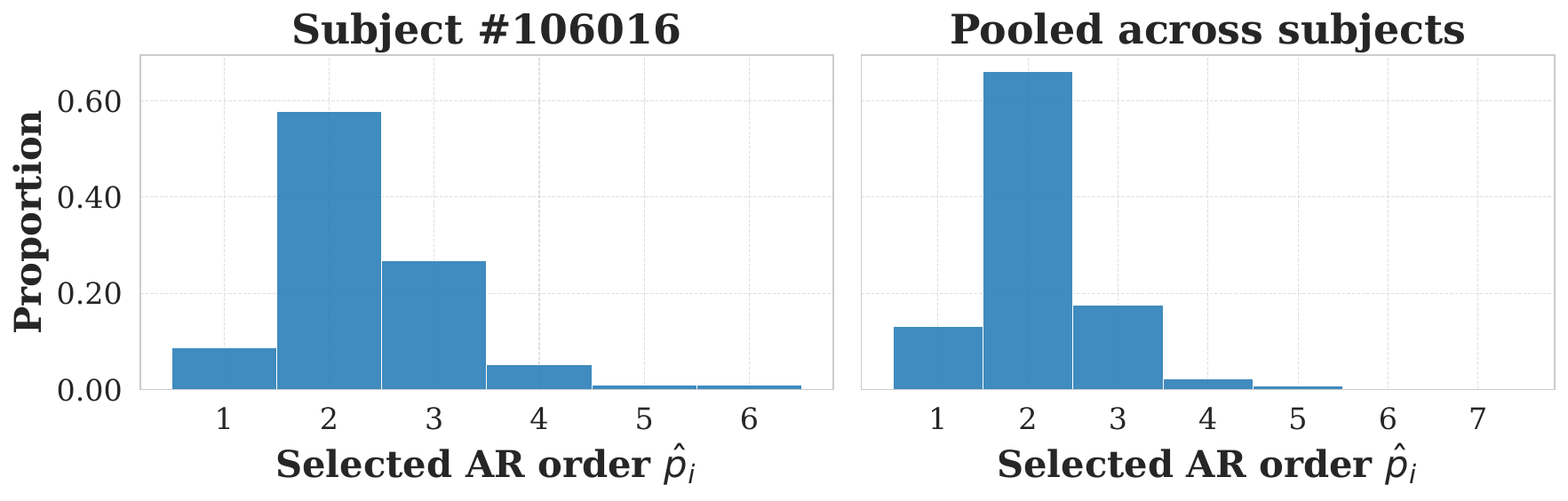}
\caption{\textbf{Distribution of the order of the autoregressive model.} Left panel: distribution of the order of the autoregressive model across the $N=116$ ROIs of the subject $\#106016$. Right panel: distribution of the order of the autoregressive model, pooled across all ROIs of our subjects. Both distributions are concentrated in the range $[2,3]$, with a tail extending to, at most, $\hat{p}_i=7$ - a result confirming that a low-order autoregressive model is powerful enough to whiten our pre-processed BOLD signals, consistently with previous findings~\cite{Woolrich2001,Arbabshirani2014}.}
\label{appC:p_opts_distr}
\end{figure*}

Figure~\ref{appC:p_opts_distr} summarises the outcome of our pre-whitening procedure: the distribution of $\hat{p}_i$ values, pooled across all ROIs of our subjects is concentrated in the range $[2,3]$, with a tail extending to, at most, $\hat{p}_i=7$ - a result confirming that a low-order autoregressive model is powerful enough to whiten our pre-processed BOLD signals, consistently with previous findings~\cite{Woolrich2001, Arbabshirani2014}.

\clearpage

\hypertarget{AppD}{}
\section{Degree distributions of cortical and subcortical regions}
\label{AppD}

\begin{figure}[t!]
\centering
\includegraphics[width=\linewidth]{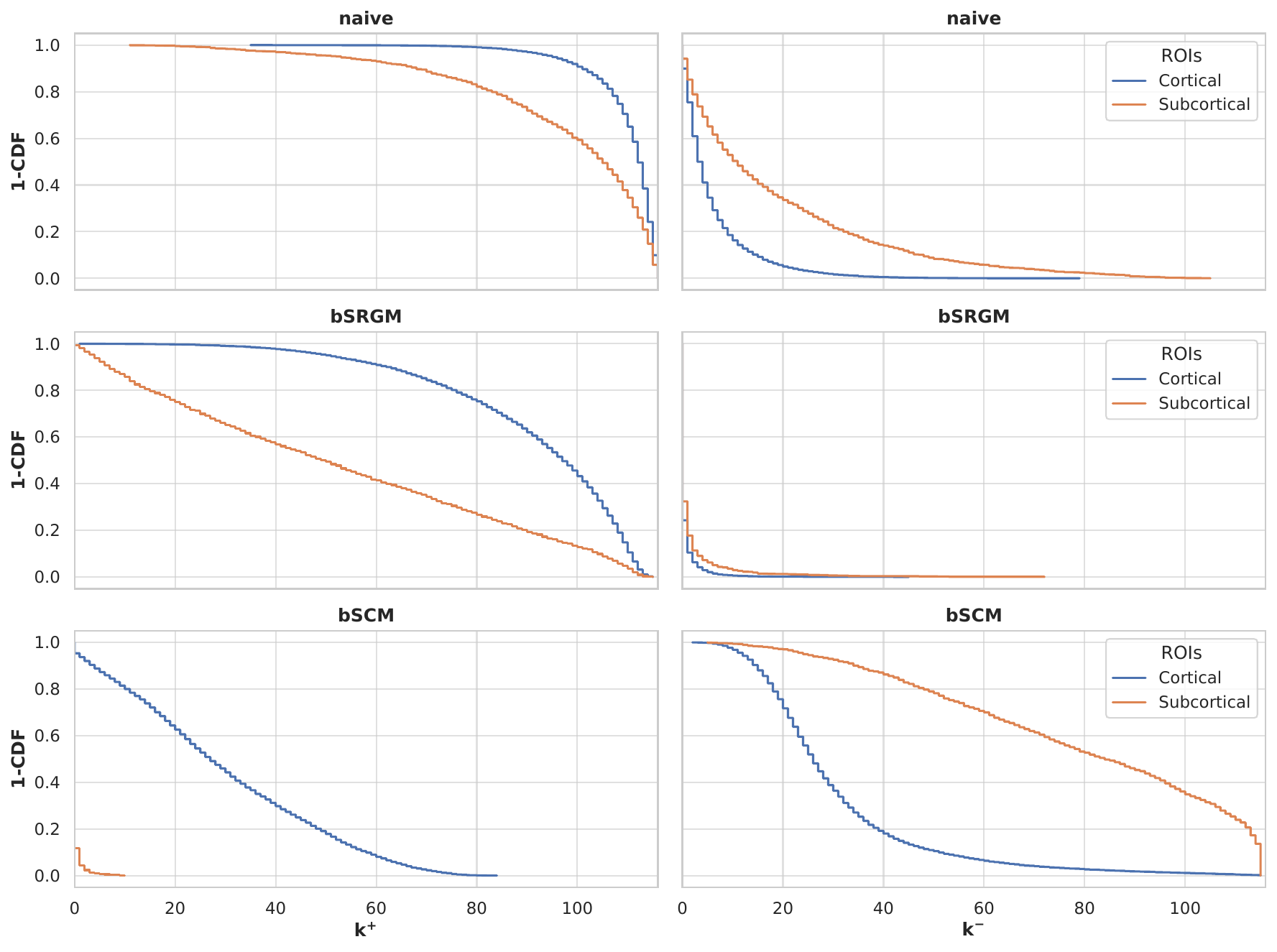}
\caption{\textbf{Cumulative distributions of positive ($k^+$) and negative ($k^-$) degrees for cortical and subcortical regions, per benchmark.} Analyses have been carried out in a pooled fashion (i.e. across the $100$ subjects considered for the present analysis).}
\label{fig:16}
\end{figure}

As anticipated in the main text, and shown in fig.~\ref{fig:16}, SUBC regions tend to have a larger negative degree than cortical regions.

\clearpage

\hypertarget{AppE}{}
\section{A comparison with Pearson's test of hypothesis}
\label{AppE}

\begin{figure}[t!]
\centering
\includegraphics[width=\linewidth]{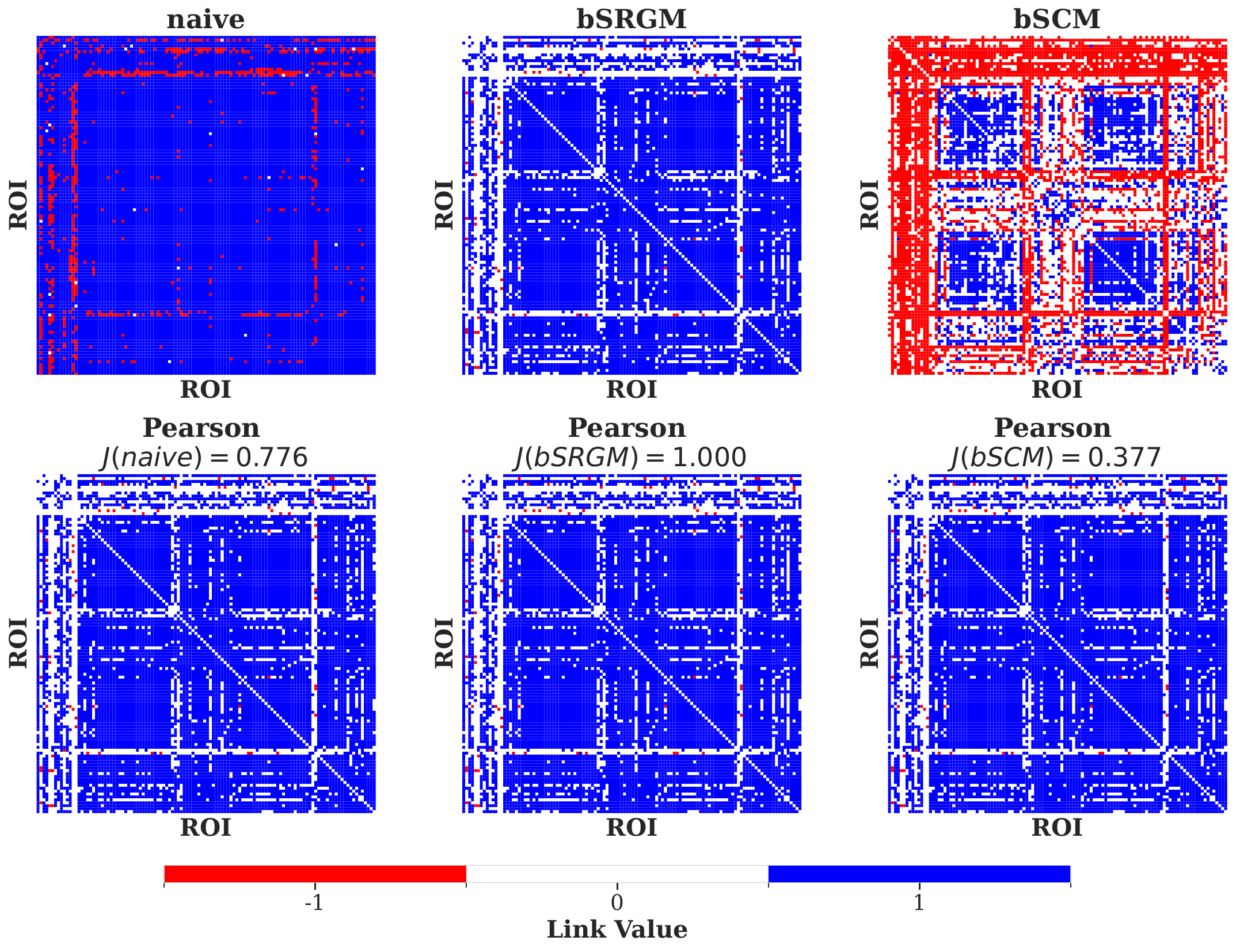}
\caption{\textbf{A comparison between our tests of hypothesis and Pearson's one.} As evident from the panels above, referring to subject $\#106016$, and further confirmed by the value of the Jaccard similarity, employing Pearson's test of hypothesis to project our biadjacency matrices is equivalent at projecting them by adopting the bSRGM as a benchmark.}
\label{fig:17}
\end{figure}

Since the signature $S_{ij}$ retains the sign of the Pearson correlation coefficient between the time series $i$ and $j$, across all subjects, to a large extent, one may, now, wonder the extent to which our tests of hypothesis overlap with the Pearson's one. Let us remind what it states: in the case of zero correlation, $T$ pairs of values from a bivariate normal distribution cause the variable

\begin{equation}
t=r{\sqrt {\frac{T-2}{1-r^{2}}}}
\end{equation}
obey a Student's t-distribution with $T-2$ degrees of freedom; one can, thus, check if the empirical value of the Pearson correlation coefficient, $r$, is significantly different from zero in a two-sided fashion. An alternative route is that of considering the Fisher transformation

\begin{equation}
F(r)=\frac{1}{2}\ln\left[\frac{1+r}{1-r}\right],
\end{equation}
which obeys a normal distribution with $\mu=F(\rho)$ and $\sigma^2=1/(N-3)$: the test concerning the value $\rho=0$ can, thus, be conducted on a single distribution (normal, in this case) as above.\\

Figure~\ref{fig:17} summarises the outcome of such a comparison on the subject $\#106016$: as evident, Pearson-induced projection overlaps to a large extent with the bSRGM-induced one: such a result, probably due to the observed balance between the number of positive and negative signs populating our biadjacency matrices, holds true for all subjects.

\end{document}